\documentclass[submit]{epsv8}

\usepackage{rotating,latexsym}
\usepackage{amssymb}
\usepackage{color}
\usepackage{topcapt}

\title{Dust measurements with the Mars Dust Counter on board \\ Nozomi (PLANET-B)}
\author{
Harald Kr\"uger, MPI f\"ur Sonnensystemforschung, G\"ottingen, Germany; 
 Planetary Exploration Research Center, Chiba Institute of Technology, Narashino, Chiba, Japan, krueger@mps.mpg.de \\
Masanori Kobayashi,  Planetary Exploration Research Center, Chiba Institute of Technology, Narashino, Chiba, Japan \\
Hiroshi Kimura,   Planetary Exploration Research Center, Chiba Institute of Technology, Narashino, Chiba, Japan \\
Tomoko Arai,     Planetary Exploration Research Center, Chiba Institute of Technology, Narashino, Chiba, Japan \\
H{\aa}kan Svedhem, Delft University of Technology, Delft, The Netherlands \\
Sho Sasaki, Graduate School of Science, Department of Earth and Space Science, Osaka University,  Osaka, Japan \\
}

\abstract{
Nozomi was Japan's first  space mission to Mars, launched on 3 July 1998 UT. 
It was equipped with the Mars Dust Counter (MDC) which was an impact ionisation dust detector. 
MDC detected 96 dust particle impacts when the spacecraft was in Earth orbit and later in interplanetary space, before its
operation ended in April 2002 due to a technical failure on board. 
We compare the Nozomi dust measurements with the dust
measurements obtained with the dust detector on board the Ulysses spacecraft. 
Impact speeds and masses of dust particles measured by Nozomi MDC are overall consistent 
with the measurements obtained by Ulysses in the same region of interplanetary space. 
Based on the  impact speeds measured while Nozomi was 
in Earth orbit, MDC detected neither dust particles of natural origin that were bound 
to the Earth nor space debris. 
The dust impact rate measured in interplanetary space varied by approximately a factor of 2, 
consistent with theoretical predictions by the Interplanetary Meteoroid Engineering Model. 
The particle 
impact direction was concentrated towards the ecliptic plane, in agreement with an
interplanetary origin of the majority of the measured dust particles. No impacts of cometary trail 
particles could positively be identified during known cometary trail crossings of Nozomi. 
The Nozomi dust data may become a valuable reference for the dust measurements to be 
obtained in the same region of interplanetary space with future space missions like, for example,  
MMX and DESTINY$^+$.
}


\begin{document}

\maketitle

\section{Introduction}

The Japanese spacecraft Nozomi, initially named PLANET-B, was launched on 3 July 1998 UT, 
heading towards Mars 
\citep{yamamoto1998b}.  The mission's main scientific objective was the study of the 
structure and dynamics of the upper Martian atmosphere and its interaction with the solar wind \citep{nakatani1995}. An additional
objective was the search for a dust ring surrounding Mars \citep{ishimoto1997,sasaki1999b} which had been 
predicted by several authors 
\citep[e.g.,][]{soter1971,hamilton1996b,krivov1997,zakharov2014} but remains as yet undiscovered \citep{showalter2006,showalter2017}. 
More recent studies of the particle dynamics in the Martian rings were presented by \citet{makuch2005,krivov2006,liu2021}. To this end, Nozomi was equipped 
with the Mars Dust Counter (MDC) which was an impact ionisation dust detector \citep{igenbergs1996,igenbergs1998},
with strong heritage from earlier instruments 
\citep[BREMSAT, Hiten, Galileo, Ulysses;][]{igenbergs1991,iglseder1993a,gruen1992a,gruen1992b}. 

Although Nozomi could not fulfil its primary mission goals at Mars due to technical failures on board 
(see Section~2), the MDC 
measurements of cosmic dust in the vicinity of the Earth and the Moon and later in  interplanetary space 
provided new 
scientific results. MDC detected 96 dust particles between 1998 and 2002. For 79 of them the impact speed
and particle mass could be derived from the measured charge signals. Most of the detected particles 
were interplanetary dust orbiting the Sun but MDC also detected several particles of 
interstellar origin \citep{sasaki1999a,sasaki2002a,sasaki2002b,sasaki2007,senger2007}. Finally, an enhanced 
dust impact rate was recorded in November 1998 when the spacecraft -- still being in Earth orbit --  crossed the  
Leonids meteoroid stream.

The objectives of this paper are threefold. First, we want to make the Nozomi dust data set  in electronic 
form available to the scientific community so that it can easily be accessed for further scientific investigations. 
To our knowledge, the Nozomi dust data are not electronically available so far. Similarly,  20 years 
of dust  measurements collected with the DEBris In orbit Evaluator 1 (DEBIE-1) dust impact detector on board the Project for On-Board Autonomy (PROBA-1) 
mission in Low Earth orbit was also recently  reanalysed and became available to the scientific community \citep{azzi2025}.
Second, new modelling tools 
for the dynamics of interplanetary dust have become available since the MDC data were 
collected about 
25 years ago, and we want to compare the predictions of those models with the MDC measurements. 
Finally, we review the Nozomi dust measurements
in view of the upcoming Martian Moons eXploration (MMX) and Demonstration and Experiment of Space Technology for 
INterplanetary voYage Phaethon fLyby and dUst Science (DESTINY$^+$)   
space missions, to be launched in 2026 and 2028, respectively \citep{kuramoto2022,ozaki2022,arai2024}.
We expect that  the Nozomi data will become a valuable reference for both upcoming space missions, given 
that both  of them will be equipped with in situ dust instruments and 
traverse the same regions of space between Earth and Mars as Nozomi did.

In Section~2 we give a brief overview of the Nozomi mission, and in Section~3 we describe the
Mars Dust Counter (MDC) on board the spacecraft. Section~4 gives an overview of the Nozomi dust data set and in
Section~5 we present a new comprehensive analysis of the Nozomi dust data. 
Section~6 is a discussion, and in Section~7 we summarise our conclusions. 

\section{Nozomi Mission}

\label{sec:mission}

	\begin{figure}[tb]
	\centering
	\vspace{-1.cm}
		\hspace{-0.2cm}
		\includegraphics[width=0.9\textwidth]{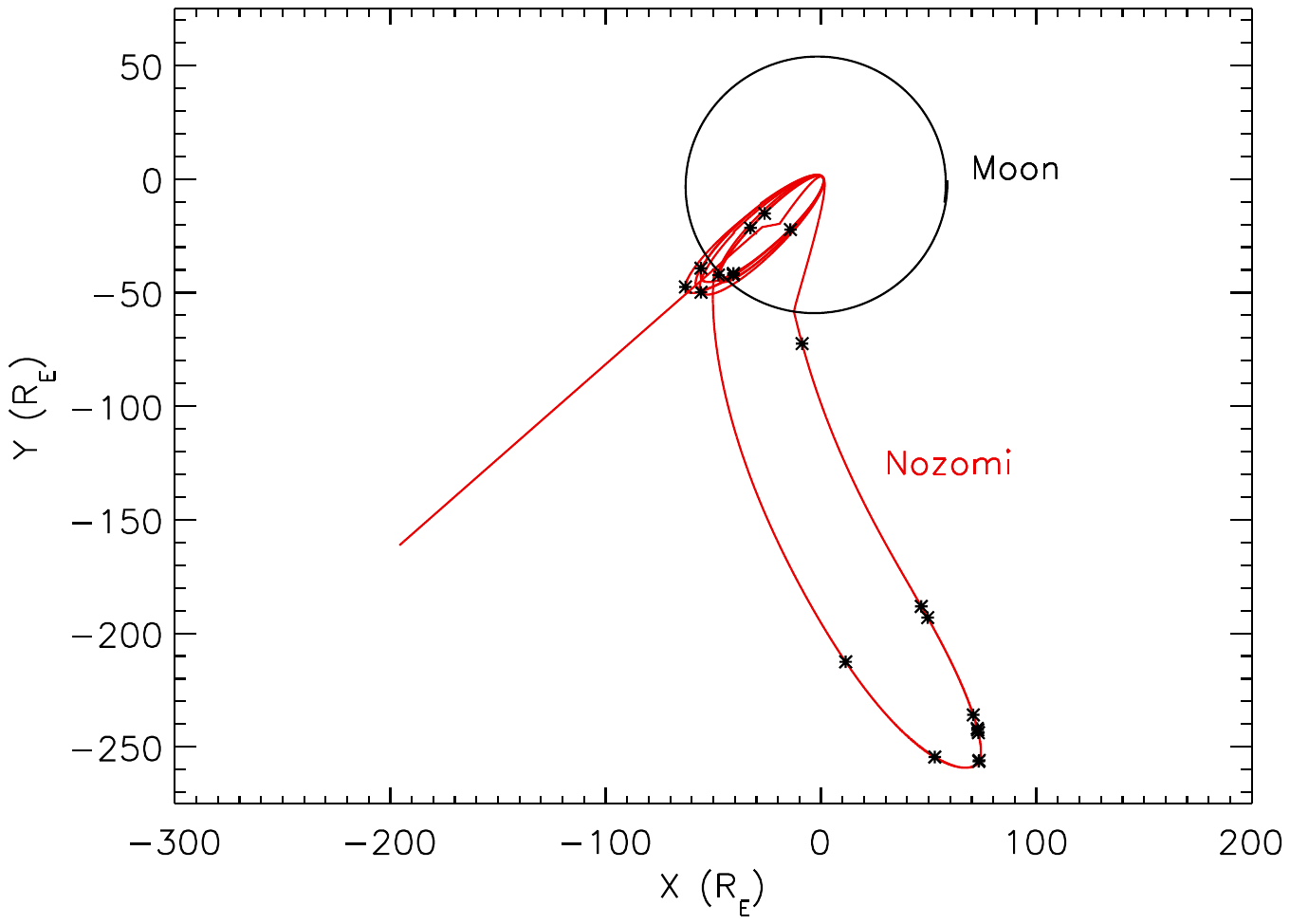}
		\vspace{-1cm}
	\caption{Nozomi trajectory when the spacecraft was in Earth orbit between launch on 3 July 1998 and 25 December 1998 (red), projected onto the ecliptic plane (vernal equinox is to the right). The  orbit of the Moon is shown in black, and the measured dust impacts are superimposed (asterisks). 
	The units are mean Earth radii ($\mathrm{R_E=6,371~km}$). The Nozomi trajectory data were  provided by ISAS/JAXA.
 }
	\label{fig:earthorbit}
\end{figure}

\begin{figure}[tb]
	\centering
	\vspace{-1.cm}
		\hspace{-0.2cm}
		\includegraphics[width=0.9\textwidth]{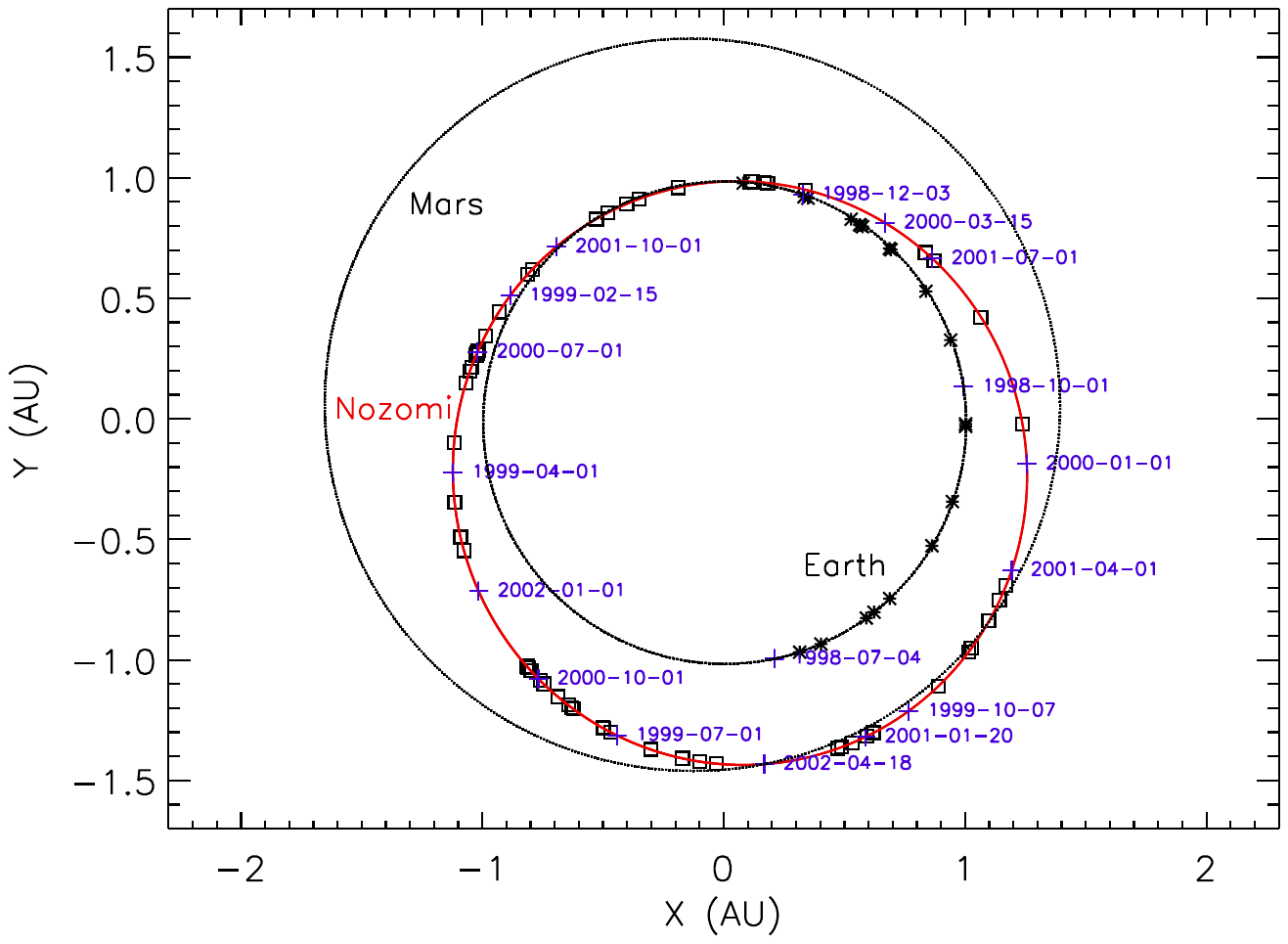}
		\vspace{-1cm}
	\caption{Nozomi trajectory projected onto the ecliptic plane (vernal equinox is to the right). The interplanetary trajectory of the
	spacecraft is shown in red. The measured particle impacts are superimposed for the Earth orbiting phase 
	(asterisks) and for the interplanetary trajectory (squares). The locations of the spacecraft at specific 
	times are indicated in blue. The Nozomi trajectory data were  provided by ISAS/JAXA.
 }
	\label{fig:orbit}
\end{figure}

On 3 July 1998  UT, the  PLANET-B spacecraft was launched at Kagoshima Space Center and renamed to Nozomi, following a Japanese 
tradition. The spacecraft was initially brought onto a highly elliptical geocentric 
trajectory with apogee beyond the lunar orbit \citep{kawaguchi1995}. As the available energy provided by the launcher was 
not sufficient for a direct 
injection into a Mars transfer orbit, a few gravitational maneuvres were necessary. 
The first  swing-by at the Moon took place on 24 September 1998, and brought Nozomi to an apogee distance of 1.7 million kilometres from 
Earth (Figure~\ref{fig:earthorbit}). After a second lunar swing-by, a powered Earth swing-by on 20 December 1998 
was supposed to bring Nozomi onto its final Mars transfer trajectory. Due to a problem in the 
propulsion system 
during the Earth swing-by, however, the spacecraft did not receive enough energy to reach Mars on a 
direct trajectory. Thus,
Nozomi's interplanetary trajectory had to be changed completely \citep{yoshikawa2005},  
and the injection
into Mars orbit -- originally planned for October 1999 -- had to be postponed to December 2003. 
The interplanetary trajectory of the spacecraft is shown in Figure~\ref{fig:orbit}. 
During the interplanetary mission phase the Nozomi orbital plane was close to the ecliptic plane within a few degrees.  

Unfortunately, the mission  suffered 
from additional technical problems. Particularly severe was the damage of a power supply unit which occurred on 24 April 2002.
It was probably caused by a solar eruption which had occurred a few days earlier \citep{forbes2005} and hit the spacecraft. 
As a consequence, Nozomi could not enter Mars
orbit and passed by the red planet on 12 December 2003 at an altitude of 894~km. 
Finally, the mission was 
 declared lost on 10 January 2004 after the exact spacecraft position could not be determined, and a telemetry link  
 could not be established anymore. Nevertheless, Nozomi provided valuable in situ dust measurements in 
 the interplanetary space between Earth and Mars.

\section{Mars Dust Counter (MDC) Overview and Operation}

\label{sec:mdc}

\subsection{Instrument Overview}

\label{sec:overview}

\begin{figure}[tb]
	\vspace{-2cm}
	
	\hspace{-2cm}
	\vspace{-1.0cm}
		\includegraphics[width=1.3\textwidth]{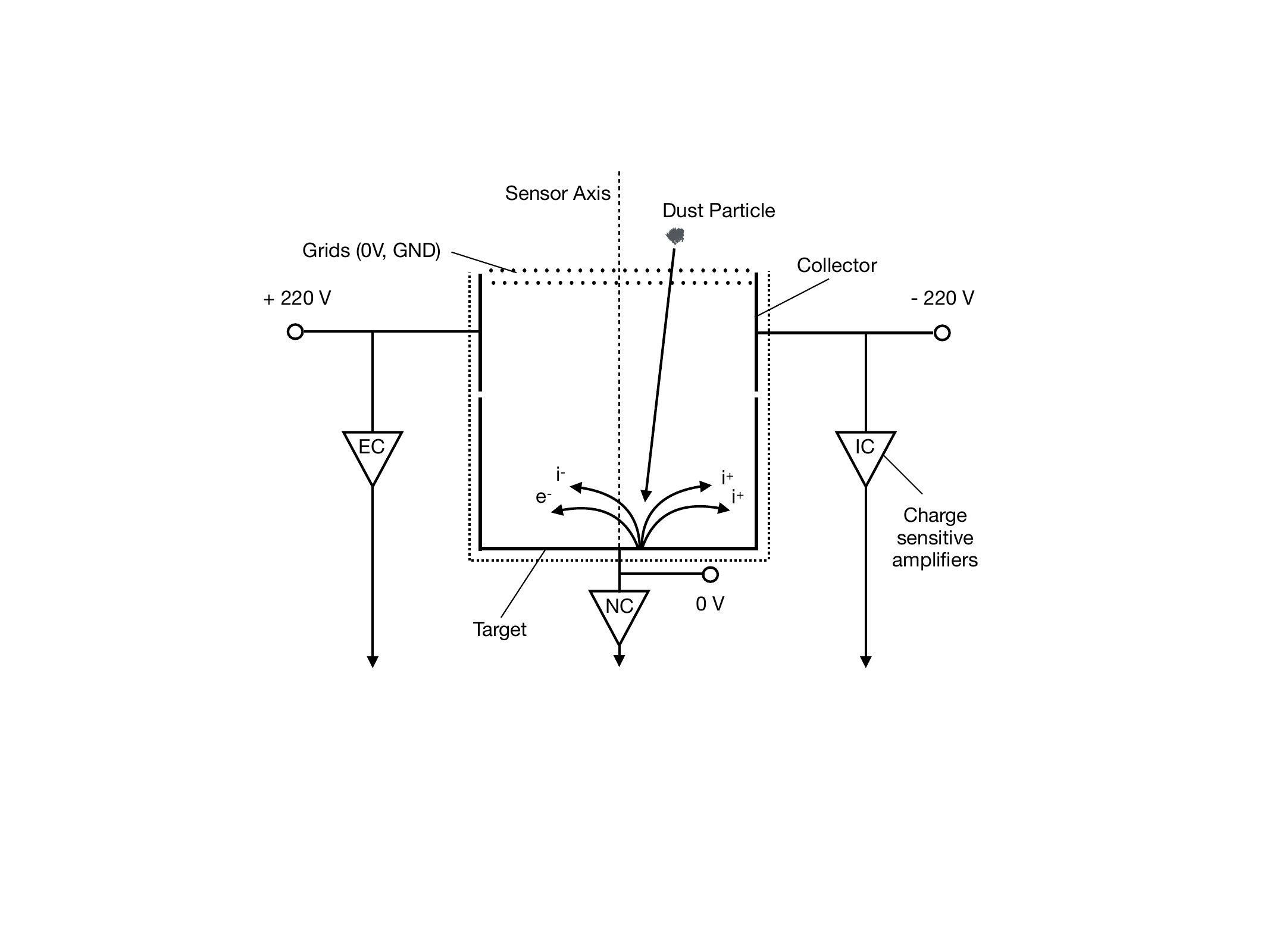}
		\vspace{-4.3cm}
	\caption{Measurement principle of MDC. Impacting dust particles vaporised when hitting the target and produced
plasma around the impact location. The charges of this plasma (negative free electrons and positive ions) were separated by the electric field inside the sensor box. Positive ions were  
	collected by the negatively biased ion channel collector and electrons were  collected by the positively biased electron channel collector.   Three charge sensitive  logarithmic amplifiers converted the measured charges 
	into voltage signals that were transformed into digital data and transmitted to Earth.
 }
	\label{fig:mdc}
\end{figure}

The Mars Dust Counter (MDC) was a light-weight impact ionisation dust detector developed mainly by the 
Technical University of Munich in Germany, and the European Space Research and Technology Centre of the 
European Space Agency \citep[ESA-ESTEC;][]{igenbergs1996,igenbergs1998}. It was an improved version of the dust detectors flown 
on board the BREMSAT and Hiten (MUSES-A) spacecraft which measured dust particles in low Earth orbit and in the Earth-Moon region, respectively 
\citep{igenbergs1991,iglseder1993a,muenzenmayer1997}. The MDC was designed to
determine not only the  flux but also the mass and speed of impacting dust particles  by measuring the ion and the
electron charges produced by high speed impacts onto gold plate targets (impact speed $v\mathrm{_{imp} > 1\,km\,s^{-1}}$). In addition, the particle impact direction could be determined from the 
sensor orientation during the impact. 

A sketch of the MDC is shown in Figure~\ref{fig:mdc}. Dust particles vaporise upon impact when they hit 
the target and produce a plasma cloud. The charges of this plasma 
are separated by the electric field inside the sensor box. Positive ions are 
	collected by the ion collector, and electrons are collected by the  electron  collector.  Up to
	three charge signals are measured for each impact.   
	 Charge sensitive amplifiers convert the charges into voltage signals that are further transformed into 
	digital data and transmitted to Earth. From the measured charge signals and the delay times 
	the location of the impact onto the sensor can be determined and subsequently used to constrain the impact 
	trajectory of the particle.  Finally, the impact speed and mass of the particle are derived from the
	charge signal's rise time, the charge amplitude and the impact position on the sensor box. The uncertainty of  
	the impact speed is a factor of 2  and that of  the mass determination is a factor of 5, respectively 
	\citep{naumann2000}. 
	
Entrance grids shield the instrument from the ambient space plasma to reduce noise signals. 
For impact angles up to $80^{\circ}$ w.r.t. the sensor axis, i.e. w.r.t. the vertical direction in Figure~\ref{fig:mdc}, 
the penetrability of the grids for particles in the size range 
 between $\mathrm{0.1\,\mu m}$ and $\mathrm{10\,\mu m}$ is nearly
constant   at about 92\% to 95\% \citep{senger2007}.

The mass of MDC was only 730\,g and, thus, much  lower  than that of comparable impact ionisation dust detectors 
on board the Galileo and Ulysses spacecraft 
\citep{gruen1992a,gruen1992b}. 
The sensor aperture was an approximate square with dimension 124 x 115 $\mathrm{mm^2}$. 
A detailed description of the sensor was given by \citet{igenbergs1998} 
and \citet{senger2007}, and the calibration of the instrument was described by \citet{naumann2000}. 
A sketch of the Nozomi spacecraft showing the MDC instrument 
is shown in Figure~\ref{fig:nozomi}, and  its detection geometry is  described in Section~3.2.

\subsection{MDC Sensor Pointing Direction}

\label{sec:antenna_direction}

\begin{figure}[tb]
	\centering
	\vspace{-0.5cm}
		\hspace{-0.2cm}
		\includegraphics[width=\textwidth]{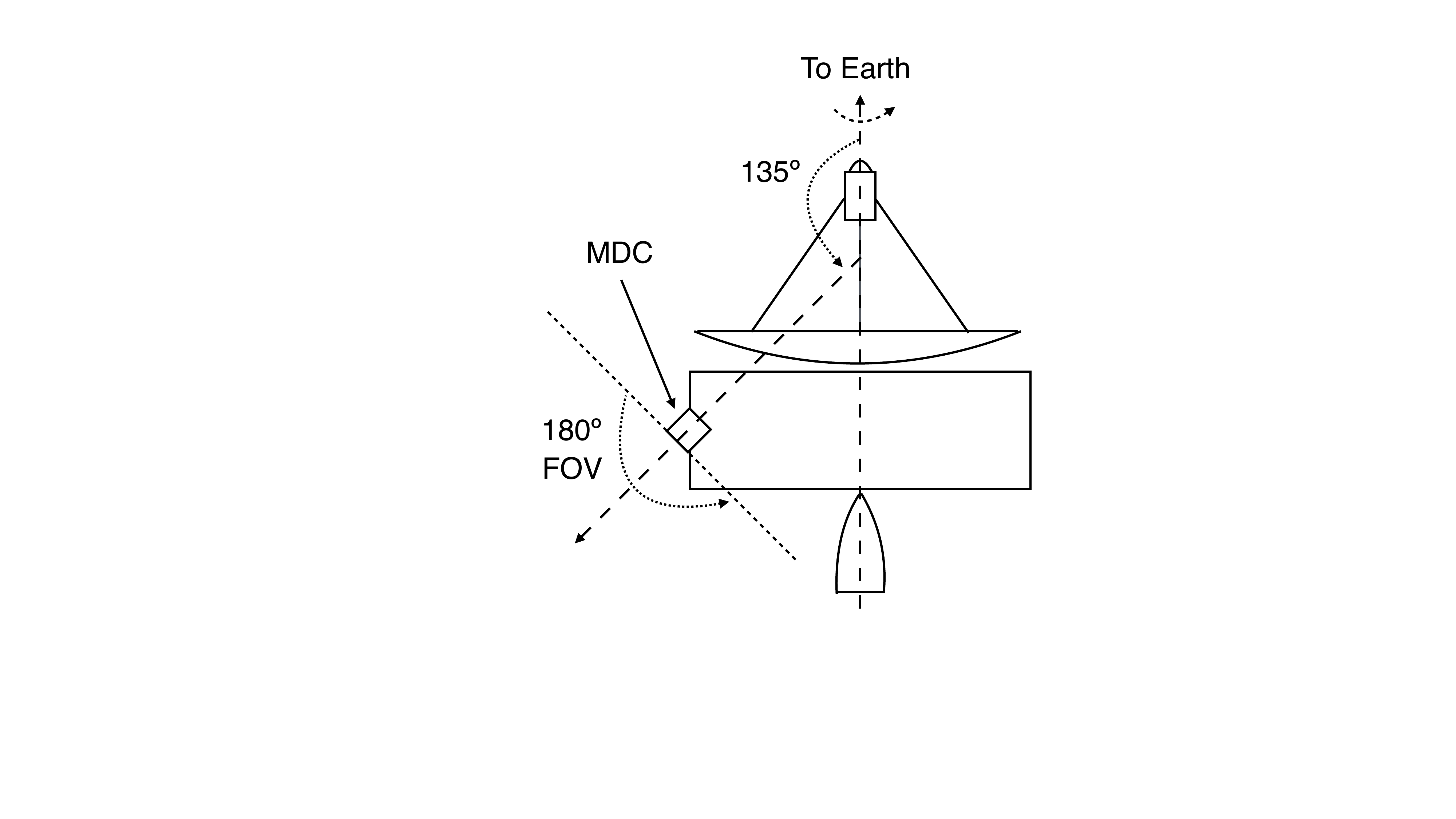}
		\vspace{-3.cm}
	\caption{Orientation of the Nozomi spacecraft and the MDC. The antenna (top) pointed towards Earth most of the time during the interplanetary mission phase, and MDC
	largely faced the anti-Earth hemisphere. The sensor axis had an angle of $135^{\circ}$ from the antenna pointing direction. During 
	one spin revolution of the spacecraft the sensor axis scanned an angle of $90^{\circ}$. MDC itself had almost $180^{\circ}$ 
	field of view (FOV).
 }
	\label{fig:nozomi}
\end{figure}

\begin{figure}[tb]
	\centering
	\vspace{-1.0cm}
		\hspace{-0.2cm}
		\includegraphics[width=0.9\textwidth]{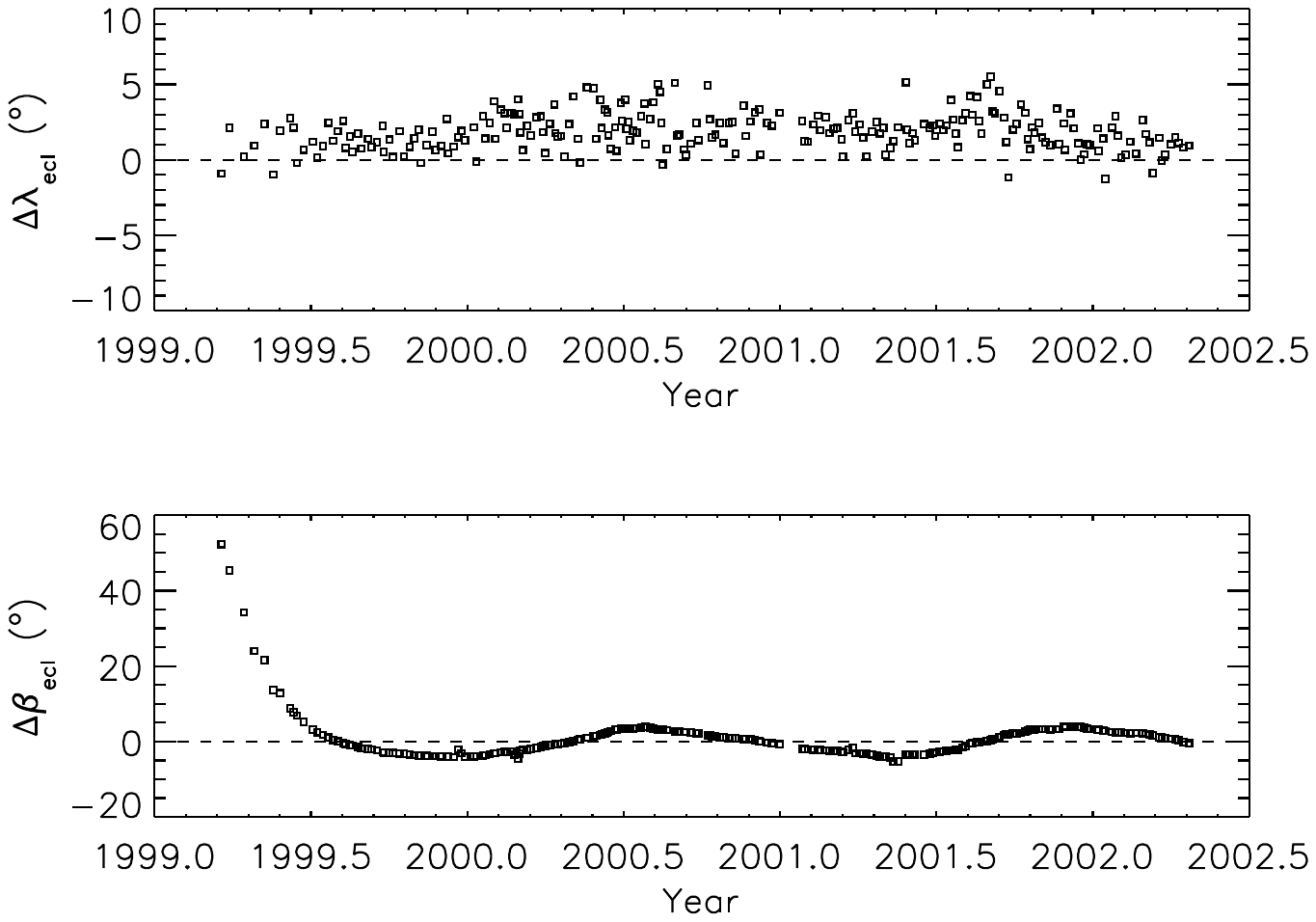}
				\vspace{-1.cm}
	\caption{Spacecraft attitude: deviation of the antenna pointing direction (i.e. positive spin axis) 
 from the nominal Earth direction during the interplanetary mission of Nozomi. The deviations are given in ecliptic 
 longitude $\Delta\lambda_{\mathrm{ecl}}$ (top) and latitude $\Delta\beta_{\mathrm{ecl}}$ (bottom, equinox J2000).
 }
	\label{fig:pointing}
\end{figure}

\begin{figure}[tb]
	\centering
	\vspace{-1.0cm}
		\hspace{-0.2cm}
		\includegraphics[width=0.9\textwidth]{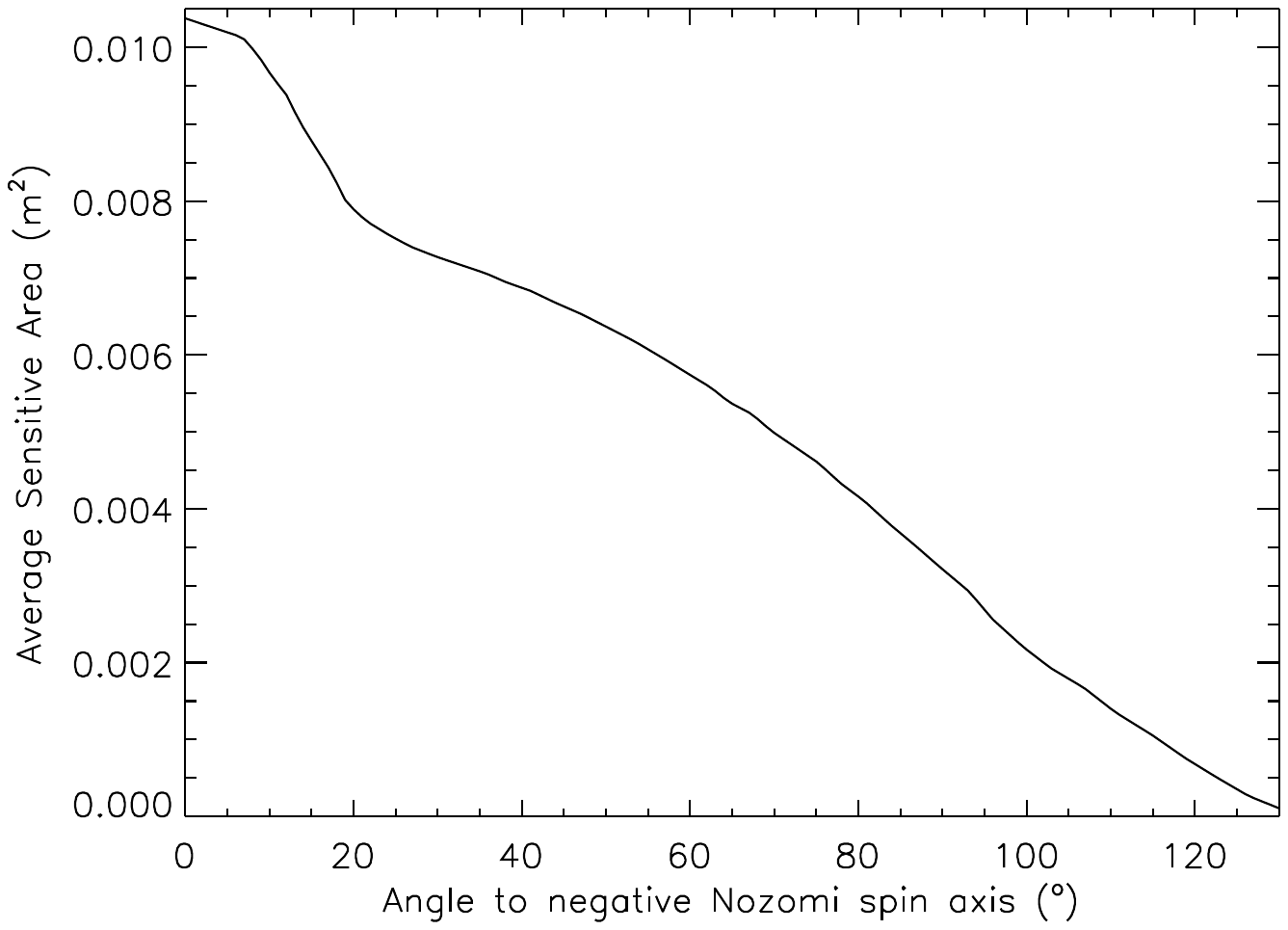}
		\vspace{-1.cm}
	\caption{MDC sensitivity profile during one spin revolution of the spacecraft, adapted from \citet{senger2007}.
	The positive spacecraft spin axis pointed close to the Earth direction during the interplanetary mission phase
	(Figure~\ref{fig:pointing}).
 }
	\label{fig:mdc_profile}
\end{figure}

	A fundamental parameter for the interpretation of  dust measurements is the sensor pointing direction
	during a dust particle impact,
	in addition to the orbital position of the spacecraft. 
MDC was mounted onto the spacecraft body of  Nozomi  at an angle of $135^{\circ}$ from the antenna 
pointing direction, as can be seen in Figure~\ref{fig:nozomi}. 	The spacecraft was spin-stabilised, rotating 
about the antenna axis with a rotation period of approximately 7 to 10 revolutions per minute.
	Most of the time when the spacecraft was in 
interplanetary space, 
the antenna pointed towards Earth for communication purposes so that MDC
	largely faced the anti-Earth hemisphere. Due its large field of view (FOV), the sensor scanned the 
	entire anti-Earth hemisphere during one spacecraft revolution, including the region around the ecliptic poles. 
	
	The antenna pointing direction in ecliptic coordinates is shown in Figure~\ref{fig:antenna_pointing} in the
	Appendix. It shows that during the early mission phase when the spacecraft was in Earth orbit, the 
	antenna pointing 
	varied significantly due to frequent spacecraft attitude changes and orbital maneouvers, and Earth and Moon 
	swing-bys.
	
		In Figure~\ref{fig:pointing} we show the deviation of the spacecraft rotation axis from the nominal Earth 
	direction during the interplanetary mission of Nozomi. Most of the time the axis pointed  within $5^{\circ}$
	of the Earth direction. This rather small deviation is negligible for the considerations and the interpretation of 
	dust measurements in this paper.

	
	\subsection{Detection Geometry}
	
	\label{sec:geometry}
	
MDC was mounted  at an angle of $135^{\circ}$ from the antenna 
pointing direction (cf. Figure~\ref{fig:nozomi}). Thus, the sensor axis scanned an angle of $90^{\circ}$ during 
one  revolution of Nozomi. MDC itself with its box-like shape had almost $180^{\circ}$ 
FOV \citep{senger2007}. 
Due to the rectangular sensor design, the effective sensor area to some extent depends  on the particle impact 
direction. In addition, the spacecraft body,  other instruments and a solar panel mounted close to 
MDC obscured the MDC's FOV. Furthermore, the penetrability of the entrance grid depends on the particle size and the
impact direction (Section~3.1). The effective sensitive area of  MDC must take these obscuring 
structures into account. The  resulting average sensitivity profile of MDC w.r.t. the negative 
spacecraft spin axis, i.e. the anti-Earth direction, is shown in Figure~\ref{fig:mdc_profile}.

\subsection{Instrument Operation}

Soon after launch, the MDC became operational and provided its first data. The failure of the power supply unit on 
24 April 2002  \citep[Section~2;][]{forbes2005} terminated the operation of all scientific instruments on board, 
 including MDC. Nevertheless, the  instrument provided 
dust measurements in the Earth environment and in the interplanetary space between Earth and Mars 
almost continuously for a total of 
approximately 3 years and 8 months.

\subsection{Instrumental Noise and Dead Time}

The MDC recorded a large number of noise events that  may have affected the instrument's detection capabilities, 
given that every single signal triggered and processed by the MDC electronics causes some instrument dead time 
and may therefore affect the analysis of the dust data. The effects and characteristics of the recorded noise rate and noise signals are discussed in detail by \citet{senger2007}. Here we give only a brief summary. 

The noise events were either  discarded by the flight software 
onboard and can only be recognised by the 
instrument's noise counter, or  passed the onboard noise recognition scheme and can be seen in the  data 
transmitted to Earth. 
Many of these noise signals were not shaped randomly as it is expected for noise caused by random 
electromagnetic background from different possible sources like radiation or the spacecraft's electrical systems. 
Rather, several nearly identical signal shapes were identified which occur in certain patterns. In addition to the
signal shape, the coincidences, i.e. delay times, between individual charge signals as well as the charge amplitudes 
are critical parameters for the separation between dust impacts and noise. 

A significant source of noise was due to the release of electrons caused by photo-ionisation at 
the inner MDC sensor box surfaces due to solar ultraviolet photons. It was a function of 
the spacecraft orientation and occurred when the Sun was illuminating the inner sensor box.  The Sun
could shine into the sensor box when 
the angle between the Nozomi rotation axis and the direction towards 
the Sun was larger than $45^{\circ}$ (cf. Figure~\ref{fig:nozomi}).
This happened mostly early in the mission when the spacecraft was orbiting the Earth,  while for most  
of the interplanetary mission after spring 1999, 
this angle was below $45^{\circ}$, and the noise usually remained low. 

Nevertheless, several  additional periods of very high 
noise rates occurred during the  interplanetary mission  phase when the Sun 
could not illuminate the inner sensor walls. 
The origin of these sporadic interferences remains unclear. 
Solar eruptions were considered as possible sources since such events cause very high proton fluxes 
and, therefore, may be recorded as noise events  by MDC, similar to the one that occurred on 24 April 2002 and was the 
likely cause of the Nozomi failure. However, no additional coincidences between periods of high
noise rates recorded by MDC and other known solar eruptions could be found. Other potential 
sources of strong noise  like the Nozomi electronics itself were not examined due to missing information  \citep{senger2007}.

Given that the high noise rates may have affected 
the detection capabilities for dust impacts, \citet{senger2007} determined 
the total detector dead time during the most significant  noise periods. 
The highest  detector dead time derived for the entire mission including the Earth orbiting phase 
reached a value of 14.5\% and occurred only 
once, namely  on 8 November 1999 for 
about six hours. During the other  high noise periods the detector dead time varied between less than 
1\% and nearly 6\%  (these values occurred during the time periods 6 to 14 June 1999, 8 to 28 Nov. 1999, 
22 Aug. to 28 Nov. 2000 and 16 Aug. 2001, respectively). During the remaining more then 90\% 
of the mission time  the 
overall dead time was significantly below 1\%. In summary, the dead time caused by high noise rates is 
 negligible for our analysis. 
 
 Even though the noise rates were very high during some intermittent 
 periods and the total number of  noise events detected during the entire mission exceeded 20,000, it is 
 expected that the number of
 erroneously discarded real dust impacts is low (see also Section~4.2). Unfortunately, 
 this value cannot be quantified because the 
 original raw data of MDC and the noise events discarded by the noise identification scheme are not available 
 anymore. Only the processed dust data presented in Tables~\ref{tab:mdcdata1} to \ref{tab:mdcdata3} are still 
 available today.

\section{Dust Impact Events}

\label{sec:data}

In this Section we give an overview of the dust measurements of MDC on board Nozomi. We extracted the 
  MDC dust data set  from the thesis of \citet{senger2007}. The data set is given in 
  Tables~\ref{tab:mdcdata1} to \ref{tab:mdcdata3} in the Appendix (R. Senger's thesis 
  also contains graphs showing the charge signal curves measured for each identified dust impact which were not 
  extracted from the thesis and not digitised). 
  An electronic version  of the MDC data is available at 

  \begin{verbatim} http://dpsc.perc.it-chiba.ac.jp/data/publication/krueger2026/krueger2026_NOZOMI_MDC_data.txt,\end{verbatim}
  and can also be obtained from the first author of this publication upon  request.

\subsection{Dust detections in Earth Orbit}

The Earth orbiting phase of Nozomi lasted from launch on 3 July 1998 UT until the Earth swing-by on 20 December 1998. About 10,000 signals were triggered by the MDC electronics, recorded, 
qualified by the on-board software and transmitted to Earth \citep{senger2007}. 20 good signals were identified as dust impacts by manual screening. 
For one of them  a correct determination of the charge rise time and  amplitude was not possible and, thus, no impact speed 
and particle mass could be derived.
Figure~\ref{fig:earthorbit} shows the Nozomi orbit during the Earth-orbiting phase with superimposed  
positions where the  particle impacts were detected. Note that all particles were detected far away from 
the Earth near the apogee of Nozomi. 
This will be discussed further in Section~5. 

\subsection{Dust Detections in Interplanetary Space}

\label{sec:interplanetary}

After leaving the Earth orbit on 20 December 1998, Nozomi was on a heliocentric orbit 
with perihelion at 1.0~AU and aphelion  at 1.44~AU, close to Mars' orbit. The MDC was in operation during the entire mission 
time in  interplanetary space until the 
system failure occurred on 24 April 2002 (cf. Section~2). Nozomi completed 3 1/2 orbits around the Sun and detected 
interplanetary and interstellar dust particles. About 11,000 signals were 
triggered, recorded, qualified and  their data sets downloaded during the interplanetary mission phase. 

In order to separate noise events from potential  signals of real dust impacts, 
a neural network was used for pattern recognition. It was used  to create a procedure that recognises 
real impact signals and separates them from noise signals produced by the MDC sensor. The input parameters must represent 
the overall characteristics of the signal in a unique way. For an optimized  performance of the pattern recognition process,  
the number of input parameters should be minimized and  any redundancies should be avoided. The network itself should 
be as slim as possible in order to reduce the required computing power, and the number of events that are falsely identified 
as real impacts should be minimal. The desired output for the 
given problem reduces to a simple ''yes/no`` answer, i.e. either the presented signal is identified as a real impact signal or not.
As input parameters, the signal itself or the whole set of digitized data, respectively, was not concerned as a practical 
approach since that would require more than 800 input neurons and lead to a very huge neural network. Instead,
a set of 56 parameters that represent the characteristics of the signal was used as initial set of input parameters. These parameters 
include a reduced set of data points of the signals, and calculated parameters like the mean pre-trigger or total amplitude of 
the signal.
In a first approach, the network was designed quite large, using all the 56 input parameters that describe the signal shape.
The  parameters were subject to optimisation that was performed by implementing different possible networks and evaluating their classification capabilities.
 Although this network showed only a weak classification performance, it was used as a basis for the following optimizations.  
 The number of input parameters was reduced to the 16 most significant parameters and 2 output parameters. There were several 
 different topologies of the neural network which could be conceived. All possible and reasonable topologies were implemented, 
 trained and validated  in order to carry out a systematic determination of the best performing network topology. As the 
 same network shows different performance for each new created instance, 5 runs for each topology were made to get a significant average. 
A training goal of 0.01, i.e. an accuracy of higher than 99\% in classification performance, was reached. 
\citep[see][for details]{senger2007}. In total, 76 dust impacts were extracted from 
 the data measured in
interplanetary space. For 16 of them no speed and mass determination was possible. 
The locations where the particles were recorded during Nozomi's 
interplanetary mission are shown in Figure~\ref{fig:orbit}.

\section{Analysis}

\label{sec:analysis}

In this Section we perform an analysis of the Nozomi MDC dust measurements. We concentrate on the 
instrument's raw data measured for each detected dust  impact, i.e. particle impact speed, 
mass, impact direction, and derived dust flux. We also perform a comparison with interplanetary dust
detections obtained with the Ulysses spacecraft \citep{krueger2010b}. Our analysis presented here 
supplements the work presented by \citet{senger2007}. 

\subsection{Dust Mass versus Impact Speed}

\label{sec:mv}

\begin{figure}[tb]
	\vspace{-3.0cm}
	
		\hspace{-1.2cm}
		\includegraphics[width=1.4\textwidth]{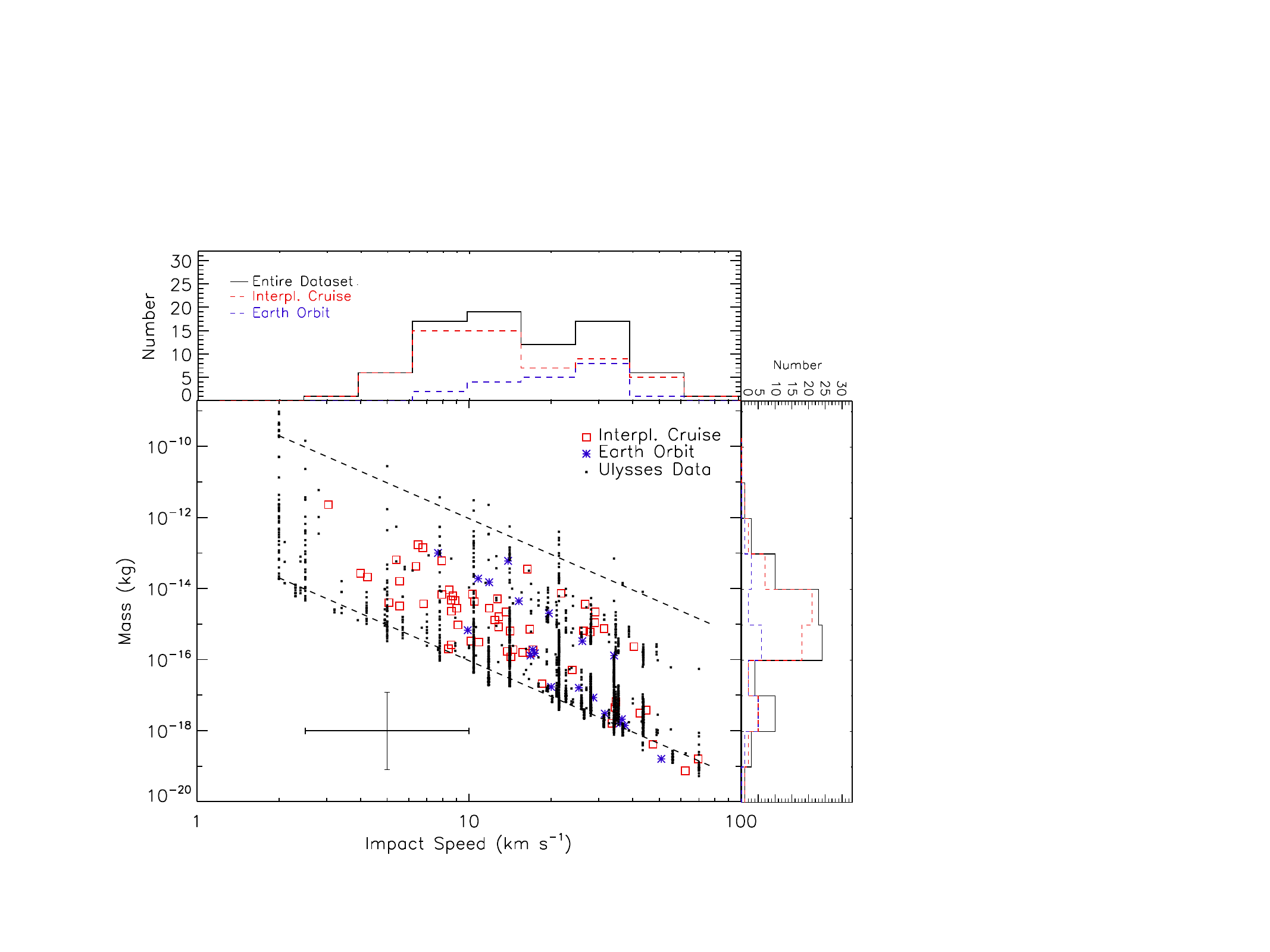}
		\vspace{-2cm}
	\caption{Particle mass vs. impact speed of the dust particles measured with MDC. Ulysses data are shown for comparison \citep{krueger2010b}. The detection limits for MDC are shown as dashed lines. They are mean values for all impact positions inside the MDC sensor box
	\citep{senger2007}. A sample error bar is shown that indicates a factor of
2 uncertainty for the impact speed and a factor of 10 for the mass determination.  Projections of the MDC data onto the mass and speed 
axes are shown as histograms (see also Figures~\ref{fig:speed_distribution} and \ref{fig:mass_distribution}).
 }
	\label{fig:mv}
\end{figure}

In Figure~\ref{fig:mv} we show the measured impact speeds and particle masses for all 79 dust particles 
 with available impact speed and mass measurement identified 
in the MDC data set \citep{senger2007}. Impact speeds occur over almost the entire calibrated range of the
instrument from $\mathrm{3\,km\,s^{-1}}$ to $\mathrm{70\,km\,s^{-1}}$. The particle masses vary over more than 7 orders of magnitude from 
approximately $\mathrm{10^{-12}\,kg}$ to $\mathrm{10^{-19}\,kg}$. The uncertainties are a factor of 2 
for the speed and a factor of 5 for the mass \citep{naumann2000}. 

For comparison, we also 
show the data measured with the  dust detector onboard the Ulysses spacecraft 
which was an impact ionisation detector of a similar design \citep{gruen1992b}. Because the 
sensitive area of the Ulysses 
detector was by a factor of ten larger than that of the MDC, and the operation time 
was much longer, Ulysses recorded a much higher number of particle impacts. From 16 years of
dust measurements between 1990 and 2007, the data sets of 6719 dust impacts onto the Ulysses sensor 
were transmitted to Earth \citep{krueger2010b}. 

The clustering in the speed values of the Ulysses data
is due to discrete steps in the charge rise time measurement but this quantisation is much smaller than the 
uncertainty in the speed measurement (due to the discrete digitisation steps in the charge and the rise time
measurement of the Ulysses instrument, each dot in Figure~\ref{fig:mv} usually corresponds to more than one dust impact.). 
Figure~\ref{fig:mv} shows that at high impact speeds above approximately $\mathrm{20\,km\,s^{-1}}$  the dust instruments on board 
both spacecraft had similar sensitivities for dust impacts. For lower speeds the Ulysses detector was a factor of 
5 to 10 more sensitive in particle mass. It is also obvious that the mass-velocity dependence of the MDC was 
somewhat steeper  than that of the Ulysses detector, as can be seen when comparing the lower boundaries 
of the Ulysses and the MDC data in Figure~\ref{fig:mv}.  These variations may be related to
differences in the instrument electronics, detector geometry or other reasons. The details, however, cannot be investigated 
anymore 30 to 40 years after the MDC and the Ulysses dust instruments were manufactured.

The 20 particles detected by MDC in Earth orbit with measured impact speed and 
particle mass are highlighted as asterisks in Figure~\ref{fig:mv}. The lowest 
impact speed was $\mathrm{7.7\,km\,s^{-1}}$, and the highest was $\mathrm{50.9\,km\,s^{-1}}$. 
From the measured impact speed and Nozomi's orbital and attitude data,
\citet{senger2007} derived the heliocentric speed of all  detected particles (Tables~\ref{tab:mdcdata1} 
to \ref{tab:mdcdata3}). 
For the particles measured in Earth orbit,  
the derived heliocentric speeds range from 17 to $\mathrm{67\,km\,s^{-1}}$. This is 
much higher than the escape speed from Earth's gravitational field at the geocentric distance 
where the particles were measured. 
Therefore, all particles detected 
during the first half year of the Nozomi mission must have been unbound to the Earth system and, thus, 
be of interplanetary or interstellar origin. This is in agreement with
the dust measurements by Hiten during its three year mission in the Earth-Moon system, which 
was equipped with a similar dust sensor \citep{iglseder1993a,svedhem1996}. 
Thus, the Nozomi MDC neither
detected dust particles of natural origin that were bound to the Earth nor  manmade space debris.

\subsection{Dust Impact Rate}

\label{sec:rate}

\begin{figure}[tb]
	\centering
	\vspace{-1.cm}
		\hspace{-0.2cm}
		\includegraphics[width=0.9\textwidth]{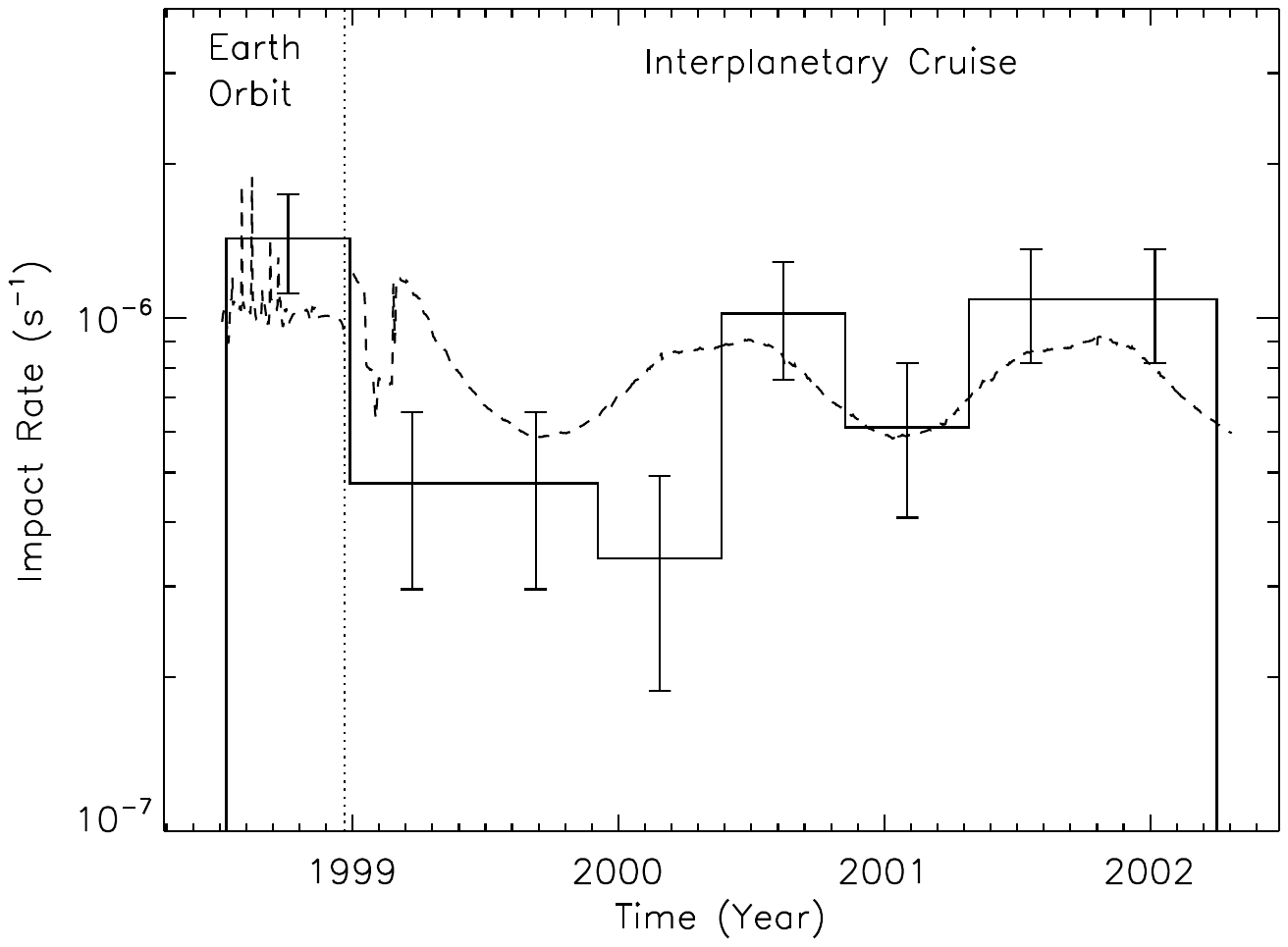}
		\vspace{-1cm}
	\caption{Dust impact rate measured by Nozomi MDC. The second Earth flyby on 20 December 1998 
	and subsequent entry into the 
	interplanetary trajectory is shown as a vertical dotted line. The error bars indicate the $\sqrt N$ variation with 
	$N$ being the number of dust impacts in the respective time bin. The dashed line shows the approximate 
	flux of interplanetary dust particles as derived from the IMEM model, see text for details.
 }
	\label{fig:rate}
\end{figure}

The dust impact rate detected by the Nozomi MDC from July 1998 until April 2002 is displayed in 
Figure~\ref{fig:rate}. The highest overall impact rate of approximately $\mathrm{1.5\times 10^{-6}\,s^{-1}}$ 
was detected during the Earth orbiting phase between July and December 1998. During the initial 
interplanetary mission phase  in 1999 and early 2000 the impact rate was about a factor of 3 lower,  
increased to about $\mathrm{1\times 10^{-6}\,s^{-1}}$ again in mid-2000, and stayed rather constant at this level 
until the end of the MDC operations in April 2002. 

In order to compare the Nozomi dust measurements with theoretical predictions by an interplanetary dust model, 
we simulated the Nozomi  measurements with the Interplanetary Meteoroid Engineering Model 
\citep[IMEM;][]{dikarev2005a,dikarev2005b}. We  assumed the  MDC sensor profile shown in 
Figure~\ref{fig:mdc_profile}, and the Nozomi antenna axis
pointing shown in Figure~\ref{fig:antenna_pointing} in the Appendix. Given that the spacecraft rotation
axis pointed close to the Earth direction after mid 1999, the MDC  scanned the anti-Earth hemisphere
 most of the time in this period. 

The dashed line in Figure~\ref{fig:rate} shows the expected dust impact rate for Nozomi 
MDC derived from the IMEM model. For the
simulations we assumed a detection threshold of MDC of $\mathrm{1.5\times10^{-16}\,kg}$  in order to
get a reasonable fit of the measured  impact rate. It corresponds to an
average dust impact speed of approximately $\mathrm{10\,km\,s^{-1}}$ (cf. Figure~\ref{fig:mv}), and it
is in overall agreement with the impact speeds measured by MDC (Figure~\ref{fig:speed_distribution}). The
upper limit of the  size range considered in the simulations is  $\mathrm{10^{-2}\,kg}$. Due to the 
steep dust size distribution 
falling off with increasing particle size,  the  choice of this latter value has no influence on the 
modelling results.  In order to assess the influence of the selected minimum mass on the predicted flux, 
we  ran the IMEM model with different mass thresholds. This showed that a factor of 2 increase or decrease 
in the lower mass cutoff changes  the average predicted flux by a factor of 1.3, a factor of 5 change 
in the mass cutoff changes the flux by a factor of 2, and a factor of 10 change in the mass cutoff leads to a 
factor of 2.6 change in the flux, respectively. The simulations were performed with a time step of 1 day, and
gravitational focussing of interplanetary dust by the Earth was taken into account in 1998 when Nozomi 
was in Earth orbit.

The strong fluctuations in the
simulated  impact rate in 1998 are caused by frequent changes in the spacecraft attitude.  On the other hand,
the variation of the  dust impact rate predicted by the IMEM model for the interplanetary mission by almost a factor of two 
is mainly caused by the variation in the dust  spatial density and impact speed between the orbits of 
Earth and Mars, and  by the changing sensor orientation w.r.t. the dust flow. 

The impact rate
predicted by the  model is in overall agreement with the MDC measurements, although the model
predicts a smaller variation of the  impact rate than is actually seen in the data. Here the relatively low
number of measured particles  in each time bin may have an influence. The measured
impact rate and the IMEM model will be discussed further in Section~6.1. 

\subsection{Impact Speed Distribution}

\label{sec:speed}

\begin{figure}[tb]
	\centering
	\vspace{-1.0cm}
		\hspace{-0.2cm}
		\includegraphics[width=0.9\textwidth]{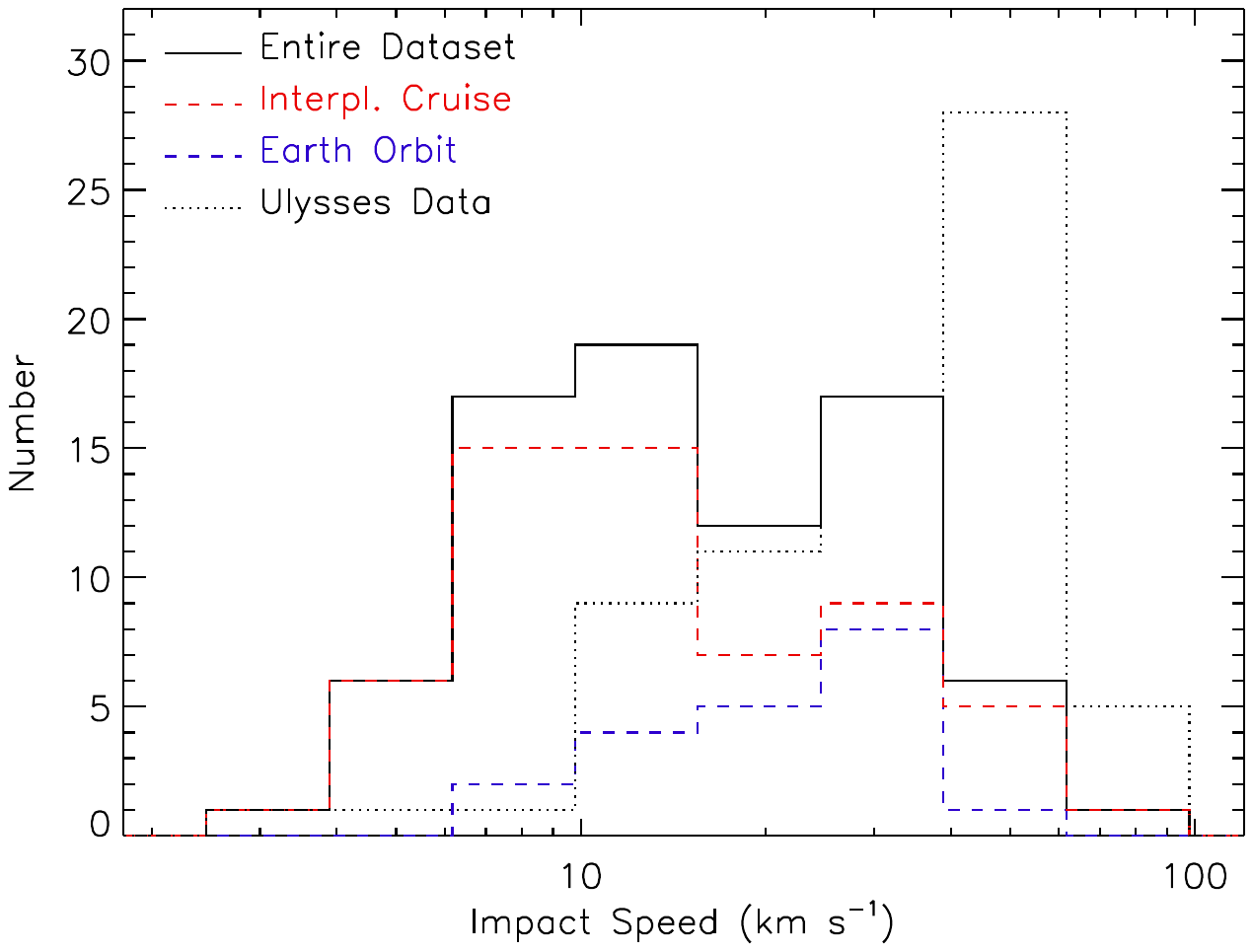}
		\vspace{-1cm}
	\caption{Impact speed distribution for the dust impacts measured by Nozomi MDC. The speed distribution measured 
	with the Ulysses dust detector in 1990 between 1 and 1.45~AU (73 dust impacts)
	is shown for comparison.
 }
	\label{fig:speed_distribution}
\end{figure}

In Figure~\ref{fig:speed_distribution} we show the impact speed distribution of the dust impacts 
measured by Nozomi MDC. The average impact speed of the 20 particles measured in Earth orbit is 
approximately $\mathrm{23\,\pm 11\,km\,s^{-1}}$,   
while that of the 80 particles detected in interplanetary space is about $\mathrm{19\,\pm 15\,km\,s^{-1}}$. 
Given that the motion of particles detected in Earth orbit is affected by 
the gravitational attraction of the Earth, they should have  higher average impact speeds than
particles
detected in interplanetary space.  Thus, the increased measured impact speeds in Earth orbit are 
in qualitative agreement with this expectation, although this is not statistically significant.

Figure~\ref{fig:speed_distribution} shows a few impacts with speeds exceeding approximately 
$\mathrm{40\,km\,s^{-1}}$. The escape speed from the gravitational attraction of the Sun at
the heliocentric distance of Nozomi (1.0 to 1.44~AU) ranges from $\mathrm{42\,km\,s^{-1}}$ to $\mathrm{35\,km\,s^{-1}}$.
Thus, based on the measured impact speeds, these high speed particles are good candidates for 
particles being unbound to the solar system. We will come back to this in Section~6.2.

Figure~\ref{fig:speed_distribution} shows the speed distribution measured with the Ulysses dust detector
for comparison \citep{gruen1995c}. We only show the  Ulysses data measured in the distance range between 1 and 1.45~AU
in the year 1990. (We used 73 Ulysses detections between launch on 6 October and 20 December 1990 when the
spacecraft was in the heliocentric distance range between 1~AU and 1.45~AU, with measured impact speeds and 
particle masses  and 
a velocity error factor $\mathrm{VEF < 10}$ \citep{krueger2010b}.). The average impact speed of the 73 particles measured 
with Ulysses in this time interval 
is $\mathrm{36\,\pm 17\,km\,s^{-1}}$ and, thus, 
much higher than the impact speeds measured with Nozomi MDC in a comparable region of space (60 particles with 
$\mathrm{19\,\pm 16\,km\,s^{-1}}$). The average impact speed predicted by the IMEM model for the MDC measurements 
 is about 
$\mathrm{12\,km\,s^{-1}}$, in agreement with the measurements. This is 
consistent with the fact that Ulysses, heading towards Jupiter on a direct trajectory, traversed 
this region of space with a much higher speed than Nozomi which had its aphelion at Mars' orbit,  leading to higher 
impact speeds in the case of Ulysses.

\subsection{Mass Distribution}

\begin{figure}[tb]
	\centering
	\vspace{-1.0cm}
		\hspace{-0.2cm}
		\includegraphics[width=0.9\textwidth]{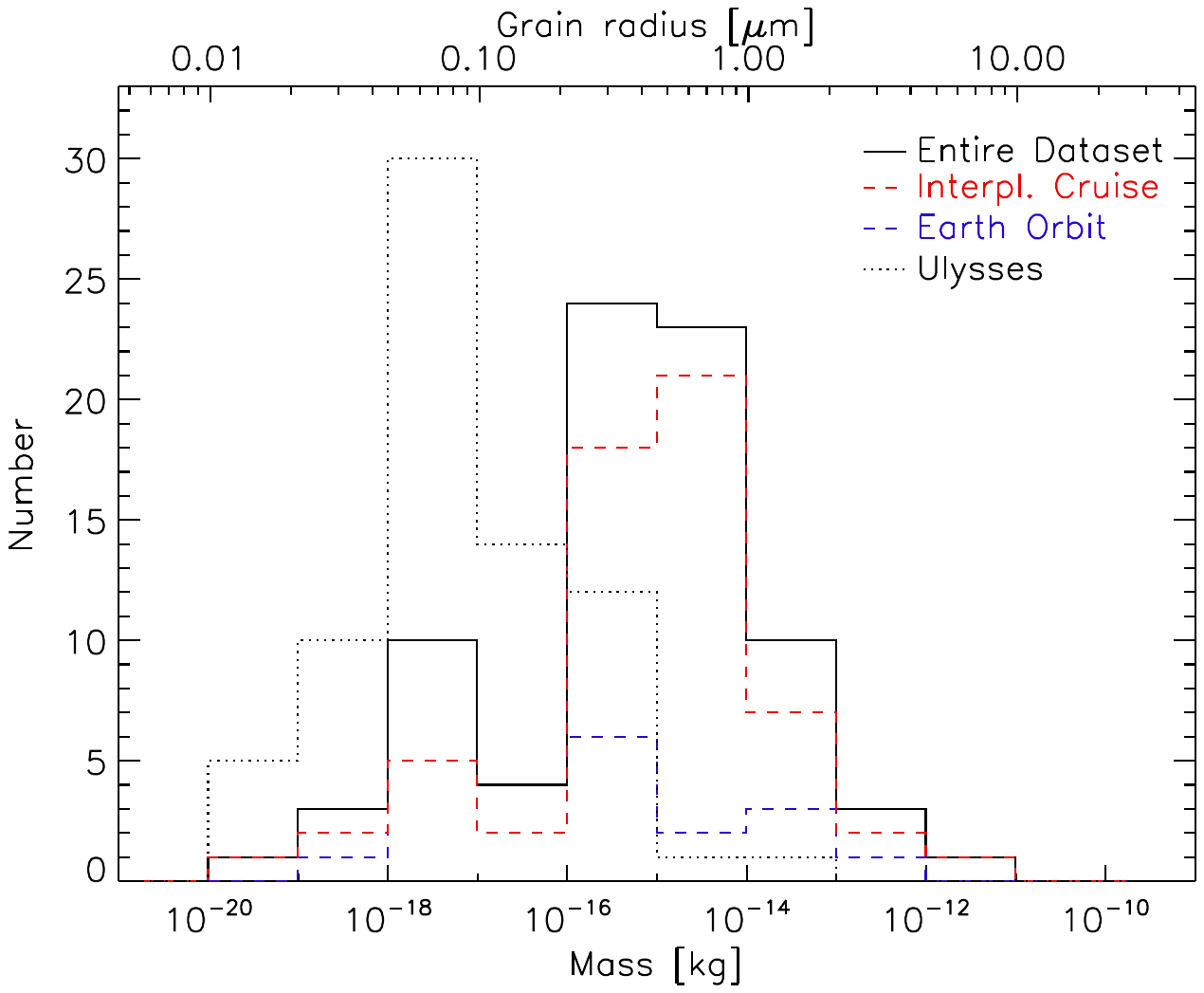}
		\vspace{-1cm}
	\caption{Mass distribution for the dust impacts measured by Nozomi MDC. The mass distribution measured 
	with the Ulysses dust detector in 1990 between 1 and 1.45~AU (73 dust impacts)
	is shown for comparison. The corresponding radius for spherical particles with density 
	$\mathrm{2,500\,kg\,m^{-3}}$ is shown at the top.
 }
	\label{fig:mass_distribution}
\end{figure}

In Figure~\ref{fig:mass_distribution} we show the mass distribution of all 79 particle impacts for which
the impact speed and particle mass are available, again separate for the Earth orbiting phase and the
interplanetary mission phase of Nozomi. Below approximately $\mathrm{10^{-15}\,kg}$ the
mass distribution is incomplete because the detection threshold varies with the impact speed (see Figure~\ref{fig:mv}). 
High impact speeds lead to a detection threshold shifted towards smaller particles and vice versa. Furthermore, 
the detection threshold of the instrument is not a fixed and 
well-defined value  but rather a  range in  impact charge. This was also seen in the Ulysses 
and Galileo dust instrument data \citep[e.g.][their Fig.~4]{krueger2010b}. 

The mass distribution measured with Ulysses shows a maximum at two orders of magnitude 
lower masses than the one measured with Nozomi. This is overall consistent with the higher impact speeds measured 
with Ulysses (Section~5.3), given that the 
detection threshold $m_{\mathrm{thr}}$ of impact-ionisation dust detectors is a strong function
of the impact speed $v_{\mathrm{imp}}$: 
$m_{\mathrm{thr}} \approx v_{\mathrm{imp}}^{-3.5}$ \citep[][see also Figure~\ref{fig:mv}]{goeller1985,auer2001}.
Thus, an increase in impact speed by a factor of 2 shifts the detection threshold to approximately
a factor of 10 smaller  particle masses. Assuming that Nozomi and Ulysses detected the same dust populations
and had comparable detection thresholds, 
this partially explains the maximum in the Ulysses measurements at lower masses. The other factor of 10
shift is likely due to a more sensitive detection threshold of the Ulysses sensor at impact speeds below
approximately $\mathrm{10\,km\,s^{-1}}$,  as is indicated in Figure~\ref{fig:mv}.
 
\subsection{Impact Direction}

\label{sec:direction}

\begin{figure}[tb]
	\centering
	\vspace{-1.0cm}
		\hspace{-0.2cm}
		\includegraphics[width=0.9\textwidth]{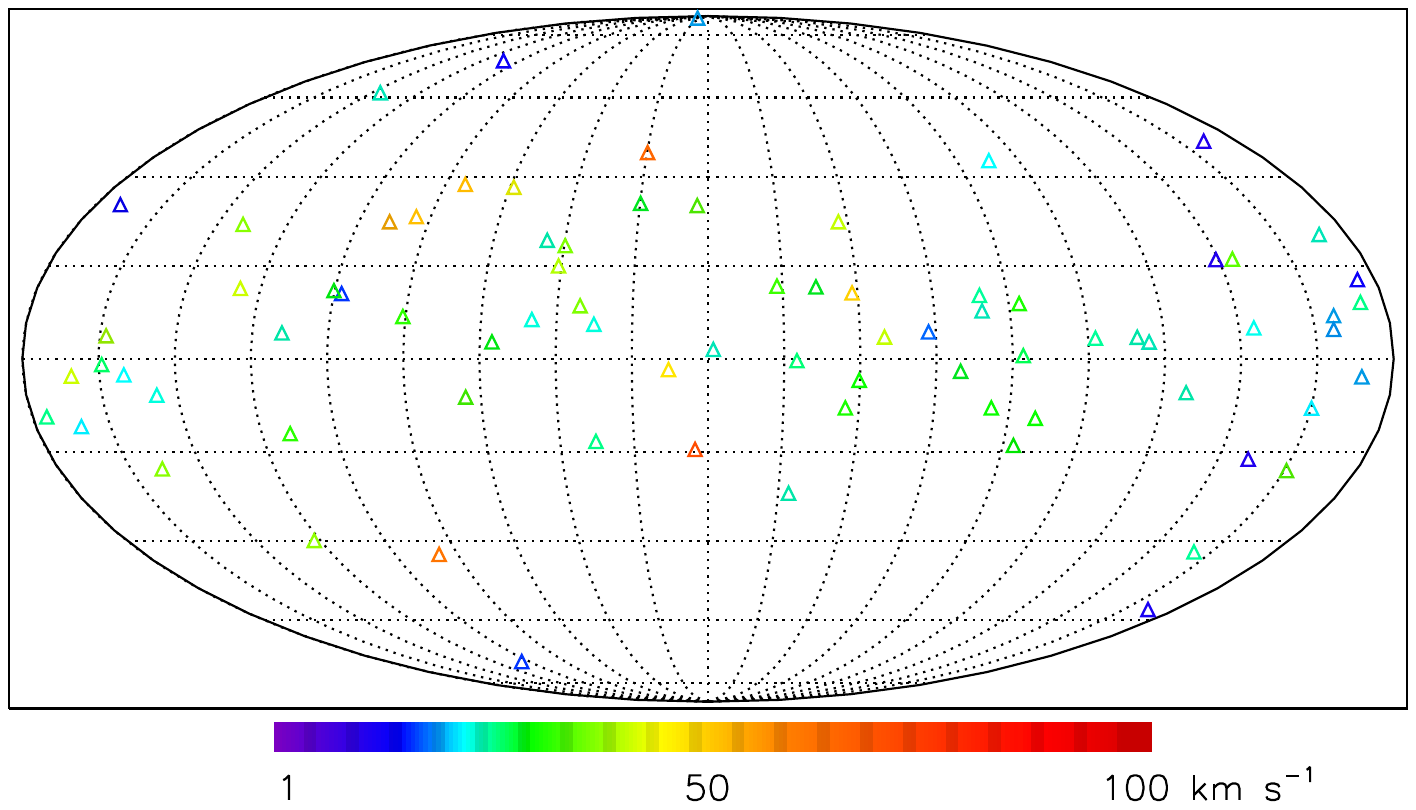}
		\vspace{-1cm}
	\caption{Dust impact directions in ecliptic coordinates showing the flow directions (downstream) at the 
	location of Nozomi of the 
	79 dust particles measured by MDC with 
	derived mass and impact speed (cf. Tables~\ref{tab:mdcdata1} to \ref{tab:mdcdata3}).  
	Each triangle denotes a dust particle impact. The colour code reflects the 
	particle's derived heliocentric speed at the location of the spacecraft where the particle was detected. 
	$0^{\circ}$ ecliptic longitude is to the left.
}
	\label{fig:skymap}
\end{figure}

\citet{senger2007} derived particle velocity vectors in  heliocentric coordinates  for all 79 particles 
for which the impact speed and impact direction were measured (Tables~\ref{tab:mdcdata1} to \ref{tab:mdcdata3}). 
The velocity vector converted 
to heliocentric ecliptic coordinates is shown in Figure~\ref{fig:skymap}. 

In his thesis, \citet{senger2007} discussed neither the uncertainties in the determination of the 
particle velocity vector nor in the particle's orbital elements. The matter is complicated because of the complex 
FOV of MDC. Due to the box shape of MDC and its inclined mounting orientation on board Nozomi, and
spacecraft structures obscuring the FOV, the accuracy 
depends on the direction (note that the sensitivity profile in Figure~\ref{fig:mdc_profile} shows an average over all directions). 
Senger gives a 
range of  $0.77\,\pi\,$sr to $0.83\,\pi\,$sr with an average of $0.81\,\pi\,$sr for the 
sensitive solid angle of MDC. 
This also illustrates the wide FOV of MDC which inevitably leads to large uncertainties in the reconstructed particle trajectories. 
Unfortunately, \citet{senger2007} did not give an estimate of the accuracy of the particle trajectories.  
According to \citet[][page 314]{gruen2001b} the dust detector on board the Hiten spacecraft could determine the trajectory
of individual dust particles with $\pm 75^{\circ}$ accuracy. Given the very similar sensor designs of both the Hiten and Nozomi 
instruments, and ignoring differences in the mounting configuration and the obscuration pattern, 
we can assume a similar accuracy for the trajectory determination for Nozomi MDC. This large uncertainty in the trajectory
determination illustrates the necessity for dust instruments with a much better spatial resolution like the DESTINY$^+$ Dust
Analyzer \citep[DDA;][]{simolka2024} in order to provide a much improved determination   of the dust particle trajectories 
(cf. Section~6.4).

The particle
distribution shows a concentration towards the ecliptic plane which is expected if the majority of the
particle detections is of interplanetary or interstellar origin. The distribution of the particle impacts 
over  the overall range in ecliptic longitudes tends to be rather smooth. To first order, this is a 
function of the overall sky coverage of the MDC FOV during the entire mission time. 
A detailed analysis showed that the entire sky was covered well by  MDC with maximum and minimum coverage 
deviating by only about 50\% from the average value \citep{senger2007}. Thus, 
a rather smooth distribution of particle impacts 
with ecliptic longitude can be expected.

\section{Discussion}

\label{sec:discussion}

In the previous Sections we  presented an overview and a re-analysis of the in situ dust measurements 
obtained by the MDC on board Nozomi in Earth orbit and in interplanetary space. In this Section we briefly 
discuss the results for three identified dust populations: interplanetary dust, interstellar dust, and dust 
detections during the Nozomi crossing of the Geminids meteroid stream. 

\subsection{Interplanetary Dust}

\label{sec:idp}

Very likely about 95\% of the 79 particles with measured mass and impact speed 
detected by Nozomi MDC are of interplanetary 
origin (see also Section~6.2). This also applies to the particles detected in Earth orbit. 

We have used the IMEM model to simulate the Nozomi measurements (cf. Section~5.2). The model was 
calibrated with infrared observations of the zodiacal cloud by the Cosmic Background Explorer (COBE) 
DIRBE instrument, in situ flux measurements by the dust detectors on board the Galileo and Ulysses spacecraft, and 
the crater size distributions on lunar rock samples retrieved by the Apollo missions. The model considers the orbital 
distributions of a number of known sources, including the asteroid belt, 
as well as comets on Jupiter-encountering 
orbits \citep{dikarev2004,dikarev2005a,dikarev2005c,dikarev2005b}. In order to calculate the particle dynamics, 
solar gravity and the velocity-dependent tangential component of radiation pressure (Poynting-Robertson effect) are 
taken into account, while the radial component of solar radiation pressure and the electromagnetic interaction of 
electrically charged dust particles are neglected.
Given that the IMEM model was calibrated with the Ulysses and Galileo in situ dust measurements, among other data sets, 
and that
the model gives overall good agreement with the Nozomi MDC measurements, we conclude that the Nozomi and the Ulysses/Galileo
measurements are in overall agreement.  The agreement between the MDC measurements and the IMEM model also
indicates that the noise identification scheme applied by \citet{senger2007} is rather reliable. In particular, a large
fraction of unrecognised noise events in the data set or, vice versa, too many real dust impacts discarded by the 
noise identification scheme, would lead to significant disagreement with the model predictions.

The so-called apex particles are a population of the interplanetary dust flux that are expected to orbit in nearly circular 
orbits with a lower circular speed than macroscopic objects due to the effect of radiation pressure. 
In order to produce sufficiently large impact charge signals and be detectable by MDC and similar 
impact ionisation dust sensors the particles should impact the sensor 
with a minimum of  $\mathrm{3\,km\,s^{-1}}$ \citep{gruen1992a,gruen1992b} which is thus an approximate value for the speed 
difference of the particles. Because of this lower speed, such particles are expected to impact preferably from the 
apex direction of the spacecraft.
\citet{senger2007} pointed out that the effective 
sensitive area of MDC for interplanetary 
particles approaching from the apex direction of Nozomi was approximately an order of magnitude lower between 
mid-1999 and mid-2000 than later in the mission. 
Thus, the minimum in the impact rate  in 1999 
and early 2000 (Figure~\ref{fig:rate}) may be related to the drop of the sensitive area in this time period, 
and the fact that radiation pressure is not  fully included in the IMEM model. 

The speed distribution of the particles detected by MDC during Nozomi's Earth orbiting phase is similar to 
the one measured by the nearly identical dust detector on board the Hiten spacecraft which was also in 
Earth orbit \citep[][not shown here]{senger2007}. In both cases  the derived {\em geocentric} speeds 
are in excess of the
local escape speed from the Earth gravitational field, indicating that by far the majority of 
the detected particles were unbound to the Earth system. Thus, Nozomi and Hiten likely 
detected similar populations of particles entering the Earth system from interplanetary space. 

Figure~\ref{fig:earthorbit} may suggest that there is a concentration of particle detections 
around the orbit of the Moon, indicating a potential dust ring around the Earth at the distance 
of the Moon.  However, \citet{senger2007} pointed out that this cannot be the case because of the 
derived  high particle speeds which imply particle orbits unbound to the Earth system. 
Furthermore, no increase in the detection rate was registered during the two lunar swingbys.

\subsection{Interstellar Dust}

\label{sec:isd}

In our local environment, the Sun and the heliosphere are surrounded by the Local Interstellar Cloud (LIC) 
of warm diffuse gas and dust where dust is assumed
to contribute about 1\% to the cloud mass \citep{mann2010,kimura2015,krueger2015a}. 
The Sun's motion with respect to this cloud causes
an inflow of interstellar matter into the heliosphere \citep{frisch1999a}.
After initial predictions based on zodiacal light measurements in the
1970s \citep{bertaux1976,may2007}, measurements with the dust detectors
on board the Ulysses and Galileo spacecraft showed that a collimated
stream of interstellar dust passes through the Solar System due to the
Sun's motion relative to the LIC \citep{gruen1993a,baguhl1995a}. The measured heliocentric speed of the dust flow is approximately $\mathrm{26\,km\,s^{-1}}$ \citep{gruen1994a,krueger2015a} 
and its flow direction is compatible with the measured direction of the inflowing interstellar
neutral helium gas \citep[$\lambda_{\mathrm{ecl}} = 75.4^{\circ}$  and $\beta_{\mathrm{ecl}}=-5^{\circ}$ downstream;][]{witte2004a,strub2015,swaczyna2023}. 

Even though most of the particles detected in 
interplanetary space by Nozomi MDC were  interplanetary dust orbiting the Sun, MDC also detected several particles of 
interstellar origin. 
From the analysis of the impact speed 
and the impact direction and derived heliocentric particle trajectories, \citet{sasaki2007} and \citet{senger2007} 
independently 
identified 3 and 5 interstellar particle candidates, respectively. 

Assuming an average of four particle impacts measured from 1999 to 2002, and considering the detection geometry 
of MDC for interstellar dust during this 
time period, we get an interstellar dust flux of approximately $\mathrm{10^{-5}\,m^{-2}\,s^{-1}}$ with roughly 
a factor of 5 uncertainty.  Given that the interstellar particles have expected impact speeds in excess of  
approximately $\mathrm{20\,km\,s^{-1}}$  this value refers to particles with masses 
above $\mathrm{10^{-17}\,kg}$ (cf. Figure~\ref{fig:mv}). This value is in overall good agreement with interstellar dust fluxes measured with
 in situ impact-ionisation dust detectors  carried by other space missions in the solar system 
 \citep[Hiten, Helios, Ulysses, Galileo, Cassini;][]{svedhem1996,krueger2019b}.
 
 The value derived from the MDC measurements  is also in  agreement
 with the dust fluxes derived from the  interstellar dust collections performed with the Stardust spacecraft 
 en route to comet 81P/Wild (also known as Wild~2). In addition to its main objective,  namely the collection of cometary
 particles during the flyby at comet 81P, Stardust also carried silica aerogel collectors
 dedicated to the collection of interstellar particles. The collectors were exposed to the  interstellar dust flow 
 for  195 days in two periods in 2000 and 2002. 
Seven dust particles  were captured in the dust  collectors and analysed on Earth \citep{westphal2012}. The estimated flux of 
these particles which have approximately $\mathrm{1\,\mu m}$ diameter is  $\mathrm{2\times 10^{-6}\,m^{-2}\,s^{-1}}$. This value
is in overall agreement with the flux derived from the MDC, in particular when considering that these 
values were derived with very different measurement techniques, namely impact ionisation detection and visual inspection of the
aerogel, respectively. A detailed re-analysis of
the interstellar particle detections by Nozomi MDC is beyond the scope of this paper. In connection with
 dust dynamical simulations, it will be 
the subject of a future publication.

\subsection{Cometary Dust Trails}

\label{sec:cometary_trails}

When a comet approaches the Sun, sublimating gases carry solid particles away from the comet's surface. 
Solar radiation pressure quickly moves the small sub-micrometre sized particles away from the comet nucleus, 
and the particles form the comet's dust tail. Larger dust particles with sizes exceeding approximately $\mathrm{10\,\mu m}$
are ejected from the cometary nucleus at lower speeds and remain very close to the comet orbit for 
several revolutions around the Sun \citep{soja2015a}. They slowly spread along the comet's orbit as a 
result of small differences in orbital period, forming a tubular structure along the orbit of the parent 
comet filled with dust. These structures are called meteoroid streams or dust trails. When the Earth intercepts 
a cometary trail, the particles can collide with the atmosphere and show up as meteors and fireballs, 
generating a meteor shower.

On 18 November 1998 the Earth crossed the trail of comet 55P/Tempel-Tuttle, leading to a very prominent 
meteor shower. For an observer on the Earth the apparent approach direction of the meteors (radiant) is 
from the  constellation Leo and they are thus called the Leonids.
Comet 55P/Tempel-Tuttle has a retrograde orbit so that the Leonid particles hit the Earth atmosphere with 
a high speed of around $\mathrm{70\,km\,s^{-1}}$.
Nozomi crossed the Leonid particle trail about 1 day after the maximum flux was observed on Earth.

Although the MDC high voltage was switched off during the encounter on 18 November, 4:00 UTC to avoid 
any damage of the 
instrument, four particle impacts  were detected between 17 and 20 November 1998 that even show a correlation in approach direction. 
However, for all four particles   
neither the derived impact speed nor the particle approach 
direction fit to the Leonid dust  stream (cf. Table~\ref{tab:mdcdata1}). Instead, the 
detected particles approached approximately from the opposite direction. Thus, they cannot have directly 
originated from the Leonid dust trail. Nevertheless, a significant 
influence of the Leonid particle stream on the lunar sodium tail during the 1998 encounter was observed \citep{smith1999}. 

An asymmetric dust cloud surrounding the Moon was  detected by the Lunar Dust EXperiment (LDEX) on board 
the Lunar Atmosphere and Dust Environment Explorer (LADEE) mission, and enhanced dust count rates were registered 
during periods when the Earth-Moon system crossed well-known meteor streams \citep[Geminids, Northern Taurids, Quadrantids and Omicron Centaurids;][]{horanyi2015,szalay2018}. \citet{yang2022} studied 
the distribution of dust ejected from the lunar surface  by hypervelocity impacts of micrometeoroids and the 
structure of the resulting steady-state dust cloud in the Earth-Moon system. Particles ejected from the lunar surface escape the gravity of the Moon, and they form an asymmetric torus between the Earth and the Moon in the geocentric distance 
range $\mathrm{[10\,R_E,50\,R_E]}$ (mean Earth radius $\mathrm{R_E=6,371~km}$).

When considering the impacts measured by MDC at the time of the Leonids trail crossing around 18 November 1998, 
one has to consider Nozomi's distance from the Moon and the travel time of dust particles 
released from the lunar surface. 
During the crossing of the Leonids meteor stream Nozomi was far outside the Earth-Moon system at a geocentric 
distance of approximately 1.6 million kilometres or approximately $\mathrm{250\,R_E}$. Thus, even for very high ejection speeds of 
$\mathrm{1\,km\,s^{-1}}$, ejecta from impacts of Leonid particles onto the lunar surface  would require
approximately 9~days to reach Nozomi. This makes an explanation of the Nozomi detections by direct ejecta from the lunar
surface unlikely. We therefore conclude that the detected impacts likely originate neither from the Leonid 
stream directly
nor  from impacts of Leonid meteoroids onto the lunar surface, and their origin remains elusive. 

In order to search for particle impacts in the MDC data set detected during crossings of other cometary trails
during the interplanetary mission of Nozomi, we performed a statistical analysis of the entire MDC 
 data set comprising 96 particle impacts. The technique was described by \citet{gruen2001b} and successfully 
applied to the measurements obtained with the Ulysses dust detector  \citep{krueger2024b},
leading to the identification of a few likely cometary trail crossings of Ulysses. It turned out
that due to the much lower number of particle detections  as compared to Ulysses no statistical significant 
particle concentration  could be identified in the Nozomi dust data. 

\subsection{Future In Situ Dust Measurements}

\label{sec:future}

The negative outcome of the dust trail analysis described in the previous Section emphasises the necessity 
for large area dust detectors in order to derive statistically meaningful results from this kind of analysis. 
In the near 
future the Martian Moons eXploration (MMX) 
mission to the Martian moons Phobos and Deimos \citep{kuramoto2022}  will be equipped with the 
Circum-Martian Dust Monitor \citep[CMDM;][]{kobayashi2018a}. 
CMDM is a large area dust detector with $\mathrm{1\,m^2}$ sensor area. For dust detection the instrument 
will use the outermost polyimide film of the spacecraft multilayer insulation (MLI), which is a thermal 
blanket of the spacecraft. The sensor will use the MLI on one flat side of the spacecraft. Additionally, 
piezoelectric elements will be mounted on the MLI surface. A dust particle hitting or penetrating the 
MLI  generates a stress wave that subsequently propagates through the MLI film
and can be detected by one or more of the piezoelectric sensors. The MMX spacecraft 
 is scheduled for launch in 2026, and the large area CMDM sensor will likely  detect  cometary 
 dust trail particles en route to Mars \citep{krueger2021},  as well as the predicted Martian dust ring.

Furthermore, the DESTINY$^+$ spacecraft -- to be launched in 2028 -- will be 
equipped with the DESTINY$^+$ Dust Analyzer \citep[DDA;][]{simolka2024}. DDA is an impact-ionisation 
time-of-flight  mass 
spectrometer with strong heritage from the Cassini Cosmic Dust Analyzer which very successfully measured 
dust throughout the Saturnian system \citep[CDA;][]{srama2004,srama2006,srama2011}. DDA has 
a sensitive area of $\mathrm{0.03\,m^2}$, i.e. it is three times  larger than the Nozomi MDC. 
It will be equipped with a trajectory
sensor to allow for the determination of the particle's velocity vector with an accuracy of about 10\% for the impact speed and  an angular accuracy of about 
$10^{\circ}$. This highly improved accuracy of the  velocity measurement 
as compared to earlier instruments like, for example, the Ulysses dust detector or Nozomi MDC will 
open the possibility to trace each individual detected dust particle back to its source object, 
rather than relying on statistical techniques  as applied in this work. Together with its improved 
particle composition measurement ($m/\Delta m \approx 100-150$, as compared to $m/\Delta m \approx 30-50$ 
in the case of CDA),  DDA  will open a new window 
onto the composition and origin of the particles' source objects.  

\section{Conclusions}

\label{sec:conclusions}

The Nozomi spacecraft, launched on 3 July 1998 UT, spent about six months in Earth orbit before 
being injected into an interplanetary trajectory. It traversed the spatial region between the orbits of Earth and 
Mars repeatedly, and the Mars Dust Counter (MDC) on board successfully measured interplanetary and interstellar 
dust until April 2002. A total of 96 dust impacts were extracted from the MDC data, 20 from the Earth
orbiting phase and 76 from the interplanetary mission \citep{senger2007}. 

We analysed the   MDC data measured during each  dust  impact, i.e. particle impact speed, 
mass, impact direction, and derived dust flux. Furthermore, we compared the Nozomi results with in situ dust
measurements obtained with the dust detector on board the Ulysses spacecraft \citep{krueger2010b}. Our analysis presented here 
supplements earlier work by \citet{sasaki2007} and \citet{senger2007}, and we make the Nozomi dust data set in electronic form 
available to the scientific community. Our results can be summarised as follows:

\begin{itemize}
\item Based on the measured particle impact speeds, all impacts detected when Nozomi was in Earth orbit 
must have been due to  particles arriving from interplanetary space. Thus, while being 
in Earth orbit  MDC detected neither dust particles of natural origin that were bound to the Earth nor space debris
 \citep{senger2007}.
\item
The dust impact rate measured in interplanetary space varied by approximately a factor of 2 and is overall
consistent with theoretical predictions by the Interplanetary Meteoroid Engineering Model (IMEM) 
model \citep{dikarev2005a,dikarev2005b}. 
\item
The particle impact direction was concentrated towards the ecliptic plane, consistent with the majority 
of the dust impacts being of  interplanetary origin.
\item
The impact speeds and masses of dust particles measured by Nozomi MDC are overall consistent with the measurements 
obtained by Ulysses in the same region of space. 
\item
No impacts of particles originating from cometary trails could be identified during known cometary trail crossings of Nozomi.
A few impacts  
detected during the  crossing of the dust trail of comet 55P/Tempel-Tuttle in November 1998 could not 
be assigned a cometary trail origin, and their origin remains elusive. 
 \end{itemize}
 
We expect that this review of the Nozomi dust measurements will be beneficial 
for future dust investigations with the upcoming Martian Moons eXploration (MMX) and DESTINY$^+$   
space missions, to be launched in 2026 and 2028, respectively. Both missions will traverse the same region
of space as Nozomi, and the MDC  data, together with the Ulysses dust measurements, will likely serve as a valuable 
reference in the future. 

\bigskip
\bigskip
\noindent
\section*{Acknowledgements}
HK gratefully acknowledges support for a research visit at Planetary Exploration Research Center (PERC), Chiba 
Institute of Technology, Narashino, Chiba, Japan,  where part of the work for this publication was done. 
The authors wish to thank Robert Senger for his comprehensive analysis of the Nozomi dust measurements 
published
in his PhD thesis which formed the basis for the analysis presented in this paper. In particular, the Nozomi MDC 
dust data set, which is now available in electronic form, was extracted from the thesis. 
 We thank two anonymous
referees whose comments improved the presentation of our results.

\clearpage

\section*{Appendix}
  
\begin{figure}[tbh]
	\centering
	\vspace{-1.0cm}
		\hspace{-0.2cm}
		\includegraphics[width=0.9\textwidth]{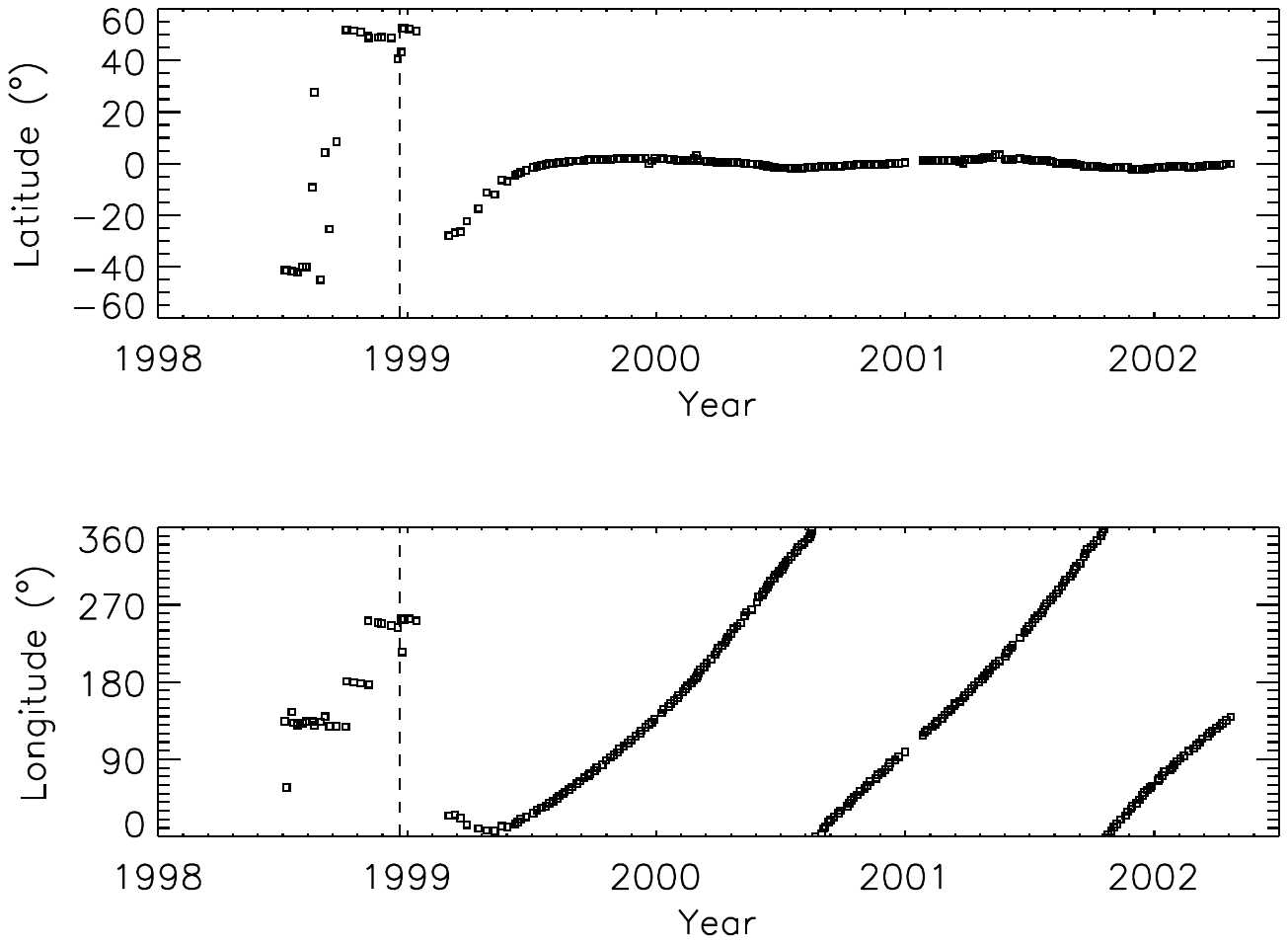}
		\vspace{-1cm}
	\caption{Antenna pointing direction in ecliptic latitude (top) and longitude (bottom). The vertical dashed
	line shows the time when Nozomi left the Earth orbit. Adapted from \citet{senger2007}.
 }
	\label{fig:antenna_pointing}
\end{figure}

  The antenna pointing direction in ecliptic longitude and latitude is shown in Figure~\ref{fig:antenna_pointing}.  The ecliptic latitude of the antenna pointing varied between approximately $+55^{\circ}$ and $-45^{\circ}$ during the first months of the mission and stabilised at $+3^{\circ}$ for the rest of the mission time, 
	while the longitude covered the entire range from $0^{\circ}$ to $360^{\circ}$.
	
	Tables~\ref{tab:mdcdata1} to \ref{tab:mdcdata3} list the entire dust data set of Nozomi MDC extracted from the
	thesis of \citet{senger2007}.

\begin{sidewaystable}[ht] 
 \setlength\tabcolsep{2.4pt}
  \tiny
      \caption{Nozomi MDC data. Col. (1): Impact Number, (2) Impact identifier, (3) and (4): Impact day and time (UTC), (5): Particle mass (kg), (6): Partile impact speed ($\mathrm{km\,s^{-1}}$), (7): Impact position on 
    dust sensor  (EC: electron channel, EC\_COL: electron collector, IC: ion channel, IC\_COL: ion collector, GRID\_EC: entrance grid, MLI\_A/B: MLI outside sensor box), (8): Spacecraft spin angle during dust particle impact ($^{\circ}$, at $0^{\circ}$ MDC points closest to the Sun direction, i.e. close to the ecliptic plane), (9): Quality parameter of dust impact measurement, (10) to (12): Spacecraft position vector in heliocentric coordinate system (km), 
    (13): Heliocentric spacecraft distance during dust impact (AU), (14) to (16): heliocentric dust particle velocity vector (km/s), (17): heliocentric speed of dust particle ($\mathrm{km\,s^{-1}}$), 
    (18) and (19): Inclination and eccentricity of dust particle trajectory.  See text for details.
      }
   \begin{tabular}{ccccccccccccccccccc}
      \hline
 Num. & Imp. ID    &   Date  &   Time &   Mass &Speed&ImpPos &                                                       Spinph&Qua&                  PosX  &    PosY &    PosZ & Rad& AbsvX &AbsvY& AbsvZ& Absv &  Incl&  Ecc \\
 (1) &(2) &   (3)  &   (4)  & (5) &  (6)  &                                                         (7) &(8)&  (9)                   &    (10) &    (11) &(12)& (13)  & (14)& (15) & (16) &  (17)&  (18) & (19)\\
      \hline
 1&98071101-26&10-Jul-1998&15:08:21&3.0E-15&31.57&IC\_COL &331.04&189&  47071711&-144794138&    28784 &1.02&  1.32&  5.19& 16.83&17.66&  80.36& 0.67\\
 2&98071601-3 &16-Jul-1998&03:52:01&4.4E-12&15.24& EC     &107.53&165&  60278018&-139746168&    63454 &1.02& 42.13&  9.46&  2.77&43.27&   3.73& 1.14\\
 3&98073001-5 &28-Jul-1998&23:23:54&1.0E-10& 7.69&EC\_COL &211.03&174&  88381925&-123617666&    59061 &1.02& 26.34& 10.01&  0.60&28.18&   1.26& 0.27\\
 4&98073101-29&31-Jul-1998&06:37:46&1.3E-13&34.14& IC     &270.13&180&  93191670&-119958656&    49421 &1.02& 17.86& -8.56& 18.23&26.92&  64.09& 0.67\\
 5&98080701-3 & 5-Aug-1998&11:14:40&1.7E-14&20.04&IC\_COL &335.83&188& 102884667&-111488040&    27355 &1.01& 23.22& 25.17&-19.06&39.19&  29.18& 0.76\\
 &&&&&&&&&&&&&&&&&& \\[-1.5ex]
 6&98082201-8 &22-Aug-1998&00:48:36&1.6E-14&25.34& EC     & 23.55&154& 129049547& -78765918&    55978 &1.01& 16.05& 25.81& 25.42&39.62&  39.91& 0.79\\
 7&98090501-14& 2-Sep-1998&21:28:35&6.7E-13& 9.89&EC\_COL & 56.12&196& 141712211& -51326819&    44935 &1.01& 12.42& 30.13&  9.35&33.91&  16.03& 0.31\\
 8&98092201-9 &21-Sep-1998&10:30:04&1.4E-15&37.41& IC     &139.35&125& 149955432&  -4648768&   -13433 &1.00& 31.36& 17.20& 15.68&39.05&  40.81& 0.91\\
 9&98092201-23&22-Sep-1998&00:23:40&1.6E-15&35.85& IC     &321.15&149& 149929104&  -3207880&    -4758 &1.00& -0.64&  0.30&-21.42&21.43&  89.22& 0.48\\
10&98101401-21&13-Oct-1998&03:45:06&2.1E-15&36.46&IC\_C   &304.89&189& 140661832&  48872548&   470810 &1.00&-11.78& 41.10&-33.67&54.42&  38.27& 2.32\\
 &&&&&&&&&&&&&&&&&& \\[-1.5ex]
11&98102701-37&26-Oct-1998&10:07:48&8.5E-15&28.64&IC\_COL &302.12&161& 125317767&  79079706&   597923 &0.99&-18.69& 30.97&-27.82&45.63&  37.57& 1.32\\
12&98110901-29& 8-Nov-1998&11:38:43&1.3E-13&16.82&IC\_C   &  4.92&179& 103639158& 104886737&   624512 &0.99&-27.04& 22.46&-15.91&38.58&  24.41& 0.66\\
13&98110901-17& 8-Nov-1998&23:02:03&1.9E-13&17.15&MLI\_A  &342.11&148& 102744228& 105734878&   623866 &0.99&-29.88& 25.74&-14.29&41.95&  19.94& 0.96\\
14&98111001-1 & 9-Nov-1998&10:35:00&   N/A & N/A &GRID\_EC&331.88&153& 101823105& 106580827&   623077 &0.99&  N/A &  N/A &  N/A & N/A &   N/A &  N/A\\
15&98111701-8 &17-Nov-1998&04:23:19&1.5E-11&11.86&EC\_C   &164.93&176&  86086726& 119183754&   591866 &0.98&-16.68& 25.48& -1.92&30.51&   3.95& 0.36\\
 &&&&&&&&&&&&&&&&&& \\[-1.5ex]
16&98111801-1 &17-Nov-1998&17:02:59&6.0E-11&13.92&IC\_COL & 75.37&184&  84951085& 119965169&   588452 &0.98&-16.51& 16.49&-11.15&25.86&  25.90& 0.30\\
17&98111801-6 &18-Nov-1998&00:35:19&3.3E-13&26.09&EC\_C   &154.95&197&  84273081& 120428155&   586334 &0.98& -5.42& 33.13& -5.33&33.99&  13.04& 0.73\\
18&98112101-6 &20-Nov-1998&10:39:41&1.9E-11&10.80&EC\_C   &305.90&208&  78952092& 123855597&   568035 &0.98&-31.47& 21.21& -7.59&38.70&  11.32& 0.66\\
19&98120201-17& 1-Dec-1998&18:58:50&1.5E-13&17.46&MLI\_A  & 25.05&144&  52215696& 136913926&   431437 &0.98&-23.47&  8.47&-16.74&30.04&  33.86& 0.02\\
20&98120301-9 & 2-Dec-1998&18:41:39&1.6E-16&50.90&IC\_C   &329.59&151&  49768815& 137804201&   415244 &0.98&-44.21& 15.97&-48.35&67.43&  45.80& 4.02\\
 &&&&&&&&&&&&&&&&&& \\[-1.5ex]
21&98121801-35&17-Dec-1998&10:17:34&2.0E-12&19.65&IC\_COL &335.96&214&  12054841& 146232969&    65009 &0.98&-30.29&  1.67&-19.97&36.32&  33.37& 0.46\\
22&99011201-43&11-Jan-1999&17:36:43&4.6E-12& 8.89&MLI\_B  &213.79&153& -60090702& 133262786&  3301958 &0.98&-25.75&-19.00&  5.91&32.54&  11.03& 0.26\\
23&99030104-29& 1-Mar-1999&09:40:43&1.6E-16&69.58&IC\_C   &343.43&193&-152913439&  40469814&  7845675 &1.06&-68.94&  4.25& 24.34&73.23&  57.20& 2.07\\
24&99030901-3 & 5-Mar-1999&22:01:49&4.1E-16&47.41& IC     &122.56&159&-157266956&  29376881&  8067958 &1.07& 31.34&-11.02& 14.82&36.37&  72.88& 0.93\\
25&99030901-31& 8-Mar-1999&19:07:38&6.8E-12& 7.95&MLI\_A  &335.04& 85&-159641635&  22221957&  8188819 &1.08& -7.80&-24.77& -6.26&26.71&  14.46& 0.20\\
      \hline
   \end{tabular}  
   \label{tab:mdcdata1}
\end{sidewaystable}
\begin{sidewaystable}[ht] 
 \setlength\tabcolsep{2.4pt}
  \tiny
      \caption{Nozomi MDC data. Table~\ref{tab:mdcdata1} continued.} 
   \begin{tabular}{ccccccccccccccccccc}
      \hline
 Num. & Imp. ID    &   Date  &   Time &   Mass &Speed&ImpPos &                                                       Spinph&Qua&                  PosX  &    PosY &    PosZ & Rad& AbsvX &AbsvY& AbsvZ& Absv &  Incl&  Ecc \\
 (1) &(2) &   (3)  &   (4)  & (5) &  (6)  &                                                         (7) &(8)&  (9)                   &    (10) &    (11) &(12)& (13)  & (14)& (15) & (16) &  (17)&  (18) & (19)\\
      \hline
26&99042001-29&20-Apr-1999&09:14:18&3.3E-13&10.18&EC\_COL &197.18&151&-160847494& -81802403&  8235959 &1.21& 15.15&-20.76&  3.25&25.91&   8.13& 0.17\\
27&99052701-39&27-May-1999&00:50:57&1.5E-13&16.78& EC     & 65.13&135&-121929404&-153506611&  6228099 &1.31& 27.58&-20.13&-13.39&36.68&  21.43& 0.99\\
28&99060201-15& 1-Jun-1999&18:55:49&2.1E-11& 4.23&MLI\_B  &143.57&150&-113496858&-162463142&  5794150 &1.33& 19.41&-14.17& -2.78&24.19&   6.83& 0.13\\
29&99071201-29&11-Jul-1999&08:00:29&1.3E-12&12.45&EC\_C   &233.57&152& -45209701&-205043548&  2284434 &1.40& 26.48&  1.79&  5.83&27.18&  13.09& 0.32\\
30&99072401-34&21-Jul-1999&16:10:04&4.2E-11& 6.38& EC     & 78.79&151& -25529578&-210559673&  1273877 &1.42& 26.26& -2.93& -5.64&27.02&  12.05& 0.17\\
 &&&&&&&&&&&&&&&&&& \\[-1.5ex]
31&99091201-12&11-Sep-1999&13:53:33&2.1E-14&18.55&EC\_COL & 11.78&122&  73154780&-203522168& -3790628 &1.45& 37.73& 14.96& -3.48&40.74&   5.03& 1.70\\
32&99092401-7 &15-Sep-1999&05:27:15&1.4E-10& 6.76&EC\_C   &242.93&172&  79662563&-200929557& -4124835 &1.45& 20.13& 13.83&  3.32&24.65&   7.77& 0.22\\
33&99101901-45&18-Oct-1999&09:37:27&3.8E-15&44.83&MLI\_B  &118.35&194& 132995420&-165734631& -6863033 &1.42&  1.00& 48.32&-27.55&55.63&  42.84& 3.08\\
34&99112001-13&20-Nov-1999&03:51:16&2.3E-09& 3.04&EC\_C   & 72.77&131& 170868118&-112651281& -8803966 &1.37&  9.71& 23.82& -2.49&25.84&   6.59& 0.19\\
35&99112501-3 &24-Nov-1999&23:36:37&3.6E-12&26.72&IC\_COL &193.32&222& 174817334&-103541015& -9005852 &1.36&-15.37& 31.53&  5.02&35.43&  11.19& 0.98\\
 &&&&&&&&&&&&&&&&&& \\[-1.5ex]
36&00040501-19& 2-Apr-2000&15:23:17&1.9E-13&17.18&MLI\_A  &357.98&143&  50731385& 142029583& -2592619 &1.01&-48.18&  8.82&  1.96&49.02&   2.38& 1.72\\
37&00041101-7 & 5-Apr-2000&16:51:25&   N/A &  N/A&GRID\_EC&206.63&107&  42452439& 143886305& -2166598 &1.00&  N/A &  N/A &  N/A & N/A &   N/A &  N/A\\
38&00041601-8 &12-Apr-2000&01:28:10&4.5E-15&34.44& IC     &167.12&160&  24907116& 146475307& -1264008 &0.99&-29.54&-31.05& -3.49&42.99&   9.02& 1.02\\
39&00052301-36&17-May-2000&07:38:15&5.1E-14&23.97& IC     & 58.23&155& -71787710& 127846302&  3706283 &0.98&-39.43&-31.41&-12.65&51.97&  14.11& 1.97\\
40&00052301-29&20-May-2000&04:29:12&1.6E-11& 5.55&IC\_COL &198.85&169& -78913331& 123954539&  4072224 &0.98&-24.52&-20.40&  2.54&32.00&   5.04& 0.18\\
 &&&&&&&&&&&&&&&&&& \\[-1.5ex]
41&00061301-19& 9-Jun-2000&20:39:49&   N/A &  N/A&EC\_COL &176.00&183&-123385600&  87376625&  6354660 &1.01&  N/A &  N/A &  N/A & N/A &   N/A &  N/A\\
42&00062101-42&17-Jun-2000&07:46:12&   N/A &  N/A&GRID\_EC& 41.84&148&-135938575&  71223534&  6998044 &1.03&  N/A &  N/A &  N/A & N/A &   N/A &  N/A\\
43&00062101-37&19-Jun-2000&08:42:39&4.0E-12& 5.07&IC\_C   &192.79&108&-139013539&  66608656&  7155521 &1.03&-12.12&-27.67&  1.39&30.24&   3.80& 0.07\\
44&00062101-4 &19-Jun-2000&11:51:26&3.1E-13&10.88&EC\_C   &223.90&151&-139206255&  66309094&  7165354 &1.03& -7.96&-30.02&  5.73&31.58&  11.41& 0.23\\
45&00070701-22&30-Jun-2000&05:52:23&5.1E-12&12.71&MLI\_A  &230.14&147&-152766946&  40869640&  7859392 &1.06& -1.49&-30.94&  6.98&31.75&  13.84& 0.28\\
 &&&&&&&&&&&&&&&&&& \\[-1.5ex]
46&00070701-26&30-Jun-2000&18:13:38&2.6E-13& 8.60& EC     &175.17&102&-153307441&  39615177&  7886932 &1.06& -3.62&-28.61& -0.34&28.84&   2.87& 0.13\\
47&00070701-29& 3-Jul-2000&19:07:25&4.3E-12&10.46&EC\_COL &110.52&126&-156281040&  32162643&  8038908 &1.07& -4.06&-31.92& -6.72&32.88&  11.96& 0.31\\
48&00072601-17&22-Jul-2000&12:16:06&2.8E-12& 9.02&MLI\_B  &164.36&199&-167002683& -14687530&  8584081 &1.12&  5.54&-26.32& -2.07&26.97&   5.02& 0.15\\
49&00081301-3 & 6-Aug-2000&12:18:13&2.0E-13& 8.39&EC\_COL &165.72&127&-166716711& -51676975&  8564725 &1.17&  9.95&-23.56& -2.06&25.66&   5.18& 0.17\\
50&00081801-37&15-Aug-2000&18:50:25&1.7E-10& 6.49& EC     &168.49&174&-163003944& -73553807&  8370791 &1.20& 11.00&-22.61& -1.57&25.19&   4.40& 0.15\\
 &&&&&&&&&&&&&&&&&& \\[-1.5ex]
51&00092901-3 &25-Sep-2000&15:04:09&3.5E-11&16.39&EC\_C   &194.85&178&-121162129&-154420890&  6208188 &1.31& 20.83& -2.97&  1.41&21.08&   5.75& 0.58\\
52&00101201-16& 7-Oct-2000&17:52:39&7.3E-12&21.79&EC\_C   &209.09&182&-102758167&-172353058&  5259521 &1.34& 21.82&  4.72&  5.76&23.05&  20.73& 0.67\\
53&00101501-37&13-Oct-2000&14:04:38&2.3E-12& 8.62&MLI\_B  &243.85&155& -93199460&-179901458&  4766798 &1.35& 21.72& -7.86&  4.29&23.49&  10.87& 0.20\\
54&00103001-22&24-Oct-2000&14:48:28&3.7E-12& 6.81&IC\_C   &153.98&146& -74153607&-192156991&  3785745 &1.38& 20.10& -5.18& -3.36&21.03&   9.19& 0.33\\
55&00103001-13&27-Oct-2000&00:45:25&3.2E-12& 5.55& EC     &164.71&140& -69838166&-194489332&  3563473 &1.38& 19.99& -5.41& -2.27&20.84&   6.28& 0.33\\
 &&&&&&&&&&&&&&&&&& \\[-1.5ex]
56&01012501-8 & 9-Jan-2001&06:44:50&2.7E-11& 3.99& EC     &220.25&155&  70699447&-204417263& -3665361 &1.45& 21.30& 10.61&  1.42&23.84&   3.45& 0.15\\
57&01012501-19&19-Jan-2001&13:17:01&7.4E-17&62.40&IC\_C   & 85.03&153&  88861867&-196719539& -4598394 &1.44& -3.31& 46.49& 43.78&63.94&  69.39& 4.10\\
58&01012501-29&21-Jan-2001&19:41:04&6.5E-13&26.36&EC\_COL &214.72&175&  92751876&-194750035& -4798212 &1.44& 24.30& 34.16&-11.27&43.41&  17.79& 1.89\\
59&01030701-32& 2-Mar-2001&17:23:13&3.1E-15&42.48&IC\_C   &250.97&163& 151659347&-144638115& -7821226 &1.40& -4.70& 45.71&-27.42&53.51&  43.93& 2.80\\
60&01030701-42& 4-Mar-2001&04:39:46&2.8E-12&11.86& EC     &338.64&162& 153407310&-142285263& -7910657 &1.40&  2.37& 16.94& -4.25&17.62&  18.35& 0.69\\
      \hline
   \end{tabular}  
   \label{tab:mdcdata2}
\end{sidewaystable}
\begin{sidewaystable}[ht] 
 \setlength\tabcolsep{2.4pt}
  \tiny
      \caption{Nozomi MDC data. Table~\ref{tab:mdcdata1} continued.} 
   \begin{tabular}{ccccccccccccccccccc}
      \hline
 Num. & Imp. ID    &   Date  &   Time &   Mass &Speed&ImpPos &                                                       Spinph&Qua&                  PosX  &    PosY &    PosZ & Rad& AbsvX &AbsvY& AbsvZ& Absv &  Incl&  Ecc \\
 (1) &(2) &   (3)  &   (4)  & (5) &  (6)  &                                                         (7) &(8)&  (9)                   &    (10) &    (11) &(12)& (13)  & (14)& (15) & (16) &  (17)&  (18) & (19)\\
      \hline
61&01031401-28&14-Mar-2001&03:56:34&1.6E-12&12.86& IC     &238.27&156& 164272640&-125429220& -8467062 &1.38&  5.26& 29.21& -7.72&30.67&  17.54& 0.60\\
62&01031401-30&14-Mar-2001&04:32:27&  N/A  & N/A &GRID\_EC&317.20&127& 164297309&-125385114& -8468330 &1.38&  N/A &  N/A &  N/A & N/A &   N/A &  N/A\\
63&01042801-49&13-Apr-2001&05:21:54&  N/A  & N/A &GRID\_EC& 48.74&161& 185210475& -66819656& -9536246 &1.32&  N/A &  N/A &  N/A & N/A &   N/A &  N/A\\
64&01042801-43&25-Apr-2001&18:20:58&  N/A  & N/A &GRID\_EC& 47.68&154& 187835874& -39757497& -9667764 &1.29&  N/A &  N/A &  N/A & N/A &   N/A &  N/A\\
65&01051701-6 &12-May-2001&03:53:34&2.3E-13&40.38& EC     & 24.35&131& 185013498&  -3160239& -9517613 &1.24&-20.86& -8.85& 12.60&25.93& 128.56& 0.82\\
 &&&&&&&&&&&&&&&&&& \\[-1.5ex]
66&01070201-6 &11-Jun-2001&06:08:30&6.0E-11& 7.93&EC\_C   &307.05&161& 159342208&  63034534& -8188596 &1.15&-16.68& 17.61& -3.57&24.52&  10.43& 0.42\\
67&01070201-50&21-Jun-2001&12:59:31&  N/A  & N/A & EC     &348.35&107& 144147630&  83721769& -7404466 &1.12&  N/A &  N/A &  N/A & N/A &   N/A &  N/A\\
68&01070201-40&22-Jun-2001&08:06:52&  N/A  & N/A &GRID\_EC&350.95&148& 142835370&  85253451& -7336723 &1.11&  N/A &  N/A &  N/A & N/A &   N/A &  N/A\\
69&01070201-1 &29-Jun-2001&09:35:55&7.0E-12&10.29&EC\_C   & 72.20&200& 130348109&  98285898& -6692786 &1.09&-22.57& 13.19&  8.33&27.44&  18.30& 0.39\\
70&01071101-3 & 2-Jul-2001&05:38:10&6.2E-12& 8.72&MLI\_A  &315.84&140& 124917369& 103219225& -6412911 &1.08&-21.18& 12.28& -3.16&24.69&   8.93& 0.42\\
 &&&&&&&&&&&&&&&&&& \\[-1.5ex]
71&01071101-14& 3-Jul-2001&08:32:53&  N/A  & N/A &MLI\_A  &  5.27&123& 122701224& 105114802& -6298786 &1.08&  N/A &  N/A &  N/A & N/A &   N/A &  N/A\\
72&01080701-35& 6-Aug-2001&13:30:43&  N/A  & N/A &IC\_C   &338.07&162&  40249764& 144297129& -2053286 &1.00&  N/A &  N/A &  N/A & N/A &   N/A &  N/A\\
73&01081601-39&11-Aug-2001&06:49:33&4.7E-12& 8.59&MLI\_A  &335.70&145&  27219203& 146222479& -1382973 &0.99&-25.03& -0.60& -0.88&25.05&   2.22& 0.36\\
74&01081601-8 &14-Aug-2001&17:42:07&7.1E-13&16.72&EC\_COL &249.57&165&  17568823& 147008013&  -886524 &0.99&-31.95&-10.70& -9.37&34.97&  17.25& 0.53\\
75&01081601-14&14-Aug-2001&23:33:07&1.6E-15&33.47&EC\_COL &254.80&187&  16885348& 147047013&  -851363 &0.99&-29.44&-22.47&-21.12&42.63&  38.56& 1.02\\
 &&&&&&&&&&&&&&&&&& \\[-1.5ex]
76&01091201-33&31-Aug-2001&01:13:26&2.2E-12&13.66& EC     &272.86&163& -28290410& 143408893&  1471372 &0.98&-25.41&-14.07& -7.99&30.13&  15.94& 0.30\\
77&01091201-24& 8-Sep-2001&22:14:27&6.4E-11& 5.40& EC     &336.09&161& -52485293& 136371100&  2714743 &0.98&-25.69&-11.09&  0.19&27.98&   1.15& 0.14\\
78&01091201-35&10-Sep-2001&04:43:43&  N/A  & N/A &GRID\_EC&251.15&113& -55854359& 135073123&  2887877 &0.98&  N/A &  N/A &  N/A & N/A &   N/A &  N/A\\
79&01101501-15& 7-Oct-2001&06:07:38&1.1E-12&28.96& IC     &297.95&207&-118445072&  92784313&  6100892 &1.01&  0.11&-17.25&-16.57&23.92&  50.08& 0.57\\
80&01101501-37& 8-Oct-2001&20:27:08&1.7E-13&13.84&MLI\_A  &338.82&108&-121445047&  89542857&  6254968 &1.01&-10.17&-16.33& -1.93&19.33&   6.05& 0.58\\
 &&&&&&&&&&&&&&&&&& \\[-1.5ex]
81&01102601-47&25-Oct-2001&20:22:45&1.2E-13&14.26&EC\_COL &105.12&165&-147692924&  51559764&  7599835 &1.05& -3.70&-29.54&  9.99&31.40&  19.65& 0.24\\
82&01110301-43&29-Oct-2001&16:13:37&9.2E-12& 8.45&MLI\_A  &302.39&113&-152106974&  42349726&  7825425 &1.06& -7.44&-23.83& -4.51&25.37&  10.70& 0.23\\
83&01110301-50&30-Oct-2001&06:59:05&6.5E-15&34.64&IC\_C   &336.93&144&-152762671&  40854760&  7858999 &1.06&  7.91& -1.62& -9.16&12.20&  93.10& 0.91\\
84&01110301-8 & 1-Nov-2001&18:22:40&  N/A  & N/A &GRID\_EC&335.30&112&-155260426&  34800106&  7986757 &1.06&  N/A &  N/A &  N/A & N/A &   N/A &  N/A\\
85&01112101-3 &16-Nov-2001&12:16:52&  N/A  & N/A &GRID\_EC&263.72& 93&-165364919&  -2030892&  8501495 &1.11&  N/A &  N/A &  N/A & N/A &   N/A &  N/A\\
 &&&&&&&&&&&&&&&&&& \\[-1.5ex]
86&01120901-40& 2-Dec-2001&22:06:35&  N/A  & N/A &GRID\_EC&346.16&143&-167460239& -42873428&  8603767 &1.16&  N/A &  N/A &  N/A & N/A &   N/A &  N/A\\
87&01121801-31&15-Dec-2001&17:18:10&6.0E-13&27.87&EC\_C   & 37.13&173&-163047897& -73317671&  8373146 &1.20&  7.11& -1.35& 11.18&13.32&  70.08& 0.81\\
88&02020501-2 &26-Jan-2002&20:50:42&6.4E-13&14.15&MLI\_B  &339.83&123&-119420615&-156340981&  6118480 &1.32&  8.61& -7.41& -4.43&12.20&  21.46& 0.78\\
89&02020501-30& 1-Feb-2002&04:10:21&1.9E-13&14.51&MLI\_B  & 33.71&152&-111524003&-164433827&  5711383 &1.33&  9.78& -6.04&  4.54&12.35&  21.68& 0.77\\
90&02021501-31&10-Feb-2002&19:09:15&2.2E-12&29.10& EC     &320.80&180& -96222343&-177623575&  4922522 &1.35&  1.55&  4.60&-13.95&14.77&  93.42& 0.72\\
 &&&&&&&&&&&&&&&&&& \\[-1.5ex]
91&02021501-14&12-Feb-2002&00:27:34&1.6E-13&15.72& EC     &247.28&147& -94200049&-179153738&  4818354 &1.35& 22.52& -3.10&-11.50&25.47&  27.92& 0.31\\
92&02022501-25&23-Feb-2002&05:15:40&7.4E-13&31.35&EC\_C   &288.19&178& -74917678&-191727021&  3825042 &1.38&  8.84&  8.41&-22.29&25.41&  76.82& 0.45\\
93&02032301-8 &19-Mar-2002&17:27:36&  N/A  & N/A &EC\_COL &337.50&104& -29441565&-209681925&  1483941 &1.42&  N/A &  N/A &  N/A & N/A &   N/A &  N/A\\
94&02032301-12&22-Mar-2002&20:37:50&  N/A  & N/A &GRID\_EC& 16.17&145& -23430977&-211023843&  1174685 &1.42&  N/A &  N/A &  N/A & N/A &   N/A &  N/A\\
95&02033101-7 &27-Mar-2002&05:49:45&9.5E-13& 9.11&IC\_C   &331.52&153& -14980509&-212550436&   739853 &1.42& 14.11& -1.18& -4.31&14.80&  16.92& 0.65\\
 &&&&&&&&&&&&&&&&&& \\[-1.5ex]
96&02041401-34& 1-Apr-2002&13:17:15&8.4E-13&12.84& EC     &308.41&119&  -4703522&-213835193&   211179 &1.43& 12.48&  1.60& -8.34&15.10&  33.82& 0.64\\
      \hline
   \end{tabular}  
   \label{tab:mdcdata3}
\end{sidewaystable}


\begin{thebibliography}{69}
\expandafter\ifx\csname natexlab\endcsname\relax\def\natexlab#1{#1}\fi
\providecommand{\url}[1]{\texttt{#1}}
\providecommand{\href}[2]{#2}
\providecommand{\path}[1]{#1}
\providecommand{\DOIprefix}{doi:}
\providecommand{\ArXivprefix}{arXiv:}
\providecommand{\URLprefix}{URL: }
\providecommand{\Pubmedprefix}{pmid:}
\providecommand{\doi}[1]{\href{http://dx.doi.org/#1}{\path{#1}}}
\providecommand{\Pubmed}[1]{\href{pmid:#1}{\path{#1}}}
\providecommand{\bibinfo}[2]{#2}
\ifx\xfnm\relax \def\xfnm[#1]{\unskip,\space#1}\fi
\bibitem[{{Arai} and {Destiny$^+$ Team}(2024)}]{arai2024}
\bibinfo{author}{{Arai}, T.}, \bibinfo{author}{{Destiny$^+$ Team}},
  \bibinfo{year}{2024}.
\newblock \bibinfo{title}{{Current Status of DESTINY+ and Flyby Observation
  Plan of (3200) Phaethon}}, in: \bibinfo{booktitle}{55th Lunar and Planetary
  Science Conference}, p. \bibinfo{pages}{1781}.
\bibitem[{{Auer}(2001)}]{auer2001}
\bibinfo{author}{{Auer}, S.}, \bibinfo{year}{2001}.
\newblock \bibinfo{title}{{Instrumentation}}, in: \bibinfo{editor}{{Gr{\"u}n},
  E.}, \bibinfo{editor}{{Gustafson}, B.A.S.}, \bibinfo{editor}{{Dermott},
  S.F.}, \bibinfo{editor}{{Fechtig}, H.} (Eds.),
  \bibinfo{booktitle}{Interplanetary Dust}, \bibinfo{publisher}{Springer
  Verlag, Berlin Heidelberg New York}. pp. \bibinfo{pages}{385--444}.
\bibitem[{{Azzi} et~al.(2025){Azzi}, {Oikonomidou} and {Lemmens}}]{azzi2025}
\bibinfo{author}{{Azzi}, S.}, \bibinfo{author}{{Oikonomidou}, X.},
  \bibinfo{author}{{Lemmens}, S.}, \bibinfo{year}{2025}.
\newblock \bibinfo{title}{{The space environment particle density in Low Earth
  Orbit based on two decades of in situ observation}}.
\newblock \bibinfo{journal}{Advances in Space Research} \bibinfo{volume}{75},
  \bibinfo{pages}{6394--6405}.
\newblock \DOIprefix\doi{10.1016/j.asr.2025.02.028},
  \href{http://arxiv.org/abs/2409.13794}{{\tt arXiv:2409.13794}}.
\bibitem[{{Baguhl} et~al.(1995){Baguhl}, {Gr{\"u}n}, {Hamilton}, {Linkert},
  {Riemann} and {Staubach}}]{baguhl1995a}
\bibinfo{author}{{Baguhl}, M.}, \bibinfo{author}{{Gr{\"u}n}, E.},
  \bibinfo{author}{{Hamilton}, D.P.}, \bibinfo{author}{{Linkert}, G.},
  \bibinfo{author}{{Riemann}, R.}, \bibinfo{author}{{Staubach}, P.},
  \bibinfo{year}{1995}.
\newblock \bibinfo{title}{{The flux of interstellar dust observed by Ulysses
  and Galileo}}.
\newblock \bibinfo{journal}{Space Science Reviews} \bibinfo{volume}{72},
  \bibinfo{pages}{471--476}.
\bibitem[{{Bertaux} and {Blamont}(1976)}]{bertaux1976}
\bibinfo{author}{{Bertaux}, J.L.}, \bibinfo{author}{{Blamont}, J.F.},
  \bibinfo{year}{1976}.
\newblock \bibinfo{title}{{Possible evidence for penetration of interstellar
  dust into the solar system}}.
\newblock \bibinfo{journal}{Nature} \bibinfo{volume}{262},
  \bibinfo{pages}{263--266}.
\bibitem[{{Dikarev} et~al.(2004){Dikarev}, {Gr{\"u}n}, {Baggaley}, {Galligan*},
  {Landgraf} and {Jehn}}]{dikarev2004}
\bibinfo{author}{{Dikarev}, V.}, \bibinfo{author}{{Gr{\"u}n}, E.},
  \bibinfo{author}{{Baggaley}, J.}, \bibinfo{author}{{Galligan*}, D.},
  \bibinfo{author}{{Landgraf}, M.}, \bibinfo{author}{{Jehn}, R.},
  \bibinfo{year}{2004}.
\newblock \bibinfo{title}{{Modeling the Sporadic Meteoroid Background Cloud}}.
\newblock \bibinfo{journal}{Earth Moon and Planets} \bibinfo{volume}{95},
  \bibinfo{pages}{109--122}.
\newblock \DOIprefix\doi{10.1007/s11038-005-9017-y}.
\bibitem[{{Dikarev} et~al.(2005a){Dikarev}, {Gr{\"u}n}, {Baggaley}, {Galligan},
  {Landgraf} and {Jehn}}]{dikarev2005a}
\bibinfo{author}{{Dikarev}, V.}, \bibinfo{author}{{Gr{\"u}n}, E.},
  \bibinfo{author}{{Baggaley}, J.}, \bibinfo{author}{{Galligan}, D.},
  \bibinfo{author}{{Landgraf}, M.}, \bibinfo{author}{{Jehn}, R.},
  \bibinfo{year}{2005}a.
\newblock \bibinfo{title}{{The new ESA meteoroid model}}.
\newblock \bibinfo{journal}{Advances in Space Research} \bibinfo{volume}{35},
  \bibinfo{pages}{1282--1289}.
\newblock \DOIprefix\doi{10.1016/j.asr.2005.05.014}.
\bibitem[{{Dikarev} et~al.(2005b){Dikarev}, {Gr\"un}, {Baggaley}, {Galligan},
  {Landgraf} and {Jehn}}]{dikarev2005c}
\bibinfo{author}{{Dikarev}, V.}, \bibinfo{author}{{Gr\"un}, E.},
  \bibinfo{author}{{Baggaley}, W.J.}, \bibinfo{author}{{Galligan}, D.P.},
  \bibinfo{author}{{Landgraf}, M.}, \bibinfo{author}{{Jehn}, R.},
  \bibinfo{year}{2005}b.
\newblock \bibinfo{title}{{A single physical model for diverse meteoroid data
  sets}}.
\newblock \bibinfo{type}{Technical Report}. https://arxiv.org/abs/1902.02977.
\bibitem[{{Dikarev} et~al.(2005c){Dikarev}, {Gr{\"u}n}, {Landgraf} and
  {Jehn}}]{dikarev2005b}
\bibinfo{author}{{Dikarev}, V.}, \bibinfo{author}{{Gr{\"u}n}, E.},
  \bibinfo{author}{{Landgraf}, M.}, \bibinfo{author}{{Jehn}, R.},
  \bibinfo{year}{2005}c.
\newblock \bibinfo{title}{{Update of the ESA Meteoroid Model}}, in:
  \bibinfo{editor}{{Danesy}, D.} (Ed.), \bibinfo{booktitle}{4th European
  Conference on Space Debris}, p. \bibinfo{pages}{271}.
\bibitem[{{Forbes} et~al.(2005){Forbes}, {Lu}, {Bruinsma}, {Nerem} and
  {Zhang}}]{forbes2005}
\bibinfo{author}{{Forbes}, J.M.}, \bibinfo{author}{{Lu}, G.},
  \bibinfo{author}{{Bruinsma}, S.}, \bibinfo{author}{{Nerem}, S.},
  \bibinfo{author}{{Zhang}, X.}, \bibinfo{year}{2005}.
\newblock \bibinfo{title}{{Thermosphere density variations due to the 15-24
  April 2002 solar events from CHAMP/STAR accelerometer measurements}}.
\newblock \bibinfo{journal}{Journal of Geophysical Research (Space Physics)}
  \bibinfo{volume}{110}, \bibinfo{pages}{A12S27}.
\newblock \DOIprefix\doi{10.1029/2004JA010856}.
\bibitem[{{Frisch} et~al.(1999){Frisch}, {Dorschner}, {Gei{\ss}}, {Greenberg},
  {Gr{\"u}n}, {Landgraf}, {Hoppe}, {Jones}, {Kr{\"a}tschmer}, {Linde},
  {Morfill}, {Reach}, {Slavin}, {Svestka}, {Witt} and {Zank}}]{frisch1999a}
\bibinfo{author}{{Frisch}, P.C.}, \bibinfo{author}{{Dorschner}, J.},
  \bibinfo{author}{{Gei{\ss}}, J.}, \bibinfo{author}{{Greenberg}, J.M.},
  \bibinfo{author}{{Gr{\"u}n}, E.}, \bibinfo{author}{{Landgraf}, M.},
  \bibinfo{author}{{Hoppe}, P.}, \bibinfo{author}{{Jones}, A.P.},
  \bibinfo{author}{{Kr{\"a}tschmer}, W.}, \bibinfo{author}{{Linde}, T.J.},
  \bibinfo{author}{{Morfill}, G.E.}, \bibinfo{author}{{Reach}, W.T.},
  \bibinfo{author}{{Slavin}, J.}, \bibinfo{author}{{Svestka}, J.},
  \bibinfo{author}{{Witt}, A.}, \bibinfo{author}{{Zank}, G.P.},
  \bibinfo{year}{1999}.
\newblock \bibinfo{title}{{Dust in the Local Interstellar Wind}}.
\newblock \bibinfo{journal}{Astrophysical Journal} \bibinfo{volume}{525},
  \bibinfo{pages}{492--516}.
\bibitem[{{G{\"o}ller} and {Gr{\"u}n}(1985)}]{goeller1985}
\bibinfo{author}{{G{\"o}ller}, J.R.}, \bibinfo{author}{{Gr{\"u}n}, E.},
  \bibinfo{year}{1985}.
\newblock \bibinfo{title}{{Calibration of the GALILEO/ISPM Dust Detectors with
  Iron Particles}}, in: \bibinfo{editor}{{Giese}, R.H.},
  \bibinfo{editor}{{Lamy}, P.} (Eds.), \bibinfo{booktitle}{{Properties and
  Interaction of Interplanetary Dust}}, \bibinfo{publisher}{Reidel, Dordrecht}.
  pp. \bibinfo{pages}{113--115}.
\bibitem[{{Gr{\"u}n} et~al.(1995){Gr{\"u}n}, {Baguhl}, {Divine}, {Fechtig},
  {Hamilton}, {Hanner}, {Kissel}, {Lindblad}, {Linkert}, {Linkert}, {Mann},
  {McDonnell}, {Morfill}, {Polanskey}, {Riemann}, {Schwehm}, {Siddique},
  {Staubach} and {Zook}}]{gruen1995c}
\bibinfo{author}{{Gr{\"u}n}, E.}, \bibinfo{author}{{Baguhl}, M.},
  \bibinfo{author}{{Divine}, N.}, \bibinfo{author}{{Fechtig}, H.},
  \bibinfo{author}{{Hamilton}, D.P.}, \bibinfo{author}{{Hanner}, M.S.},
  \bibinfo{author}{{Kissel}, J.}, \bibinfo{author}{{Lindblad}, B.A.},
  \bibinfo{author}{{Linkert}, D.}, \bibinfo{author}{{Linkert}, G.},
  \bibinfo{author}{{Mann}, I.}, \bibinfo{author}{{McDonnell}, J.A.M.},
  \bibinfo{author}{{Morfill}, G.E.}, \bibinfo{author}{{Polanskey}, C.},
  \bibinfo{author}{{Riemann}, R.}, \bibinfo{author}{{Schwehm}, G.H.},
  \bibinfo{author}{{Siddique}, N.}, \bibinfo{author}{{Staubach}, P.},
  \bibinfo{author}{{Zook}, H.A.}, \bibinfo{year}{1995}.
\newblock \bibinfo{title}{{Two years of Ulysses dust data}}.
\newblock \bibinfo{journal}{Planetary and Space Science} \bibinfo{volume}{43},
  \bibinfo{pages}{971--999}.
\bibitem[{{Gr{\"u}n} et~al.(2001){Gr{\"u}n}, {Baguhl}, {Svedhem} and
  {Zook}}]{gruen2001b}
\bibinfo{author}{{Gr{\"u}n}, E.}, \bibinfo{author}{{Baguhl}, M.},
  \bibinfo{author}{{Svedhem}, H.}, \bibinfo{author}{{Zook}, H.A.},
  \bibinfo{year}{2001}.
\newblock \bibinfo{title}{{In situ measurements of cosmic dust}}, in:
  \bibinfo{editor}{{Gr{\"u}n}, E.}, \bibinfo{editor}{{Gustafson}, B.A.S.},
  \bibinfo{editor}{{Dermott}, S.F.}, \bibinfo{editor}{{Fechtig}, H.} (Eds.),
  \bibinfo{booktitle}{Interplanetary Dust}, \bibinfo{publisher}{Springer
  Verlag, Berlin Heidelberg New York}. pp. \bibinfo{pages}{295--346}.
\bibitem[{{Gr{\"u}n} et~al.(1992a){Gr{\"u}n}, {Fechtig}, {Hanner}, {Kissel},
  {Lindblad}, {Linkert}, {Maas}, {Morfill} and {Zook}}]{gruen1992a}
\bibinfo{author}{{Gr{\"u}n}, E.}, \bibinfo{author}{{Fechtig}, H.},
  \bibinfo{author}{{Hanner}, M.S.}, \bibinfo{author}{{Kissel}, J.},
  \bibinfo{author}{{Lindblad}, B.A.}, \bibinfo{author}{{Linkert}, D.},
  \bibinfo{author}{{Maas}, D.}, \bibinfo{author}{{Morfill}, G.E.},
  \bibinfo{author}{{Zook}, H.A.}, \bibinfo{year}{1992}a.
\newblock \bibinfo{title}{{The Galileo dust detector}}.
\newblock \bibinfo{journal}{Space Science Reviews} \bibinfo{volume}{60},
  \bibinfo{pages}{317--340}.
\bibitem[{{Gr{\"u}n} et~al.(1992b){Gr{\"u}n}, {Fechtig}, {Kissel}, {Linkert},
  {Maas}, {McDonnell}, {Morfill}, {Schwehm}, {Zook} and {Giese}}]{gruen1992b}
\bibinfo{author}{{Gr{\"u}n}, E.}, \bibinfo{author}{{Fechtig}, H.},
  \bibinfo{author}{{Kissel}, J.}, \bibinfo{author}{{Linkert}, D.},
  \bibinfo{author}{{Maas}, D.}, \bibinfo{author}{{McDonnell}, J.A.M.},
  \bibinfo{author}{{Morfill}, G.E.}, \bibinfo{author}{{Schwehm}, G.H.},
  \bibinfo{author}{{Zook}, H.A.}, \bibinfo{author}{{Giese}, R.H.},
  \bibinfo{year}{1992}b.
\newblock \bibinfo{title}{{The Ulysses dust experiment}}.
\newblock \bibinfo{journal}{Astronomy and Astrophysics, Supplement}
  \bibinfo{volume}{92}, \bibinfo{pages}{411--423}.
\bibitem[{{Gr{\"u}n} et~al.(1994){Gr{\"u}n}, {Gustafson}, {Mann}, {Baguhl},
  {Morfill}, {Staubach}, {Taylor} and {Zook}}]{gruen1994a}
\bibinfo{author}{{Gr{\"u}n}, E.}, \bibinfo{author}{{Gustafson}, B.E.},
  \bibinfo{author}{{Mann}, I.}, \bibinfo{author}{{Baguhl}, M.},
  \bibinfo{author}{{Morfill}, G.E.}, \bibinfo{author}{{Staubach}, P.},
  \bibinfo{author}{{Taylor}, A.}, \bibinfo{author}{{Zook}, H.A.},
  \bibinfo{year}{1994}.
\newblock \bibinfo{title}{Interstellar dust in the heliosphere}.
\newblock \bibinfo{journal}{Astronomy and Astrophysics} \bibinfo{volume}{286},
  \bibinfo{pages}{915--924}.
\bibitem[{{Gr{\"u}n} et~al.(1993){Gr{\"u}n}, {Zook}, {Baguhl}, {Balogh},
  {Bame}, {Fechtig}, {Forsyth}, {Hanner}, {Hor\'anyi}, {Kissel}, {Lindblad},
  {Linkert}, {Linkert}, {Mann}, {McDonnell}, {Morfill}, {Phillips},
  {Polanskey}, {Schwehm}, {Siddique}, {Staubach}, {Svestka} and
  {Taylor}}]{gruen1993a}
\bibinfo{author}{{Gr{\"u}n}, E.}, \bibinfo{author}{{Zook}, H.A.},
  \bibinfo{author}{{Baguhl}, M.}, \bibinfo{author}{{Balogh}, A.},
  \bibinfo{author}{{Bame}, S.J.}, \bibinfo{author}{{Fechtig}, H.},
  \bibinfo{author}{{Forsyth}, R.}, \bibinfo{author}{{Hanner}, M.S.},
  \bibinfo{author}{{Hor\'anyi}, M.}, \bibinfo{author}{{Kissel}, J.},
  \bibinfo{author}{{Lindblad}, B.A.}, \bibinfo{author}{{Linkert}, D.},
  \bibinfo{author}{{Linkert}, G.}, \bibinfo{author}{{Mann}, I.},
  \bibinfo{author}{{McDonnell}, J.A.M.}, \bibinfo{author}{{Morfill}, G.E.},
  \bibinfo{author}{{Phillips}, J.L.}, \bibinfo{author}{{Polanskey}, C.},
  \bibinfo{author}{{Schwehm}, G.H.}, \bibinfo{author}{{Siddique}, N.},
  \bibinfo{author}{{Staubach}, P.}, \bibinfo{author}{{Svestka}, J.},
  \bibinfo{author}{{Taylor}, A.}, \bibinfo{year}{1993}.
\newblock \bibinfo{title}{{Discovery of Jovian dust streams and interstellar
  grains by the Ulysses spacecraft}}.
\newblock \bibinfo{journal}{Nature} \bibinfo{volume}{362},
  \bibinfo{pages}{428--430}.
\bibitem[{{Hamilton}(1996)}]{hamilton1996b}
\bibinfo{author}{{Hamilton}, D.P.}, \bibinfo{year}{1996}.
\newblock \bibinfo{title}{The asymmetric time-variable rings of mars}.
\newblock \bibinfo{journal}{Icarus} \bibinfo{volume}{119},
  \bibinfo{pages}{153--172}.
\bibitem[{{Hor{\'a}nyi} et~al.(2015){Hor{\'a}nyi}, {Szalay}, {Kempf},
  {Schmidt}, {Gr{\"u}n}, {Srama} and {Sternovsky}}]{horanyi2015}
\bibinfo{author}{{Hor{\'a}nyi}, M.}, \bibinfo{author}{{Szalay}, J.R.},
  \bibinfo{author}{{Kempf}, S.}, \bibinfo{author}{{Schmidt}, J.},
  \bibinfo{author}{{Gr{\"u}n}, E.}, \bibinfo{author}{{Srama}, R.},
  \bibinfo{author}{{Sternovsky}, Z.}, \bibinfo{year}{2015}.
\newblock \bibinfo{title}{{A permanent, asymmetric dust cloud around the
  Moon}}.
\newblock \bibinfo{journal}{Nature} \bibinfo{volume}{522},
  \bibinfo{pages}{324--326}.
\newblock \DOIprefix\doi{10.1038/nature14479}.
\bibitem[{{Igenbergs} et~al.(1991){Igenbergs}, {H{\"u}dephol}, {Uesugi},
  {Hayashi}, {Svedhem}, {Iglseder}, {Koller}, {Glasmachers}, {Grun}, {Schwehm},
  {Mizutani}, {Yamamoto}, {Fugimura}, {Ishii}, {Araki}, {Yamakoshi} and
  {Nogami}}]{igenbergs1991}
\bibinfo{author}{{Igenbergs}, E.}, \bibinfo{author}{{H{\"u}dephol}, A.},
  \bibinfo{author}{{Uesugi}, K.}, \bibinfo{author}{{Hayashi}, T.},
  \bibinfo{author}{{Svedhem}, H.}, \bibinfo{author}{{Iglseder}, H.},
  \bibinfo{author}{{Koller}, G.}, \bibinfo{author}{{Glasmachers}, A.},
  \bibinfo{author}{{Grun}, E.}, \bibinfo{author}{{Schwehm}, G.},
  \bibinfo{author}{{Mizutani}, H.}, \bibinfo{author}{{Yamamoto}, T.},
  \bibinfo{author}{{Fugimura}, A.}, \bibinfo{author}{{Ishii}, N.},
  \bibinfo{author}{{Araki}, H.}, \bibinfo{author}{{Yamakoshi}, K.},
  \bibinfo{author}{{Nogami}, K.}, \bibinfo{year}{1991}.
\newblock \bibinfo{title}{{The Present Status of the Munich Dust Counter
  Experiment on Board of the HITEN Spacecraft (invited Contribution)}}, in:
  \bibinfo{editor}{{Levasseur-Regoud}, A.C.}, \bibinfo{editor}{{Hasegawa}, H.}
  (Eds.), \bibinfo{booktitle}{IAU Colloq. 126: Origin and Evolution of
  Interplanetary Dust}, p.~\bibinfo{pages}{15}.
\newblock \DOIprefix\doi{10.1007/978-94-011-3640-2\_3}.
\bibitem[{{Igenbergs} et~al.(1996){Igenbergs}, {Sasaki}, {Farber}, {Fisher},
  {Munzenmayer}, {Fujiwara}, {Iglseder}, {Glasmachers}, {Grun}, {Mukai},
  {Nogami}, {Ohashi}, {Schwehm}, {Svedhem} and {Yamakoshi}}]{igenbergs1996}
\bibinfo{author}{{Igenbergs}, E.}, \bibinfo{author}{{Sasaki}, S.},
  \bibinfo{author}{{Farber}, G.}, \bibinfo{author}{{Fisher}, F.},
  \bibinfo{author}{{Munzenmayer}, R.}, \bibinfo{author}{{Fujiwara}, A.},
  \bibinfo{author}{{Iglseder}, H.}, \bibinfo{author}{{Glasmachers}, A.},
  \bibinfo{author}{{Grun}, E.}, \bibinfo{author}{{Mukai}, T.},
  \bibinfo{author}{{Nogami}, K.I.}, \bibinfo{author}{{Ohashi}, H.},
  \bibinfo{author}{{Schwehm}, G.}, \bibinfo{author}{{Svedhem}, H.},
  \bibinfo{author}{{Yamakoshi}, K.}, \bibinfo{year}{1996}.
\newblock \bibinfo{title}{{Mars Dust Counter on Board Isas Planet-B}}, in:
  \bibinfo{editor}{{Gustafson}, B.A.S.}, \bibinfo{editor}{{Hanner}, M.S.}
  (Eds.), \bibinfo{booktitle}{IAU Colloq. 150: Physics, Chemistry, and Dynamics
  of Interplanetary Dust}, p. \bibinfo{pages}{233}.
\bibitem[{{Igenbergs} et~al.(1998){Igenbergs}, {Sasaki}, {M{\"u}nzenmayer},
  {Ohashi}, {F{\"a}rber}, {Fischer}, {Fujiwara}, {Glasmachers}, {Gr{\"u}n},
  {Hamabe}, {Iglseder}, {Klinge}, {Maiyamoto}, {Mukai}, {Naumann}, {Nogami},
  {Schwehm}, {Svedhem} and {Yamakoshi}}]{igenbergs1998}
\bibinfo{author}{{Igenbergs}, E.}, \bibinfo{author}{{Sasaki}, S.},
  \bibinfo{author}{{M{\"u}nzenmayer}, R.}, \bibinfo{author}{{Ohashi}, H.},
  \bibinfo{author}{{F{\"a}rber}, G.}, \bibinfo{author}{{Fischer}, F.},
  \bibinfo{author}{{Fujiwara}, A.}, \bibinfo{author}{{Glasmachers}, A.},
  \bibinfo{author}{{Gr{\"u}n}, E.}, \bibinfo{author}{{Hamabe}, Y.},
  \bibinfo{author}{{Iglseder}, H.}, \bibinfo{author}{{Klinge}, D.},
  \bibinfo{author}{{Maiyamoto}, H.}, \bibinfo{author}{{Mukai}, T.},
  \bibinfo{author}{{Naumann}, W.}, \bibinfo{author}{{Nogami}, K.I.},
  \bibinfo{author}{{Schwehm}, G.H.}, \bibinfo{author}{{Svedhem}, H.},
  \bibinfo{author}{{Yamakoshi}, K.}, \bibinfo{year}{1998}.
\newblock \bibinfo{title}{Mars dust counter}.
\newblock \bibinfo{journal}{Earth Planets Space} \bibinfo{volume}{50},
  \bibinfo{pages}{241--245}.
\bibitem[{{Iglseder} et~al.(1993){Iglseder}, {M{\"u}nzenmayer}, {Svedhem} and
  {Gr{\"u}n}}]{iglseder1993a}
\bibinfo{author}{{Iglseder}, H.}, \bibinfo{author}{{M{\"u}nzenmayer}, R.},
  \bibinfo{author}{{Svedhem}, H.}, \bibinfo{author}{{Gr{\"u}n}, E.},
  \bibinfo{year}{1993}.
\newblock \bibinfo{title}{{Cosmic dust and space debris measurements with the
  Munich dust counter on board the satellites hiten and brem-sat}}.
\newblock \bibinfo{journal}{Advances in Space Research} \bibinfo{volume}{13},
  \bibinfo{pages}{129--132}.
\newblock \DOIprefix\doi{10.1016/0273-1177(93)90579-Z}.
\bibitem[{{Ishimoto} et~al.(1997){Ishimoto}, {Kimura}, {Nakagawa} and
  {Mukai}}]{ishimoto1997}
\bibinfo{author}{{Ishimoto}, H.}, \bibinfo{author}{{Kimura}, H.},
  \bibinfo{author}{{Nakagawa}, N.}, \bibinfo{author}{{Mukai}, T.},
  \bibinfo{year}{1997}.
\newblock \bibinfo{title}{{Planned observation of phobos/deimos dust rings by
  PLANET-B}}.
\newblock \bibinfo{journal}{Advances in Space Research} \bibinfo{volume}{19},
  \bibinfo{pages}{123--126}.
\newblock \DOIprefix\doi{10.1016/S0273-1177(96)00126-3}.
\bibitem[{{Kawaguchi} et~al.(1995){Kawaguchi}, {Yamakawa}, {Uesugi} and
  {Matsuo}}]{kawaguchi1995}
\bibinfo{author}{{Kawaguchi}, J.}, \bibinfo{author}{{Yamakawa}, H.},
  \bibinfo{author}{{Uesugi}, T.}, \bibinfo{author}{{Matsuo}, H.},
  \bibinfo{year}{1995}.
\newblock \bibinfo{title}{{On making use of lunar and solar gravity assists in
  LUNAR-A, PLANET-B missions.}}
\newblock \bibinfo{journal}{Acta Astronautica} \bibinfo{volume}{35},
  \bibinfo{pages}{633--642}.
\newblock \DOIprefix\doi{10.1016/0094-5765(95)00013-P}.
\bibitem[{{Kimura}(2015)}]{kimura2015}
\bibinfo{author}{{Kimura}, H.}, \bibinfo{year}{2015}.
\newblock \bibinfo{title}{{Interstellar dust in the Local Cloud surrounding the
  Sun}}.
\newblock \bibinfo{journal}{Monthly Notices of the Royal Astro. Soc.}
  \bibinfo{volume}{449}, \bibinfo{pages}{2250--2258}.
\newblock \DOIprefix\doi{10.1093/mnras/stv427}.
\bibitem[{{Kobayashi} et~al.(2018){Kobayashi}, {Kr{\"u}ger}, {Senshu}, {Wada},
  {Okudaira}, {Sasaki} and {Kimura}}]{kobayashi2018a}
\bibinfo{author}{{Kobayashi}, M.}, \bibinfo{author}{{Kr{\"u}ger}, H.},
  \bibinfo{author}{{Senshu}, H.}, \bibinfo{author}{{Wada}, K.},
  \bibinfo{author}{{Okudaira}, O.}, \bibinfo{author}{{Sasaki}, S.},
  \bibinfo{author}{{Kimura}, H.}, \bibinfo{year}{2018}.
\newblock \bibinfo{title}{{In situ observation of dust particles of Martian
  dust belts by a large-sensitive-area dust sensor}}.
\newblock \bibinfo{journal}{Planetary and Space Science} \bibinfo{volume}{156},
  \bibinfo{pages}{41--46}.
\newblock \DOIprefix\doi{10.1016/j.pss.2017.12.011}.
\bibitem[{{Krivov} et~al.(2006){Krivov}, {Feofilov} and {Dikarev}}]{krivov2006}
\bibinfo{author}{{Krivov}, A.V.}, \bibinfo{author}{{Feofilov}, A.G.},
  \bibinfo{author}{{Dikarev}, V.V.}, \bibinfo{year}{2006}.
\newblock \bibinfo{title}{{Search for the putative dust belts of Mars: The late
  2007 opportunity}}.
\newblock \bibinfo{journal}{Planetary and Space Science} \bibinfo{volume}{54},
  \bibinfo{pages}{871--878}.
\newblock \DOIprefix\doi{10.1016/j.pss.2006.05.007}.
\bibitem[{{Krivov} and {Hamilton}(1997)}]{krivov1997}
\bibinfo{author}{{Krivov}, A.V.}, \bibinfo{author}{{Hamilton}, D.P.},
  \bibinfo{year}{1997}.
\newblock \bibinfo{title}{{Martian dust belts: Waiting for discovery}}.
\newblock \bibinfo{journal}{Icarus} \bibinfo{volume}{128},
  \bibinfo{pages}{335--353}.
\bibitem[{{Kr{\"u}ger} et~al.(2010){Kr{\"u}ger}, {Dikarev}, {Anweiler},
  {Dermott}, {Graps}, {Gr{\"u}n}, {Gustafson}, {Hamilton}, {Hanner},
  {Hor\'anyi}, {Kissel}, {Linkert}, {Linkert}, {Mann}, {McDonnell}, {Morfill},
  {Polanskey}, {Schwehm} and {Srama}}]{krueger2010b}
\bibinfo{author}{{Kr{\"u}ger}, H.}, \bibinfo{author}{{Dikarev}, V.},
  \bibinfo{author}{{Anweiler}, B.}, \bibinfo{author}{{Dermott}, S.F.},
  \bibinfo{author}{{Graps}, A.L.}, \bibinfo{author}{{Gr{\"u}n}, E.},
  \bibinfo{author}{{Gustafson}, B.A.}, \bibinfo{author}{{Hamilton}, D.P.},
  \bibinfo{author}{{Hanner}, M.M.S.}, \bibinfo{author}{{Hor\'anyi}, M.},
  \bibinfo{author}{{Kissel}, J.}, \bibinfo{author}{{Linkert}, D.},
  \bibinfo{author}{{Linkert}, G.}, \bibinfo{author}{{Mann}, I.},
  \bibinfo{author}{{McDonnell}, J.A.M.}, \bibinfo{author}{{Morfill}, G.E.},
  \bibinfo{author}{{Polanskey}, C.}, \bibinfo{author}{{Schwehm}, G.H.},
  \bibinfo{author}{{Srama}, R.}, \bibinfo{year}{2010}.
\newblock \bibinfo{title}{{Three years of Ulysses dust data: 2005 to 2007}}.
\newblock \bibinfo{journal}{Planetary and Space Science} \bibinfo{volume}{58},
  \bibinfo{pages}{951--964}.
\bibitem[{{Kr{\"u}ger} et~al.(2021){Kr{\"u}ger}, {Kobayashi}, {Strub},
  {Klostermeyer}, {Sommer}, {Kimura}, {Gr{\"u}n} and {Srama}}]{krueger2021}
\bibinfo{author}{{Kr{\"u}ger}, H.}, \bibinfo{author}{{Kobayashi}, M.},
  \bibinfo{author}{{Strub}, P.}, \bibinfo{author}{{Klostermeyer}, G.M.},
  \bibinfo{author}{{Sommer}, M.}, \bibinfo{author}{{Kimura}, H.},
  \bibinfo{author}{{Gr{\"u}n}, E.}, \bibinfo{author}{{Srama}, R.},
  \bibinfo{year}{2021}.
\newblock \bibinfo{title}{{Modelling cometary meteoroid stream traverses of the
  Martian Moons eXploration (MMX) spacecraft en route to Phobos}}.
\newblock \bibinfo{journal}{Earth Planets and Space} \bibinfo{volume}{73},
  \bibinfo{pages}{93}.
\newblock \DOIprefix\doi{10.1186/s40623-021-01412-5}.
\bibitem[{{Kr{\"u}ger} et~al.(2019){Kr{\"u}ger}, {Strub}, {Altobelli},
  {Sterken}, {Srama} and {Gr\"un}}]{krueger2019b}
\bibinfo{author}{{Kr{\"u}ger}, H.}, \bibinfo{author}{{Strub}, P.},
  \bibinfo{author}{{Altobelli}, N.}, \bibinfo{author}{{Sterken}, V.},
  \bibinfo{author}{{Srama}, R.}, \bibinfo{author}{{Gr\"un}, E.},
  \bibinfo{year}{2019}.
\newblock \bibinfo{title}{Interstellar dust in the solar system: model versus
  in-situ spacecraft data}.
\newblock \bibinfo{journal}{Astronomy and Astrophysics} \bibinfo{volume}{626},
  \bibinfo{pages}{A37}.
\newblock \DOIprefix\doi{doi.org/10.1051/0004-6361/201834316}.
\bibitem[{{Kr{\"u}ger} et~al.(2024){Kr{\"u}ger}, {Strub} and
  {Gr{\"u}n}}]{krueger2024b}
\bibinfo{author}{{Kr{\"u}ger}, H.}, \bibinfo{author}{{Strub}, P.},
  \bibinfo{author}{{Gr{\"u}n}, E.}, \bibinfo{year}{2024}.
\newblock \bibinfo{title}{{Ulysses spacecraft in situ detections of cometary
  dust trails}}.
\newblock \bibinfo{journal}{Philosophical Transactions of the Royal Society A}
  \bibinfo{volume}{382}, \bibinfo{pages}{20230200}.
\newblock \DOIprefix\doi{10.1098/rsta.2023.0200}.
\bibitem[{{Kr{\"u}ger} et~al.(2015){Kr{\"u}ger}, {Strub}, {Sterken} and
  {Gr\"un}}]{krueger2015a}
\bibinfo{author}{{Kr{\"u}ger}, H.}, \bibinfo{author}{{Strub}, P.},
  \bibinfo{author}{{Sterken}, V.J.}, \bibinfo{author}{{Gr\"un}, E.},
  \bibinfo{year}{2015}.
\newblock \bibinfo{title}{Sixteen years of ulysses interstellar dust
  measurements in the solar system: I. mass distribution and gas-to-dust mass
  ratio}.
\newblock \bibinfo{journal}{Astrophysical Journal} \bibinfo{volume}{812},
  \bibinfo{pages}{139}.
\newblock \DOIprefix\doi{doi:10.1088/0004-637X/812/2/139}.
\bibitem[{{Kuramoto} et~al.(2022){Kuramoto}, {Kawakatsu}, {Fujimoto}, {Araya},
  {Barucci}, {Genda}, {Hirata}, {Ikeda}, {Imamura}, {Helbert}, {Kameda},
  {Kobayashi}, {Kusano}, {Lawrence}, {Matsumoto}, {Michel}, {Miyamoto},
  {Morota}, {Nakagawa}, {Nakamura}, {Ogawa}, {Otake}, {Ozaki}, {Russell},
  {Sasaki}, {Sawada}, {Senshu}, {Tachibana}, {Terada}, {Ulamec}, {Usui},
  {Wada}, {Watanabe} and {Yokota}}]{kuramoto2022}
\bibinfo{author}{{Kuramoto}, K.}, \bibinfo{author}{{Kawakatsu}, Y.},
  \bibinfo{author}{{Fujimoto}, M.}, \bibinfo{author}{{Araya}, A.},
  \bibinfo{author}{{Barucci}, M.A.}, \bibinfo{author}{{Genda}, H.},
  \bibinfo{author}{{Hirata}, N.}, \bibinfo{author}{{Ikeda}, H.},
  \bibinfo{author}{{Imamura}, T.}, \bibinfo{author}{{Helbert}, J.},
  \bibinfo{author}{{Kameda}, S.}, \bibinfo{author}{{Kobayashi}, M.},
  \bibinfo{author}{{Kusano}, H.}, \bibinfo{author}{{Lawrence}, D.J.},
  \bibinfo{author}{{Matsumoto}, K.}, \bibinfo{author}{{Michel}, P.},
  \bibinfo{author}{{Miyamoto}, H.}, \bibinfo{author}{{Morota}, T.},
  \bibinfo{author}{{Nakagawa}, H.}, \bibinfo{author}{{Nakamura}, T.},
  \bibinfo{author}{{Ogawa}, K.}, \bibinfo{author}{{Otake}, H.},
  \bibinfo{author}{{Ozaki}, M.}, \bibinfo{author}{{Russell}, S.},
  \bibinfo{author}{{Sasaki}, S.}, \bibinfo{author}{{Sawada}, H.},
  \bibinfo{author}{{Senshu}, H.}, \bibinfo{author}{{Tachibana}, S.},
  \bibinfo{author}{{Terada}, N.}, \bibinfo{author}{{Ulamec}, S.},
  \bibinfo{author}{{Usui}, T.}, \bibinfo{author}{{Wada}, K.},
  \bibinfo{author}{{Watanabe}, S.i.}, \bibinfo{author}{{Yokota}, S.},
  \bibinfo{year}{2022}.
\newblock \bibinfo{title}{{Martian moons exploration MMX: sample return mission
  to Phobos elucidating formation processes of habitable planets}}.
\newblock \bibinfo{journal}{Earth, Planets and Space} \bibinfo{volume}{74},
  \bibinfo{pages}{12}.
\newblock \DOIprefix\doi{10.1186/s40623-021-01545-7}.
\bibitem[{{Liu} and {Schmidt}(2021)}]{liu2021}
\bibinfo{author}{{Liu}, X.}, \bibinfo{author}{{Schmidt}, J.},
  \bibinfo{year}{2021}.
\newblock \bibinfo{title}{{Configuration of the Martian dust rings: shapes,
  densities, and size distributions from direct integrations of particle
  trajectories}}.
\newblock \bibinfo{journal}{Monthly Notices of the Royal Astro. Soc.}
  \bibinfo{volume}{500}, \bibinfo{pages}{2979--2985}.
\newblock \DOIprefix\doi{10.1093/mnras/staa3084},
  \href{http://arxiv.org/abs/2201.02847}{{\tt arXiv:2201.02847}}.
\bibitem[{{Makuch} et~al.(2005){Makuch}, {Krivov} and {Spahn}}]{makuch2005}
\bibinfo{author}{{Makuch}, M.}, \bibinfo{author}{{Krivov}, A.V.},
  \bibinfo{author}{{Spahn}, F.}, \bibinfo{year}{2005}.
\newblock \bibinfo{title}{{Long-term dynamical evolution of dusty ejecta from
  Deimos}}.
\newblock \bibinfo{journal}{Planetary and Space Science} \bibinfo{volume}{53},
  \bibinfo{pages}{357--369}.
\newblock \DOIprefix\doi{10.1016/j.pss.2004.09.063}.
\bibitem[{{Mann}(2010)}]{mann2010}
\bibinfo{author}{{Mann}, I.}, \bibinfo{year}{2010}.
\newblock \bibinfo{title}{{Interstellar Dust in the Solar System}}.
\newblock \bibinfo{journal}{Annual Review of Astronomy and Astrophysics}
  \bibinfo{volume}{48}, \bibinfo{pages}{173--203}.
\newblock \DOIprefix\doi{10.1146/annurev-astro-081309-130846}.
\bibitem[{{May}(2007)}]{may2007}
\bibinfo{author}{{May}, B.H.}, \bibinfo{year}{2007}.
\newblock \bibinfo{title}{{A survey of radial velocities in the zodiacal dust
  cloud}}.
\newblock Ph.D. thesis. PhD thesis, Imperial College of Science, Technology and
  Medicine London, U.~K.
\bibitem[{{M{\"u}nzenmayer} et~al.(1997){M{\"u}nzenmayer}, {Igenbergs},
  {Iglseder} and {Svedhem}}]{muenzenmayer1997}
\bibinfo{author}{{M{\"u}nzenmayer}, R.}, \bibinfo{author}{{Igenbergs}, E.},
  \bibinfo{author}{{Iglseder}, H.}, \bibinfo{author}{{Svedhem}, H.},
  \bibinfo{year}{1997}.
\newblock \bibinfo{title}{{The munich dust counter on board the muses-a
  mission: calibration of impacts inside and in front of the detector}}.
\newblock \bibinfo{journal}{Advances in Space Research} \bibinfo{volume}{20},
  \bibinfo{pages}{1485--1488}.
\newblock \DOIprefix\doi{10.1016/S0273-1177(97)00422-5}.
\bibitem[{{Nakatani} et~al.(1995){Nakatani}, {Tsuruda} and
  {Yamamoto}}]{nakatani1995}
\bibinfo{author}{{Nakatani}, I.}, \bibinfo{author}{{Tsuruda}, K.},
  \bibinfo{author}{{Yamamoto}, T.}, \bibinfo{year}{1995}.
\newblock \bibinfo{title}{{Planet-B: Mars mission with small spacecraft but
  potentially with large science reward.}}
\newblock \bibinfo{journal}{Acta Astronautica} \bibinfo{volume}{35},
  \bibinfo{pages}{337--344}.
\bibitem[{{Naumann}(2000)}]{naumann2000}
\bibinfo{author}{{Naumann}, W.}, \bibinfo{year}{2000}.
\newblock \bibinfo{title}{{Rechnergest\"utzte Automatisierung eines Experiments
  im interplanetaren Raum}}.
\newblock Ph.D. thesis. Technische Universit\"at M\"unchen.
\bibitem[{{Ozaki} et~al.(2022){Ozaki}, {Yamamoto}, {Gonzalez-Franquesa},
  {Gutierrez-Ramon}, {Pushparaj}, {Chikazawa}, {Tos}, {{\c{C}}elik}, {Marmo},
  {Kawakatsu}, {Arai}, {Nishiyama} and {Takashima}}]{ozaki2022}
\bibinfo{author}{{Ozaki}, N.}, \bibinfo{author}{{Yamamoto}, T.},
  \bibinfo{author}{{Gonzalez-Franquesa}, F.},
  \bibinfo{author}{{Gutierrez-Ramon}, R.}, \bibinfo{author}{{Pushparaj}, N.},
  \bibinfo{author}{{Chikazawa}, T.}, \bibinfo{author}{{Tos}, D.A.D.},
  \bibinfo{author}{{{\c{C}}elik}, O.}, \bibinfo{author}{{Marmo}, N.},
  \bibinfo{author}{{Kawakatsu}, Y.}, \bibinfo{author}{{Arai}, T.},
  \bibinfo{author}{{Nishiyama}, K.}, \bibinfo{author}{{Takashima}, T.},
  \bibinfo{year}{2022}.
\newblock \bibinfo{title}{{Mission design of DESTINY+: Toward active asteroid
  (3200) Phaethon and multiple small bodies}}.
\newblock \bibinfo{journal}{Acta Astronautica} \bibinfo{volume}{196},
  \bibinfo{pages}{42--56}.
\newblock \DOIprefix\doi{10.1016/j.actaastro.2022.03.029},
  \href{http://arxiv.org/abs/2201.01933}{{\tt arXiv:2201.01933}}.
\bibitem[{{Sasaki}(1999)}]{sasaki1999b}
\bibinfo{author}{{Sasaki}, S.}, \bibinfo{year}{1999}.
\newblock \bibinfo{title}{{Dust ring/torus around Mars, waiting for detection
  by NOZOMI}}.
\newblock \bibinfo{journal}{Advances in Space Research} \bibinfo{volume}{23},
  \bibinfo{pages}{1907--1910}.
\newblock \DOIprefix\doi{10.1016/S0273-1177(99)00278-1}.
\bibitem[{{Sasaki} et~al.(2002a){Sasaki}, {Igenbergs}, {Ohashi}, {M{\"
  u}nzenmayer}, {Naumann}, {Hofschuster}, {Born}, {F{\" a}rber}, {Fischer},
  {Fujiwara}, {Glasmachers}, {Gr{\" u}n}, {Hamabe}, {Iglseder}, {Kawamura},
  {Miyamoto}, {Morishige}, {Mukai}, {Naoi}, {Nogami}, {Schwehm} and
  {Svedhem}}]{sasaki2002a}
\bibinfo{author}{{Sasaki}, S.}, \bibinfo{author}{{Igenbergs}, E.},
  \bibinfo{author}{{Ohashi}, H.}, \bibinfo{author}{{M{\" u}nzenmayer}, R.},
  \bibinfo{author}{{Naumann}, W.}, \bibinfo{author}{{Hofschuster}, G.},
  \bibinfo{author}{{Born}, M.}, \bibinfo{author}{{F{\" a}rber}, G.},
  \bibinfo{author}{{Fischer}, F.}, \bibinfo{author}{{Fujiwara}, A.},
  \bibinfo{author}{{Glasmachers}, A.}, \bibinfo{author}{{Gr{\" u}n}, E.},
  \bibinfo{author}{{Hamabe}, Y.}, \bibinfo{author}{{Iglseder}, H.},
  \bibinfo{author}{{Kawamura}, T.}, \bibinfo{author}{{Miyamoto}, H.},
  \bibinfo{author}{{Morishige}, K.}, \bibinfo{author}{{Mukai}, T.},
  \bibinfo{author}{{Naoi}, T.}, \bibinfo{author}{{Nogami}, K.},
  \bibinfo{author}{{Schwehm}, G.}, \bibinfo{author}{{Svedhem}, H.},
  \bibinfo{year}{2002}a.
\newblock \bibinfo{title}{{Observation of interplanetary and interstellar dust
  particles by Mars Dust Counter (MDC) on board NOZOMI}}.
\newblock \bibinfo{journal}{Advances in Space Research} \bibinfo{volume}{29},
  \bibinfo{pages}{1145--1153}.
\bibitem[{{Sasaki} et~al.(1999){Sasaki}, {Igenbergs}, {Ohashi},
  {M{\"u}enzenmayer}, {Naumann}, {Born}, {Farber}, {Fisher}, {Fujiwara},
  {Glasmachers}, {Gruen}, {Hamabe}, {Hofschuster}, {Iglseder}, {Miyamoto},
  {Morishige}, {Mukai}, {Nogami}, {Schwehm}, {Svedhem} and
  {Yamakoshi}}]{sasaki1999a}
\bibinfo{author}{{Sasaki}, S.}, \bibinfo{author}{{Igenbergs}, E.},
  \bibinfo{author}{{Ohashi}, H.}, \bibinfo{author}{{M{\"u}enzenmayer}, R.},
  \bibinfo{author}{{Naumann}, W.}, \bibinfo{author}{{Born}, M.},
  \bibinfo{author}{{Farber}, G.}, \bibinfo{author}{{Fisher}, F.},
  \bibinfo{author}{{Fujiwara}, A.}, \bibinfo{author}{{Glasmachers}, A.},
  \bibinfo{author}{{Gruen}, E.}, \bibinfo{author}{{Hamabe}, Y.},
  \bibinfo{author}{{Hofschuster}, G.}, \bibinfo{author}{{Iglseder}, H.},
  \bibinfo{author}{{Miyamoto}, H.}, \bibinfo{author}{{Morishige}, K.},
  \bibinfo{author}{{Mukai}, T.}, \bibinfo{author}{{Nogami}, K.},
  \bibinfo{author}{{Schwehm}, G.}, \bibinfo{author}{{Svedhem}, H.},
  \bibinfo{author}{{Yamakoshi}, K.}, \bibinfo{year}{1999}.
\newblock \bibinfo{title}{{Initial Results of Mars Dust Counter (MDC) on Board
  NOZOMI: Leonids Encounter}}, in: \bibinfo{booktitle}{Lunar and Planetary
  Science Conference}, p. \bibinfo{pages}{1581}.
\bibitem[{{Sasaki} et~al.(2002b){Sasaki}, {Igenbergs}, {Ohashi}, {Senger},
  {Hofschuster}, {M{\"u}nzenmayer}, {Naumann}, {Gr{\"u}n}, {Fujiwara},
  {Hamabe}, {Mann}, {Miyamoto}, {Mukai}, {Nogami}, {Shoji} and
  {Svedhem}}]{sasaki2002b}
\bibinfo{author}{{Sasaki}, S.}, \bibinfo{author}{{Igenbergs}, E.},
  \bibinfo{author}{{Ohashi}, H.}, \bibinfo{author}{{Senger}, R.},
  \bibinfo{author}{{Hofschuster}, G.}, \bibinfo{author}{{M{\"u}nzenmayer}, R.},
  \bibinfo{author}{{Naumann}, W.}, \bibinfo{author}{{Gr{\"u}n}, E.},
  \bibinfo{author}{{Fujiwara}, A.}, \bibinfo{author}{{Hamabe}, Y.},
  \bibinfo{author}{{Mann}, I.}, \bibinfo{author}{{Miyamoto}, H.},
  \bibinfo{author}{{Mukai}, T.}, \bibinfo{author}{{Nogami}, K.},
  \bibinfo{author}{{Shoji}, S.}, \bibinfo{author}{{Svedhem}, H.},
  \bibinfo{year}{2002}b.
\newblock \bibinfo{title}{{Interplanetary and interstellar dust observations by
  Mars dust counter on board NOZOMI: four-year operation}}, in:
  \bibinfo{editor}{{Warmbein}, B.} (Ed.), \bibinfo{booktitle}{Asteroids,
  Comets, and Meteors: ACM 2002}, pp. \bibinfo{pages}{79--82}.
\bibitem[{{Sasaki} et~al.(2007){Sasaki}, {Igenbergs}, {Ohashi}, {Senger},
  {M{\"u}nzenmayer}, {Naumann}, {Gr{\"u}n}, {Nogami}, {Mann} and
  {Svedhem}}]{sasaki2007}
\bibinfo{author}{{Sasaki}, S.}, \bibinfo{author}{{Igenbergs}, E.},
  \bibinfo{author}{{Ohashi}, H.}, \bibinfo{author}{{Senger}, R.},
  \bibinfo{author}{{M{\"u}nzenmayer}, R.}, \bibinfo{author}{{Naumann}, W.},
  \bibinfo{author}{{Gr{\"u}n}, E.}, \bibinfo{author}{{Nogami}, K.},
  \bibinfo{author}{{Mann}, I.}, \bibinfo{author}{{Svedhem}, H.},
  \bibinfo{year}{2007}.
\newblock \bibinfo{title}{{Summary of interplanetary and interstellar dust
  observation by Mars Dust Counter on board NOZOMI}}.
\newblock \bibinfo{journal}{Advances in Space Research} \bibinfo{volume}{39},
  \bibinfo{pages}{485--488}.
\newblock \DOIprefix\doi{10.1016/j.asr.2006.11.006}.
\bibitem[{{Senger}(2007)}]{senger2007}
\bibinfo{author}{{Senger}, R.}, \bibinfo{year}{2007}.
\newblock \bibinfo{title}{{Data handling and evaluation for autonomous
  experiments in interplanetary missions}}.
\newblock Ph.D. thesis. Munich University of Technology, Germany,
  https://mediatum.ub.tum.de/node?id=625799.
\bibitem[{{Showalter}(2017)}]{showalter2017}
\bibinfo{author}{{Showalter}, M.R.}, \bibinfo{year}{2017}.
\newblock \bibinfo{title}{{Dust at the Martian moons and in the circummartian
  space}}.
\newblock \bibinfo{journal}{The Dust, Atmospheres and Plasma Environment of the
  Moon and Small Bodies (DAP-2017) Workshop} ,
  \bibinfo{pages}{https://impact.colorado.edu/dap/2017/abstracts/mark\_showalter.pdf}.
\bibitem[{{Showalter} et~al.(2006){Showalter}, {Hamilton} and
  {Nicholson}}]{showalter2006}
\bibinfo{author}{{Showalter}, M.R.}, \bibinfo{author}{{Hamilton}, D.P.},
  \bibinfo{author}{{Nicholson}, P.D.}, \bibinfo{year}{2006}.
\newblock \bibinfo{title}{{A deep search for Martian dust rings and inner moons
  using the Hubble Space Telescope}}.
\newblock \bibinfo{journal}{Planetary and Space Science} \bibinfo{volume}{54},
  \bibinfo{pages}{844--854}.
\newblock \DOIprefix\doi{10.1016/j.pss.2006.05.009}.
\bibitem[{{Simolka} et~al.(2024){Simolka}, {Blanco}, {Ingerl}, {Kr\"uger},
  {Sommer}, {Srama}, {Strack}, {Wagner}, {Arai}, {Bauer}, {Fr\"ohlich},
  {Gl\"aser}, Gr\"a{\ss}lin, {Henselowsky}, {Hillier}, {Hirai}, {Ito}, {Kempf},
  {Khawaja}, {Kimura}, {Klinkner}, {Kobayashi}, {Lengowski}, {Li}, {Mocker},
  {Moragas-Klostermeyer}, {Postberg}, {Rieth}, {Sasaki}, {Schmidt}, {Sterken},
  {Sternovsky}, {Strub}, {Trieloff}, {Szalay} and {Yabuta}}]{simolka2024}
\bibinfo{author}{{Simolka}, J.}, \bibinfo{author}{{Blanco}, R.},
  \bibinfo{author}{{Ingerl}, S.}, \bibinfo{author}{{Kr\"uger}, H.},
  \bibinfo{author}{{Sommer}, M.}, \bibinfo{author}{{Srama}, R.},
  \bibinfo{author}{{Strack}, H.}, \bibinfo{author}{{Wagner}, C.},
  \bibinfo{author}{{Arai}, T.}, \bibinfo{author}{{Bauer}, M.},
  \bibinfo{author}{{Fr\"ohlich}, P.}, \bibinfo{author}{{Gl\"aser}, J.},
  \bibinfo{author}{Gr\"a{\ss}lin, M.}, \bibinfo{author}{{Henselowsky}, C.},
  \bibinfo{author}{{Hillier}, J.}, \bibinfo{author}{{Hirai}, T.},
  \bibinfo{author}{{Ito}, M.}, \bibinfo{author}{{Kempf}, S.},
  \bibinfo{author}{{Khawaja}, N.}, \bibinfo{author}{{Kimura}, H.},
  \bibinfo{author}{{Klinkner}, S.}, \bibinfo{author}{{Kobayashi}, M.},
  \bibinfo{author}{{Lengowski}, M.}, \bibinfo{author}{{Li}, Y.},
  \bibinfo{author}{{Mocker}, A.}, \bibinfo{author}{{Moragas-Klostermeyer}, G.},
  \bibinfo{author}{{Postberg}, F.}, \bibinfo{author}{{Rieth}, F.},
  \bibinfo{author}{{Sasaki}, S.}, \bibinfo{author}{{Schmidt}, J.},
  \bibinfo{author}{{Sterken}, V.J.}, \bibinfo{author}{{Sternovsky}, Z.},
  \bibinfo{author}{{Strub}, P.}, \bibinfo{author}{{Trieloff}, M.},
  \bibinfo{author}{{Szalay}, J.}, \bibinfo{author}{{Yabuta}, H.},
  \bibinfo{year}{2024}.
\newblock \bibinfo{title}{{The DESTINY+ Dust Analyser - A Dust Telescope for
  Analysing Cosmic Dust Dynamics and Composition}}.
\newblock \bibinfo{journal}{Philosophical Transactions of the Royal Society A}
  ,
  \bibinfo{pages}{382:20230199}\DOIprefix\doi{doi.org/10.1098/rsta.2023.0199}.
\bibitem[{{Smith} et~al.(1999){Smith}, {Wilson}, {Baumgardner} and
  {Mendillo}}]{smith1999}
\bibinfo{author}{{Smith}, S.M.}, \bibinfo{author}{{Wilson}, J.K.},
  \bibinfo{author}{{Baumgardner}, J.}, \bibinfo{author}{{Mendillo}, M.},
  \bibinfo{year}{1999}.
\newblock \bibinfo{title}{{Discovery of the distant lunar sodium tail and its
  enhancement following the Leonid Meteor Shower of 1998}}.
\newblock \bibinfo{journal}{Geophysical Research Letters} \bibinfo{volume}{26},
  \bibinfo{pages}{1649--1652}.
\newblock \DOIprefix\doi{10.1029/1999GL900314}.
\bibitem[{{Soja} et~al.(2015){Soja}, {Sommer}, {Herzog}, {Agarwal}, {Rodmann},
  {Srama}, {Vaubaillon}, {Strub}, {Hornig}, {Bausch} and
  {Gr{\"u}n}}]{soja2015a}
\bibinfo{author}{{Soja}, R.H.}, \bibinfo{author}{{Sommer}, M.},
  \bibinfo{author}{{Herzog}, J.}, \bibinfo{author}{{Agarwal}, J.},
  \bibinfo{author}{{Rodmann}, J.}, \bibinfo{author}{{Srama}, R.},
  \bibinfo{author}{{Vaubaillon}, J.}, \bibinfo{author}{{Strub}, P.},
  \bibinfo{author}{{Hornig}, A.}, \bibinfo{author}{{Bausch}, L.},
  \bibinfo{author}{{Gr{\"u}n}, E.}, \bibinfo{year}{2015}.
\newblock \bibinfo{title}{{Characteristics of the dust trail of
  67P/Churyumov-Gerasimenko: an application of the IMEX model}}.
\newblock \bibinfo{journal}{Astronomy and Astrophysics} \bibinfo{volume}{583},
  \bibinfo{pages}{A18}.
\newblock \DOIprefix\doi{10.1051/0004-6361/201526184}.
\bibitem[{{Soter}(1971)}]{soter1971}
\bibinfo{author}{{Soter}, S.}, \bibinfo{year}{1971}.
\newblock \bibinfo{title}{{The dust belts of Mars}}.
\newblock \bibinfo{type}{Technical Report}. Center for Radiophysics and Space
  Research Report No. 462.
\bibitem[{{Srama} et~al.(2004){Srama}, {Ahrens}, {Auer}, {Bradley}, {Dikarev},
  {Economou}, {Fechtig}, {G\"orlich}, {Grande}, {Graps}, {Gr\"un}, {Havnes},
  {Helfert}, {Hor\'anyi}, {Igenbergs}, {Je{\ss}berger}, {Johnson}, {Kempf},
  {Krivov}, {Kr\"uger}, {Moragas-Klostermeyer}, {Lamy}, {Landgraf}, {Linkert},
  {Linkert}, {Lura}, {Mocker-Ahlreep}, {McDonnell}, {M\"ohlmann}, {Morfill},
  {M\"uller}, {Roy}, {Sch\"afer}, {Schlotzhauer}, {Schwehm}, {Spahn},
  {St\"ubig}, {Svestka}, {Tschernjawski}, {Tuzzolino}, {W\"asch} and
  {Zook}}]{srama2004}
\bibinfo{author}{{Srama}, R.}, \bibinfo{author}{{Ahrens}, T. J.~{Altobelli},
  N.}, \bibinfo{author}{{Auer}, S.}, \bibinfo{author}{{Bradley}, J.
  G.~{Burton}, M.}, \bibinfo{author}{{Dikarev}, V.V.},
  \bibinfo{author}{{Economou}, T.}, \bibinfo{author}{{Fechtig}, H.},
  \bibinfo{author}{{G\"orlich}, M.}, \bibinfo{author}{{Grande}, M.},
  \bibinfo{author}{{Graps}, A.L.}, \bibinfo{author}{{Gr\"un}, E.},
  \bibinfo{author}{{Havnes}, O.}, \bibinfo{author}{{Helfert}, S.},
  \bibinfo{author}{{Hor\'anyi}, M.}, \bibinfo{author}{{Igenbergs}, E.},
  \bibinfo{author}{{Je{\ss}berger}, E.K.}, \bibinfo{author}{{Johnson}, T.V.},
  \bibinfo{author}{{Kempf}, S.}, \bibinfo{author}{{Krivov}, A.V.},
  \bibinfo{author}{{Kr\"uger}, H.}, \bibinfo{author}{{Moragas-Klostermeyer},
  G.}, \bibinfo{author}{{Lamy}, P.}, \bibinfo{author}{{Landgraf}, M.},
  \bibinfo{author}{{Linkert}, D.}, \bibinfo{author}{{Linkert}, G.},
  \bibinfo{author}{{Lura}, F.}, \bibinfo{author}{{Mocker-Ahlreep}, A.},
  \bibinfo{author}{{McDonnell}, J.A.M.}, \bibinfo{author}{{M\"ohlmann}, D.},
  \bibinfo{author}{{Morfill}, G.E.}, \bibinfo{author}{{M\"uller}, M.},
  \bibinfo{author}{{Roy}, M.}, \bibinfo{author}{{Sch\"afer}, G.},
  \bibinfo{author}{{Schlotzhauer}, G.H.}, \bibinfo{author}{{Schwehm}, G.H.},
  \bibinfo{author}{{Spahn}, F.}, \bibinfo{author}{{St\"ubig}, M.},
  \bibinfo{author}{{Svestka}, J.}, \bibinfo{author}{{Tschernjawski}, V.},
  \bibinfo{author}{{Tuzzolino}, A.J.}, \bibinfo{author}{{W\"asch}, R.},
  \bibinfo{author}{{Zook}, H.A.}, \bibinfo{year}{2004}.
\newblock \bibinfo{title}{{The Cassini Cosmic Dust Analyzer}}.
\newblock \bibinfo{journal}{Space Science Reviews} \bibinfo{volume}{114},
  \bibinfo{pages}{465--518}.
\bibitem[{{Srama} et~al.(2011){Srama}, {Kempf}, {Moragas-Klostermeyer},
  {Altobelli}, {Auer}, {Beckmann}, {Bugiel}, {Burton}, {Economou}, {Fechtig},
  {Fiege}, {Green}, {Grande}, {Havnes}, {Hillier}, {Helfert}, {Horanyi}, {Hsu},
  {Igenbergs}, {Jessberger}, {Johnson}, {Khalisi}, {Kr\"uger}, {Matt},
  {Mocker}, {Lamy}, {Linkert}, {Lura}, {M\"ohlmann}, {Morfill}, {Otto},
  {Postberg}, {Roy}, {Schmidt}, {Schwehm}, {Spahn}, {Sterken}, {Svestka},
  {Tschernjawski}, {Gr\"un} and {R\"oser}}]{srama2011}
\bibinfo{author}{{Srama}, R.}, \bibinfo{author}{{Kempf}, S.},
  \bibinfo{author}{{Moragas-Klostermeyer}, G.}, \bibinfo{author}{{Altobelli},
  N.}, \bibinfo{author}{{Auer}, S.}, \bibinfo{author}{{Beckmann}, U.},
  \bibinfo{author}{{Bugiel}, S.}, \bibinfo{author}{{Burton}, M.},
  \bibinfo{author}{{Economou}, T.}, \bibinfo{author}{{Fechtig}, H.},
  \bibinfo{author}{{Fiege}, K.}, \bibinfo{author}{{Green}, S.F.},
  \bibinfo{author}{{Grande}, M.}, \bibinfo{author}{{Havnes}, O.},
  \bibinfo{author}{{Hillier}, J.K.}, \bibinfo{author}{{Helfert}, S.},
  \bibinfo{author}{{Horanyi}, M.}, \bibinfo{author}{{Hsu}, S.},
  \bibinfo{author}{{Igenbergs}, E.}, \bibinfo{author}{{Jessberger}, E.K.},
  \bibinfo{author}{{Johnson}, T.V.}, \bibinfo{author}{{Khalisi}, E.},
  \bibinfo{author}{{Kr\"uger}, H.}, \bibinfo{author}{{Matt}, G.},
  \bibinfo{author}{{Mocker}, A.}, \bibinfo{author}{{Lamy}, P.},
  \bibinfo{author}{{Linkert}, G.}, \bibinfo{author}{{Lura}, F.},
  \bibinfo{author}{{M\"ohlmann}, D.}, \bibinfo{author}{{Morfill}, G.},
  \bibinfo{author}{{Otto}, K.}, \bibinfo{author}{{Postberg}, F.},
  \bibinfo{author}{{Roy}, M.}, \bibinfo{author}{{Schmidt}, J.},
  \bibinfo{author}{{Schwehm}, G.H.}, \bibinfo{author}{{Spahn}, F.},
  \bibinfo{author}{{Sterken}, V.}, \bibinfo{author}{{Svestka}, J.},
  \bibinfo{author}{{Tschernjawski}, V.}, \bibinfo{author}{{Gr\"un}, E.},
  \bibinfo{author}{{R\"oser}, H.P.}, \bibinfo{year}{2011}.
\newblock \bibinfo{title}{{The Cosmic Dust Analyser onboard CASSINI: Ten years
  of discoveries}}.
\newblock \bibinfo{journal}{CEAS Space Journal} \bibinfo{volume}{2},
  \bibinfo{pages}{3--16}.
\newblock \DOIprefix\doi{10.1007/s12567-011-0014-x}.
\bibitem[{{Srama} et~al.(2006){Srama}, {Kempf}, {Moragas-Klostermeyer},
  {Helfert}, {Ahrens}, {Altobelli}, {Auer}, {Beckmann}, {Bradley}, {Burton},
  {Dikarev}, {Economou}, {Fechtig}, {Green}, {Grande}, {Havnes}, {Hillier},
  {Horanyi}, {Igenbergs}, {Jessberger}, {Johnson}, {Kr{\"u}ger}, {Matt},
  {McBride}, {Mocker}, {Lamy}, {Linkert}, {Linkert}, {Lura}, {McDonnell},
  {M{\"o}hlmann}, {Morfill}, {Postberg}, {Roy}, {Schwehm}, {Spahn}, {Svestka},
  {Tschernjawski}, {Tuzzolino}, {W{\"a}sch} and {Gr{\"u}n}}]{srama2006}
\bibinfo{author}{{Srama}, R.}, \bibinfo{author}{{Kempf}, S.},
  \bibinfo{author}{{Moragas-Klostermeyer}, G.}, \bibinfo{author}{{Helfert},
  S.}, \bibinfo{author}{{Ahrens}, T.J.}, \bibinfo{author}{{Altobelli}, N.},
  \bibinfo{author}{{Auer}, S.}, \bibinfo{author}{{Beckmann}, U.},
  \bibinfo{author}{{Bradley}, J.G.}, \bibinfo{author}{{Burton}, M.},
  \bibinfo{author}{{Dikarev}, V.V.}, \bibinfo{author}{{Economou}, T.},
  \bibinfo{author}{{Fechtig}, H.}, \bibinfo{author}{{Green}, S.F.},
  \bibinfo{author}{{Grande}, M.}, \bibinfo{author}{{Havnes}, O.},
  \bibinfo{author}{{Hillier}, J.K.}, \bibinfo{author}{{Horanyi}, M.},
  \bibinfo{author}{{Igenbergs}, E.}, \bibinfo{author}{{Jessberger}, E.K.},
  \bibinfo{author}{{Johnson}, T.V.}, \bibinfo{author}{{Kr{\"u}ger}, H.},
  \bibinfo{author}{{Matt}, G.}, \bibinfo{author}{{McBride}, N.},
  \bibinfo{author}{{Mocker}, A.}, \bibinfo{author}{{Lamy}, P.},
  \bibinfo{author}{{Linkert}, D.}, \bibinfo{author}{{Linkert}, G.},
  \bibinfo{author}{{Lura}, F.}, \bibinfo{author}{{McDonnell}, J.A.M.},
  \bibinfo{author}{{M{\"o}hlmann}, D.}, \bibinfo{author}{{Morfill}, G.E.},
  \bibinfo{author}{{Postberg}, F.}, \bibinfo{author}{{Roy}, M.},
  \bibinfo{author}{{Schwehm}, G.H.}, \bibinfo{author}{{Spahn}, F.},
  \bibinfo{author}{{Svestka}, J.}, \bibinfo{author}{{Tschernjawski}, V.},
  \bibinfo{author}{{Tuzzolino}, A.J.}, \bibinfo{author}{{W{\"a}sch}, R.},
  \bibinfo{author}{{Gr{\"u}n}, E.}, \bibinfo{year}{2006}.
\newblock \bibinfo{title}{{In situ dust measurements in the inner Saturnian
  system}}.
\newblock \bibinfo{journal}{Planetary and Space Science} \bibinfo{volume}{54},
  \bibinfo{pages}{967--987}.
\newblock \DOIprefix\doi{10.1016/j.pss.2006.05.021}.
\bibitem[{{Strub} et~al.(2015){Strub}, {Kr{\"u}ger} and {Sterken}}]{strub2015}
\bibinfo{author}{{Strub}, P.}, \bibinfo{author}{{Kr{\"u}ger}, H.},
  \bibinfo{author}{{Sterken}, V.J.}, \bibinfo{year}{2015}.
\newblock \bibinfo{title}{{Sixteen years of Ulysses interstellar dust
  measurements in the Solar System: II. Fluctuations in the dust flow from the
  data}}.
\newblock \bibinfo{journal}{Astrophysical Journal} \bibinfo{volume}{812},
  \bibinfo{pages}{140}.
\newblock \DOIprefix\doi{doi:10.1088/0004-637X/812/2/140}.
\bibitem[{{Svedhem} et~al.(1996){Svedhem}, {M\"unzenmeyer} and
  {Iglseder}}]{svedhem1996}
\bibinfo{author}{{Svedhem}, H.}, \bibinfo{author}{{M\"unzenmeyer}, R.},
  \bibinfo{author}{{Iglseder}, H.}, \bibinfo{year}{1996}.
\newblock \bibinfo{title}{{Detection of possible interstellar particles by the
  Hiten spacecraft}}, in: \bibinfo{editor}{{Gustafson}, B.A.S.},
  \bibinfo{editor}{{Hanner}, M.S.} (Eds.), \bibinfo{booktitle}{IAU Colloq. 150:
  Physics, Chemistry, and Dynamics of Interplanetary Dust}, pp.
  \bibinfo{pages}{27--30}.
\bibitem[{{Swaczyna} et~al.(2023){Swaczyna}, {Bzowski}, {Heerikhuisen},
  {Kubiak}, {Rahmanifard}, {Zirnstein}, {Fuselier}, {Galli}, {McComas},
  {M{\"o}bius} and {Schwadron}}]{swaczyna2023}
\bibinfo{author}{{Swaczyna}, P.}, \bibinfo{author}{{Bzowski}, M.},
  \bibinfo{author}{{Heerikhuisen}, J.}, \bibinfo{author}{{Kubiak}, M.A.},
  \bibinfo{author}{{Rahmanifard}, F.}, \bibinfo{author}{{Zirnstein}, E.J.},
  \bibinfo{author}{{Fuselier}, S.A.}, \bibinfo{author}{{Galli}, A.},
  \bibinfo{author}{{McComas}, D.J.}, \bibinfo{author}{{M{\"o}bius}, E.},
  \bibinfo{author}{{Schwadron}, N.A.}, \bibinfo{year}{2023}.
\newblock \bibinfo{title}{{Interstellar Conditions Deduced from Interstellar
  Neutral Helium Observed by IBEX and Global Heliosphere Modeling}}.
\newblock \bibinfo{journal}{Astrophysical Journal} \bibinfo{volume}{953},
  \bibinfo{pages}{107}.
\newblock \DOIprefix\doi{10.3847/1538-4357/ace719},
  \href{http://arxiv.org/abs/2307.06694}{{\tt arXiv:2307.06694}}.
\bibitem[{{Szalay} et~al.(2018){Szalay}, {Poppe}, {Agarwal}, {Britt},
  {Belskaya}, {Hor{\'a}nyi}, {Nakamura}, {Sachse} and {Spahn}}]{szalay2018}
\bibinfo{author}{{Szalay}, J.R.}, \bibinfo{author}{{Poppe}, A.R.},
  \bibinfo{author}{{Agarwal}, J.}, \bibinfo{author}{{Britt}, D.},
  \bibinfo{author}{{Belskaya}, I.}, \bibinfo{author}{{Hor{\'a}nyi}, M.},
  \bibinfo{author}{{Nakamura}, T.}, \bibinfo{author}{{Sachse}, M.},
  \bibinfo{author}{{Spahn}, F.}, \bibinfo{year}{2018}.
\newblock \bibinfo{title}{{Dust Phenomena Relating to Airless Bodies}}.
\newblock \bibinfo{journal}{Space Science Reviews} \bibinfo{volume}{214},
  \bibinfo{pages}{98}.
\newblock \DOIprefix\doi{10.1007/s11214-018-0527-0}.
\bibitem[{{Westphal} et~al.(2012){Westphal}, {Achilles}, {Allen}, {Ansari},
  {Bajt}, {Bassim}, {Bastien}, {Bechtel}, {Borg}, {Brenker}, {Bridges},
  {Brownlee}, {Burchell}, {Burghammer}, {Butterworth}, {Changela}, {Cloetens},
  {Davis}, {Floss}, {Flynn}, {Fougeray}, {Frank}, {Gainsforth}, {Gruen},
  {Heck}, {Hillier}, {Hoppe}, {Hudson}, {Huss}, {Huth}, {Hvide}, {Kearsley},
  {King}, {Lai}, {Leitner}, {Lemelle}, {Leonard}, {Leroux}, {Lettieri},
  {Marchant}, {Nittler}, {Ogliore}, {Postberg}, {Price}, {Sandford}, {Sans
  Tresseras}, {Schmitz}, {Schoonjans}, {Schreiber}, {Silversmit},
  {Simionovici}, {Sole}, {Srama}, {Stephan}, {Sterken}, {Stodolna}, {Stroud},
  {Sutton}, {Treiloff}, {Tsou}, {Tsuchiyama}, {Tyliszczak}, {Vekemans},
  {Vincze}, {Wordsworth}, {Zevin}, {Zolensky} and {$> 30,000$ Stardust@Home
  Dusters}}]{westphal2012}
\bibinfo{author}{{Westphal}, A.J.}, \bibinfo{author}{{Achilles}, C.},
  \bibinfo{author}{{Allen}, C.}, \bibinfo{author}{{Ansari}, A.},
  \bibinfo{author}{{Bajt}, S.}, \bibinfo{author}{{Bassim}, N.},
  \bibinfo{author}{{Bastien}, R.}, \bibinfo{author}{{Bechtel}, H.A.},
  \bibinfo{author}{{Borg}, J.}, \bibinfo{author}{{Brenker}, F.E.},
  \bibinfo{author}{{Bridges}, J.}, \bibinfo{author}{{Brownlee}, D.E.},
  \bibinfo{author}{{Burchell}, M.}, \bibinfo{author}{{Burghammer}, M.},
  \bibinfo{author}{{Butterworth}, A.}, \bibinfo{author}{{Changela}, H.},
  \bibinfo{author}{{Cloetens}, P.}, \bibinfo{author}{{Davis}, A.M.},
  \bibinfo{author}{{Floss}, C.}, \bibinfo{author}{{Flynn}, G.},
  \bibinfo{author}{{Fougeray}, P.}, \bibinfo{author}{{Frank}, D.},
  \bibinfo{author}{{Gainsforth}, Z.}, \bibinfo{author}{{Gruen}, E.},
  \bibinfo{author}{{Heck}, P.R.}, \bibinfo{author}{{Hillier}, J.K.},
  \bibinfo{author}{{Hoppe}, P.}, \bibinfo{author}{{Hudson}, B.},
  \bibinfo{author}{{Huss}, G.R.}, \bibinfo{author}{{Huth}, J.},
  \bibinfo{author}{{Hvide}, B.}, \bibinfo{author}{{Kearsley}, A.},
  \bibinfo{author}{{King}, A.J.}, \bibinfo{author}{{Lai}, B.},
  \bibinfo{author}{{Leitner}, J.}, \bibinfo{author}{{Lemelle}, L.},
  \bibinfo{author}{{Leonard}, A.}, \bibinfo{author}{{Leroux}, H.},
  \bibinfo{author}{{Lettieri}, R.}, \bibinfo{author}{{Marchant}, W.},
  \bibinfo{author}{{Nittler}, L.R.}, \bibinfo{author}{{Ogliore}, R.},
  \bibinfo{author}{{Postberg}, F.}, \bibinfo{author}{{Price}, M.C.},
  \bibinfo{author}{{Sandford}, S.A.}, \bibinfo{author}{{Sans Tresseras}, J.A.},
  \bibinfo{author}{{Schmitz}, S.}, \bibinfo{author}{{Schoonjans}, T.},
  \bibinfo{author}{{Schreiber}, K.}, \bibinfo{author}{{Silversmit}, G.},
  \bibinfo{author}{{Simionovici}, A.}, \bibinfo{author}{{Sole}, V.A.},
  \bibinfo{author}{{Srama}, R.}, \bibinfo{author}{{Stephan}, T.},
  \bibinfo{author}{{Sterken}, V.}, \bibinfo{author}{{Stodolna}, J.},
  \bibinfo{author}{{Stroud}, R.M.}, \bibinfo{author}{{Sutton}, S.},
  \bibinfo{author}{{Treiloff}, M.}, \bibinfo{author}{{Tsou}, P.},
  \bibinfo{author}{{Tsuchiyama}, A.}, \bibinfo{author}{{Tyliszczak}, T.},
  \bibinfo{author}{{Vekemans}, B.}, \bibinfo{author}{{Vincze}, L.},
  \bibinfo{author}{{Wordsworth}, N.}, \bibinfo{author}{{Zevin}, D.},
  \bibinfo{author}{{Zolensky}, M.E.}, \bibinfo{author}{{$> 30,000$
  Stardust@Home Dusters}}, \bibinfo{year}{2012}.
\newblock \bibinfo{title}{{Status of the Stardust ISPE and the Origin of Four
  Interstellar Dust Candidates}}, in: \bibinfo{booktitle}{Lunar and Planetary
  Institute Science Conference Abstracts}, p. \bibinfo{pages}{2084}.
\bibitem[{{Witte}(2004)}]{witte2004a}
\bibinfo{author}{{Witte}, M.}, \bibinfo{year}{2004}.
\newblock \bibinfo{title}{{Kinetic parameters of interstellar neutral helium.
  Review of results obtained during one solar cycle with the
  Ulysses/GAS-instrument}}.
\newblock \bibinfo{journal}{Astronomy and Astrophysics} \bibinfo{volume}{426},
  \bibinfo{pages}{835--844}.
\newblock \DOIprefix\doi{10.1051/0004-6361:20035956}.
\bibitem[{{Yamamoto} and {Tsuruda}(1998)}]{yamamoto1998b}
\bibinfo{author}{{Yamamoto}, T.}, \bibinfo{author}{{Tsuruda}, K.},
  \bibinfo{year}{1998}.
\newblock \bibinfo{title}{{The PLANET-B mission}}.
\newblock \bibinfo{journal}{Earth, Planets and Space} \bibinfo{volume}{50},
  \bibinfo{pages}{175--181}.
\newblock \DOIprefix\doi{10.1186/BF03352100}.
\bibitem[{{Yang} et~al.(2022){Yang}, {Schmidt}, {Feng} and {Liu}}]{yang2022}
\bibinfo{author}{{Yang}, K.}, \bibinfo{author}{{Schmidt}, J.},
  \bibinfo{author}{{Feng}, W.}, \bibinfo{author}{{Liu}, X.},
  \bibinfo{year}{2022}.
\newblock \bibinfo{title}{{Distribution of dust ejected from the lunar surface
  into the Earth-Moon system}}.
\newblock \bibinfo{journal}{Astronomy and Astrophysics} \bibinfo{volume}{659},
  \bibinfo{pages}{A120}.
\newblock \DOIprefix\doi{10.1051/0004-6361/202140810},
  \href{http://arxiv.org/abs/2204.01040}{{\tt arXiv:2204.01040}}.
\bibitem[{{Yoshikawa} et~al.(2005){Yoshikawa}, {Kawaguchi}, {Yamakawa}, {Kato},
  {Ichikawa}, {Ohnishi} and {Ishibashi}}]{yoshikawa2005}
\bibinfo{author}{{Yoshikawa}, M.}, \bibinfo{author}{{Kawaguchi}, J.},
  \bibinfo{author}{{Yamakawa}, H.}, \bibinfo{author}{{Kato}, T.},
  \bibinfo{author}{{Ichikawa}, T.}, \bibinfo{author}{{Ohnishi}, T.},
  \bibinfo{author}{{Ishibashi}, S.}, \bibinfo{year}{2005}.
\newblock \bibinfo{title}{{Summary of the orbit determination of NOZOMI
  spacecraft for all the mission period}}.
\newblock \bibinfo{journal}{Acta Astronautica} \bibinfo{volume}{57},
  \bibinfo{pages}{510--519}.
\newblock \DOIprefix\doi{10.1016/j.actaastro.2005.03.053}.
\bibitem[{{Zakharov} et~al.(2014){Zakharov}, {Horanyi}, {Lee}, {Witasse} and
  {Cipriani}}]{zakharov2014}
\bibinfo{author}{{Zakharov}, A.}, \bibinfo{author}{{Horanyi}, M.},
  \bibinfo{author}{{Lee}, P.}, \bibinfo{author}{{Witasse}, O.},
  \bibinfo{author}{{Cipriani}, F.}, \bibinfo{year}{2014}.
\newblock \bibinfo{title}{{Dust at the Martian moons and in the circummartian
  space}}.
\newblock \bibinfo{journal}{Planetary and Space Science} \bibinfo{volume}{102},
  \bibinfo{pages}{171--175}.
\newblock \DOIprefix\doi{10.1016/j.pss.2013.12.011}.

\end{thebibliography}

\end{document}